\documentclass{aa} 
\usepackage{graphicx}
\usepackage{lscape}
\usepackage{longtable}
\usepackage{arydshln}
\usepackage{multirow}
\usepackage{enumitem}
\usepackage{epstopdf}
\usepackage{natbib}
\usepackage{txfonts}

\bibpunct{(}{)}{;}{a}{}{,} % to follow the A&A style

\begin{document}
\title{\textit{Herschel}-PACS observations of [OI] and $\rm H_{2}O$ in Cha II.\thanks{{\it Herschel} is an ESA space observatory with
    science instruments provided by European-led Principal
    Investigator consortia and with important participation from
    NASA.}}
%\subtitle{}
   \author{P. Riviere-Marichalar\inst{1,2,3}, A. Bayo\inst{4,5}, I. Kamp\inst{1}, S. Vicente\inst{1}, J. P. Williams\inst{6}, D. Barrado\inst{7}, C. Eiroa\inst{2}, G. Duch\^ene\inst{10,11}, B. Montesinos\inst{7}, G. Mathews\inst{8,9}, L. Podio\inst{12}, W. R. F. Dent\inst{13}, N. Hu\'{e}lamo\inst{7}, B. Mer\'in\inst{3}
}

   \institute{Kapteyn Astronomical Institute, University of Groningen, P.O. Box 800, 9700 AV Groningen, The Netherlands %1
                \email{riviere@astro.rug.nl}
   \and Depto. de F\'isica Te\'orica, Fac. de Ciencias, UAM Campus Cantoblanco, 28049 Madrid, Spain %2    
   \and European Space Astronomy Centre (ESA), P.O. Box 78, 28691 Villanueva de la Ca\~nada, Spain%3         
   \and Max Planck Institut f\"ur Astronomie, K\"onigstuhl 17, 69117, Heidelberg, Germany %4
   \and Instituto de F\'isica y Astronom\'ia, Facultad de Ciencias, Universidad de Valpara\'iso, Av. Gran Breta\~na 1111, Playa Ancha, Valpara\'iso, Chile %5
   \and Institute for Astronomy, University of Hawaii at Manoa, Honolulu, HI, 96822, USA %6
   \and Centro de Astrobiolog\'{\i}a (INTA--CSIC) -- Depto. Astrof\'isica, POB 78, ESAC Campus, 28691 Villanueva de la Ca\~nada, Spain %7
   \and Leiden Observatory, Leiden University, PO Box 9513, 2300 RA Leiden, The Netherlands%8
   \and University of Hawaii, Department of Physics and Astronomy, 2505 Correa Rd.  Honolulu, Hawaii 96822%9
   \and Astronomy Department, University of California, Berkeley CA 94720-3411 USA %10
   \and Univ. Grenoble Alpes, IPAG, F-38000 Grenoble, France \\ CNRS, IPAG, F-38000 Grenoble, France %11	
   \and INAF-Osservatorio Astrofisico di Arcetri, Largo E. Fermi 5, I-50125, Florence, Italy %12
   \and ALMA, Avda Apoquindo 3846, Piso 19, Edificio Alsacia, Las Condes, Santiago, Chile %13
}
   \authorrunning{Riviere-Marichalar et al.}
   \date{}

 \abstract
%context heading (optional)
{Gas plays a major role in the dynamical evolution of protoplanetary discs. Its coupling with the dust is the key to our understanding planetary formation.  Studying the gas content is therefore a crucial step towards understanding protoplanetary discs evolution. Such a study can be made through spectroscopic observations of emission lines in the far-infrared, where some of the most important gas coolants emit, such as the [OI] $\rm ^{3}P_{1} \rightarrow ^{3}P_{2}$ transition at 63.18 $\rm \mu m$. }
{We aim at characterising the gas content of protoplanetary discs in the intermediate-aged (from the perspective of the disc lifetime) Chamaeleon II (Cha II) star forming region. We also aim at characterising the gaseous detection fractions within this age range, which is an essential  step tracing gas evolution with age in different star forming regions. This evolutionary study can be used to tackle the problem of the gas dispersal timescale in future studies.} 
{We obtained \textit{Herschel}-PACS line scan spectroscopic observations at 63 $\rm \mu m$ of 19 Cha II Class I and II stars. The observations were used to trace [OI] and o-$\rm H_{2}O$ at 63 $\rm \mu m$. The analysis of the spatial distribution of [OI], when extended, can be used to understand the origin of the emission.}
{We have detected [OI] emission toward seven out of the nineteen systems observed, and o-$\rm H_{2}O$ emission at 63.32 $\rm \mu m$ in just one of them, Sz 61. Cha II members show a correlation between [OI] line fluxes and the continuum at 70 $\rm \mu m$, similar to what is observed in Taurus. We analyse the extended [OI] emission towards the star DK Cha and study its dynamical footprints in the PACS Integral Field Unit (IFU). We conclude that there is a high velocity component from a jet combined with a low velocity component with an origin that may be a combination of disc, envelope and wind emission. The stacking of spectra of objects not detected individually in [OI] leads to a marginal 2.6$\sigma$ detection that may indicate the presence of gas just below our detection limits for some, if not all, of them.}  
{}
\keywords{Stars: Circumstellar matter, Stars: evolution, astrochemistry, protoplanetary disks, \object{DK~Cha}}
   \maketitle
   \begin{table*}[!t]
\caption{Line fluxes for Cha II members observed with \textit{Herschel}-PACS. }             
\label{lineFluxes}      
\centering          
\begin{tabular}{llllllcccc}     % 6 columns 
\hline\hline       
Name & RA & DEC & Sp. Type & Class & OBSID & [OI] flux & $\rm FWHM_{[OI]}$ & $\rm o-H_{2}O$ flux & $\rm FWHM_{H_{2}O}$\\ 
 & & & & & & @ 63.185 $\rm \mu m$ & @63.324 $\rm \mu m$ & & \\
\hline
-- & -- & -- & -- & -- & -- & ($\rm 10^{-17}W/m^{2}$) & ($\rm \mu m$) & ($\rm 10^{-17}W/m^{2}$) &  ($\rm \mu m$)  \\ 
\hline                    
\object{DK~Cha} &   12 53 17.23  &  -77 07 10.7 &  F0 & I-II & 1342226006 & $\rm 190.0 \pm 0.7^{*}$& -- &  $\rm < 2.9$ & --\\ 
  &     &  &   &  &  & $\rm 384 \pm 2 ^{**}$ & & & \\
\object{IRAS~12500-7658} &    12 53 42.86   & -77 15 11.5  & M6.5 & I & 1342226005 & $\rm 5.04 \pm 0.25$ & 0.021 & $\rm < 0.55 $ & -- \\
\object{Sz~46N} & 12 56 33.66  &  -76 45 45.3 & M1 & II  &  1342226009 & $\rm < 0.75$ & -- & $\rm < 0.75 $ & --\\
\object{IRAS~12535-7623} &  12 57 11.77  &  -76 40 11.3  &   M0 & II & 1342226010 & $\rm < 0.91 $ & -- & $\rm < 0.91$ & -- \\
\object{ISO~ChaII~13}    &      12 58 06.78  & -77 09 09.4  &   M7 &  II    &  1342228419 & $\rm < 0.69$ & -- & $\rm < 0.69$ & -- \\
\object{Sz~50} &  13 00 55.36 &  -77 10 22.1   &  M3 & II & 1342226008 & $\rm < 0.65$ & -- & $\rm < 0.65$ & -- \\
\object{Sz~51}$\dagger$ & 13 01 58.94  &  -77 51 21.7  & K8.5 & II & 1342229799 & $\rm 0.43 \pm 0.14$ & 0.012 & $\rm < 0.68$ & -- \\        
\object{[VCE2001]~C50}  & 13 02 22.85  &  -77 34 49.3   &  M5 &  II  & 1342227072  & $\rm < 0.66$ & -- & $\rm < 0.66$ & -- \\     
\object{Sz~52}  & 13 04 24.92   & -77 52 30.1  & M2.5 & II & 1342229800 & $\rm 1.2 \pm 0.3$ & 0.024 & $\rm < 0.73$ & --\\    
\object{Hn~25} & 13 05 08.53 &   -77 33 42.4 &  M2.5 & II & 1342227071 & $\rm < 0.75$ & -- & $\rm < 0.75$ & -- \\        
\object{Sz~53}   & 13 05 12.69  &  -77 30 52.3  & M1 & II &   1342229829  & $\rm < 0.76$ & -- & $\rm < 0.76$ & -- \\        
\object{Sz~54}  & 13 05 20.68 &   -77 39 01.4  & K5 & II &  1342229831 & $\rm 1.5 \pm 0.4$ & 0.032 & $\rm < 0.74$  & --\\        
\object{J13052169-7738102}  & 13 05 21.66  &  -77 38 10.0   &  -- & Flat & 1342229830 & $\rm 0.81 \pm 0.23$ & 0.018 & $\rm < 0.68$ & --\\       
\object{J13052904-7741401}  & 13 05 29.04  &  -77 41 40.1  & -- & II  & 1342229832 & $\rm < 0.78$ & -- &  $\rm < 0.78$ & --\\             
\object{[VCE2001]~C62}     & 13 07 18.05&    -77 40 52.9  & M4.5 & II & 1342229833 & $\rm < 0.70$ & -- & $\rm < 0.70$ & -- \\                 
\object{Hn~26}$\rm ^{*}$ &       13 07 48.51&    -77 41 21.4  & M2 & II  & 1342228418 & $\rm < 0.71$ & -- &  $\rm < 0.71$ & --\\                        
\object{Sz~61}  & 13 08 06.28  &  -77 55 05.2 &  K5 & II  & 1342229801 & $\rm 0.87 \pm 0.22$ & 0.028  & $\rm 0.56 \pm 0.15$ & 0.012 \\                       
\object{[VCE2001]~C66}  & 13 08 27.17  &  -77 43 23.2  &  M4.5 & II & 1342228419 & $\rm < 0.72$ & -- & $\rm < 0.72$ & -- \\                        
\object{Sz~62}  &  13 09 50.38  &  -77 57 23.9  &  M2.5 & II & 1342228420 & $\rm < 0.63$ & -- & $\rm < 0.63$ & -- \\                       
\hline                  
\end{tabular}
\tablefoot{Spectral types taken from \cite{Spezzi2008}. SED classes \citep[after][]{Lada1987} from \cite{Alcala2008}. FWHMs are not corrected for instrumental resolution. (*): line flux from integration of the observed central spaxel spectrum. (**): line flux integration of the 5x5 co-added spectrum. ($\dagger$): the FWHM of the Gaussian fit is slightly smaller than the instrumental value, therefore the upper limit, computed using instrumental FWHM, is larger than the detected flux.}
\end{table*}

\section{Introduction} 
Protoplanetary discs made of gas and dust are a natural product of the star formation process, and are the birthplace of planets. Studying the evolution of gas and dust in these systems is crucial for our understanding of planetary formation. In the past 30 years, the study of protoplanetary discs has led to a deeper understanding of the dust physics \citep[see the recent review by][and references therein]{Williams2011}. Although the gas is expected to be the main mass reservoir in the disc, it is less well characterised because it is harder to detect. Besides, a profound understanding of the chemistry is needed. The complexity of chemistry models and limited coverage and sensitivity of the observations preclude reaching a complete picture for any disc. Knowledge of the gas phase is important not only for deriving the total mass of the discs:  but it has also been recently highlighted by \cite{Pinte2014} that the gas-to-dust ratio can influence the dust distribution in the disc and the process of grain growth and planet formation. Furthermore, the evolution of the gas content with age can help us to derive the gas-clearing timescale, putting crucial constraints on theories of planet formation.

Young clusters and stellar associations are the natural laboratories for studying the evolution of protoplanetary discs. Their dust dissipates in the first 10 Myr \citep[see e. g.][]{Haisch2001,Mamajek2009}, and therefore observations of protoplanetary discs are restricted to young stellar associations and star forming regions. Besides, the age of their members is well known, so that evolutionary studies can be performed. We can use associations at different ages to get snapshots of the gas and dust properties with age, and put them together to get a picture of gas and dust evolution.

Cha II is a low-mass star forming region placed 178$\rm \pm$18 pc away from the Sun \citep{Whittet1997}, with an age of $\rm 4\pm2~ Myr$ \citep{Spezzi2008}. It is one of the major star forming regions within 200 pc, with a disc fraction of  73\% \citep{Evans2009}. The most recent census of Cha II members by \cite{Alcala2008} has extended the mass range down to $\rm  0.03~M_{\odot}$. A detailed description of the different spectroscopic and photometric surveys of Cha II members can be found in \cite{Spezzi2008}. The region characteristics make it, together with Taurus, a good probe for studying the decline of gas and dust contents in the crucial period between 1 and 5 Myr \citep[see e. g][]{Sicilia2006,Howard2013}.

The \textit{Herschel Space Observatory} \citep{Pilbratt2010} provided a unique opportunity to study protoplanetary discs, allowing us to systematically study  gas emission in the far-infrared (FIR). The \textit{Photodetector Array Camera \& Spectrometer} \citep[PACS,][]{Poglitsch2010} was widely used to survey gas in protoplanetary discs, typically focusing on detecting the [OI] $\rm ^{3}P_{1} \rightarrow ^{3}P_{2}$ transition at 63.18 $\rm \mu m$, the strongest far-IR line observed towards these systems. 

In this study we present PACS spectroscopic observations of 19 Cha II members to detect [OI] emission at 63.18 $\rm \mu m$ as part of the GASPS programme  \citep[from GAs Survey of Protoplanetary Systems,][]{Dent2013}. The paper is structured as follows. In Sec. \ref{sec:Sample} we describe the  sample. In Sec. \ref{sec:obsDat} we describe our observations and the techniques used to reduce the data. In Sec. \ref{sec:resDisc} we present our results and discuss the implications. Finally in Sec. \ref{ref:SumConc} we give an overview of the contents and results of this work. 

\section{Sample}\label{sec:Sample}
The \textit{Herschel Space Observatory} Open Time Key Programme GASPS targeted more than 250 stars in seven associations with ages in the range 1--30 Myr to perform an evolutionary study of gas and dust properties in circumstellar environments.  As part of the project, a total of 19 Cha II members were observed with PACS in LineScan mode intending to detect [OI] emission at 63.18 $\rm \mu m$ (see Table \ref{lineFluxes} for a list of the observed stars). The age of the Cha II star forming region is intermediate between Taurus and Upper Scorpius, filling a particularly interesting period in disc evolution, where a rapid decline in gas content is expected. This decline is observed when the [OI] detection ratios for Taurus \citep[$\rm \sim 57\%$,][]{Howard2013} and Upper Scorpius \citep[$\rm \sim$11\%, see][]{Mathews2013} are compared. 

Our source list was chosen from the \textit{Spitzer Legacy} cores-to-discs (c2d) survey of the region \citep{Evans2009} and contains young stellar objects with strong infrared excess in the 2--25 $\rm \mu m$ range. The sources were chosen to be  characterised well through photometry from X-rays to the far-infrared and optical spectra \citep{Alcala2008} to allow for detailed characterisation of the disc in future studies. They represent more than 40\% of the total population of Cha II members with discs. The number of selected sources is low due to time limitations in the GASPS programme. We selected mainly Class II objects to study the gas content in their discs. According to the classification by  \cite{Alcala2008}, there is one Class I object, namely IRAS 12500-7658, one flat-spectrum source, J13052169-7738102, one object in the transition from Class I to Class II, DK Cha, plus  16 Class II objects. Of the total sample, seven sources have been confirmed as members by \cite{LopezMarti2013} through a study of their proper motions. Those sources are Sz 46N, IRAS 12535-7623, Sz 50, Sz 51, Sz 54, Sz 61 and Sz 62. Objects lacking a proper motion study are most likely members given their strong mid- and far-IR excess.

\section{Observations and data reduction}\label{sec:obsDat}

\begin{figure*}[!t]
\begin{center}
%   \centering
\includegraphics[scale=0.3,trim=8mm 24mm 0mm 0mm,clip]{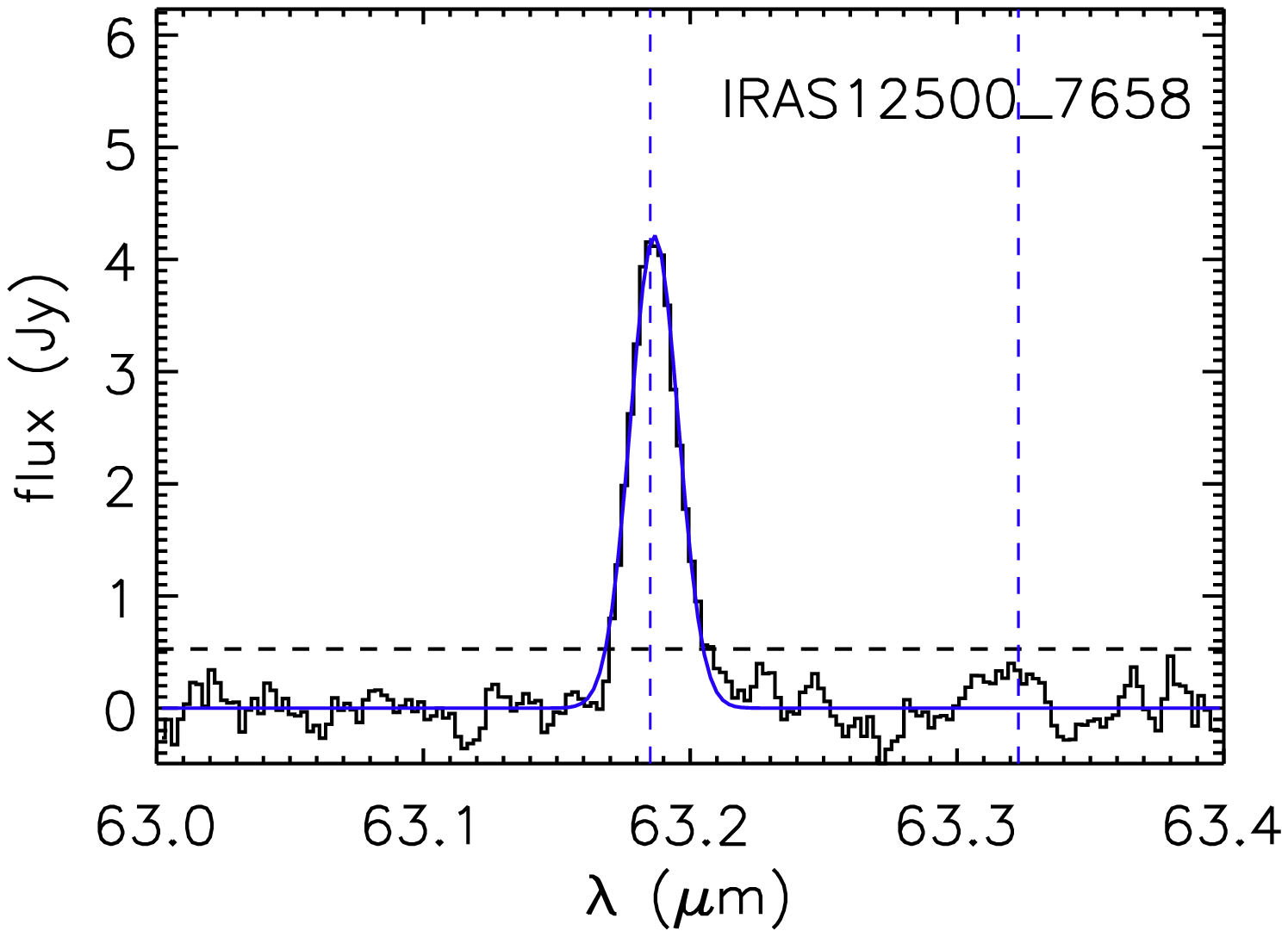}
\includegraphics[scale=0.3,trim=24mm 24mm 0mm 0mm,clip]{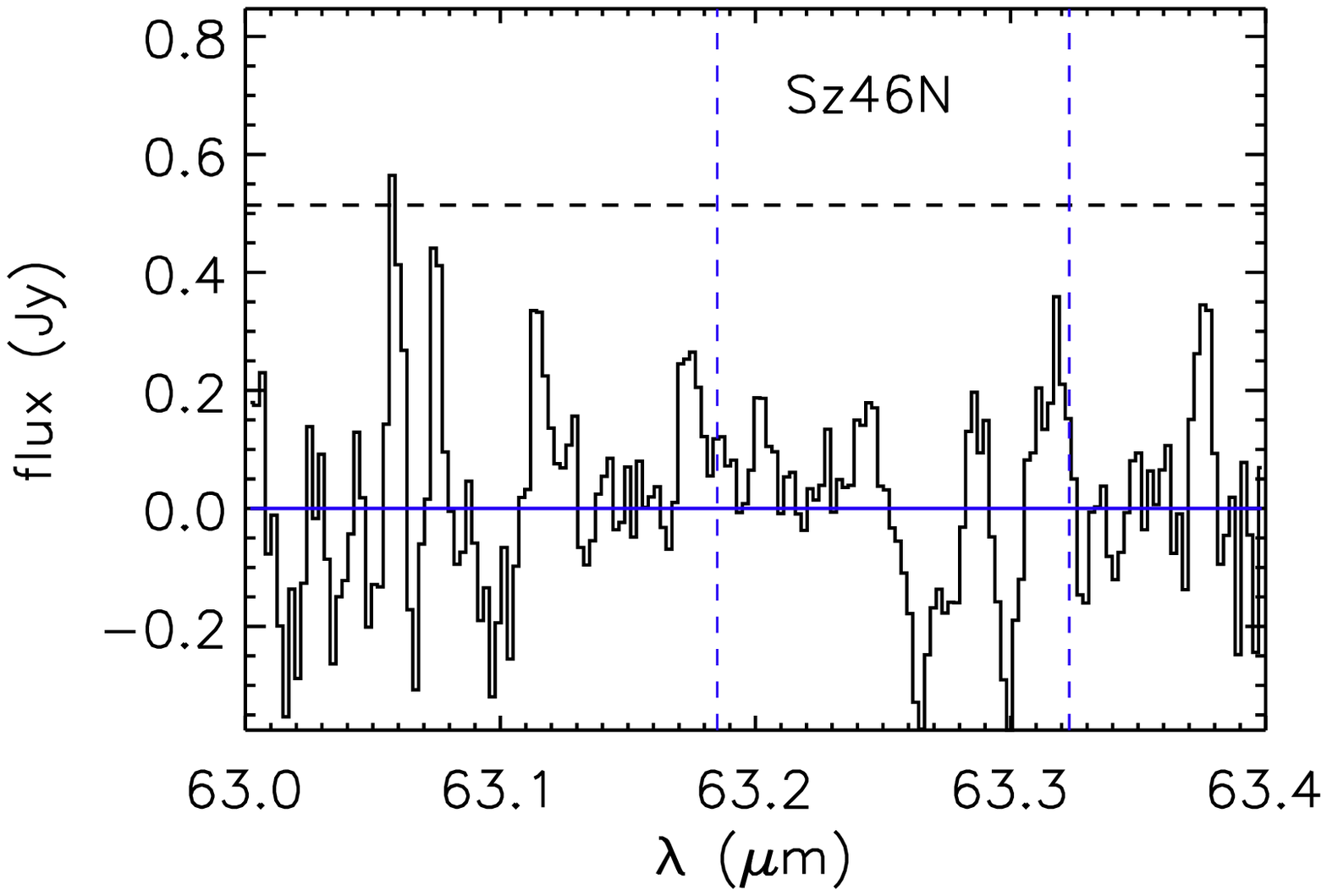}
\includegraphics[scale=0.3,trim=24mm 24mm 0mm 0mm,clip]{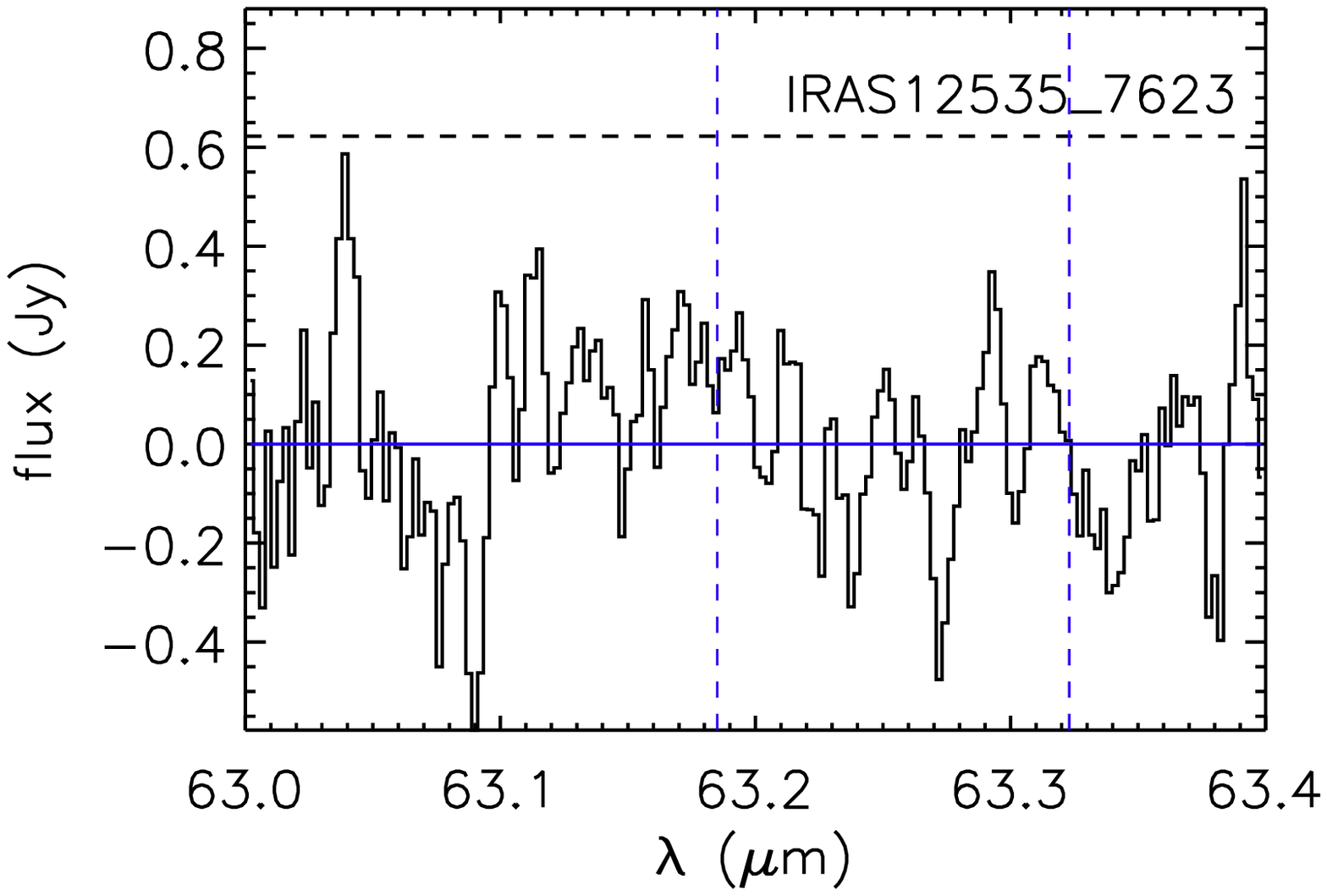}
\includegraphics[scale=0.3,trim=8mm 24mm 0mm 10mm,clip]{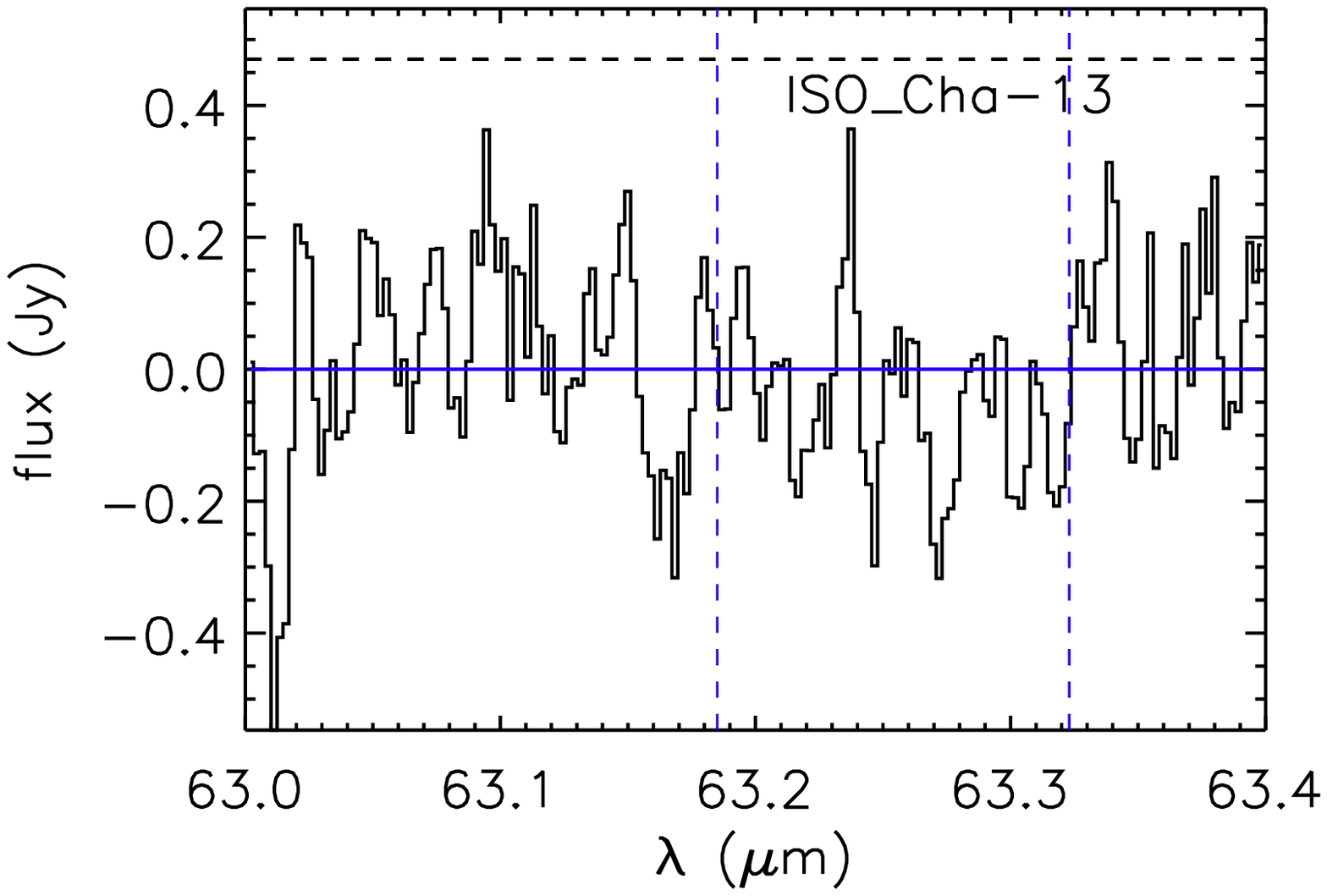}
\includegraphics[scale=0.3,trim=24mm 24mm 0mm 10mm,clip]{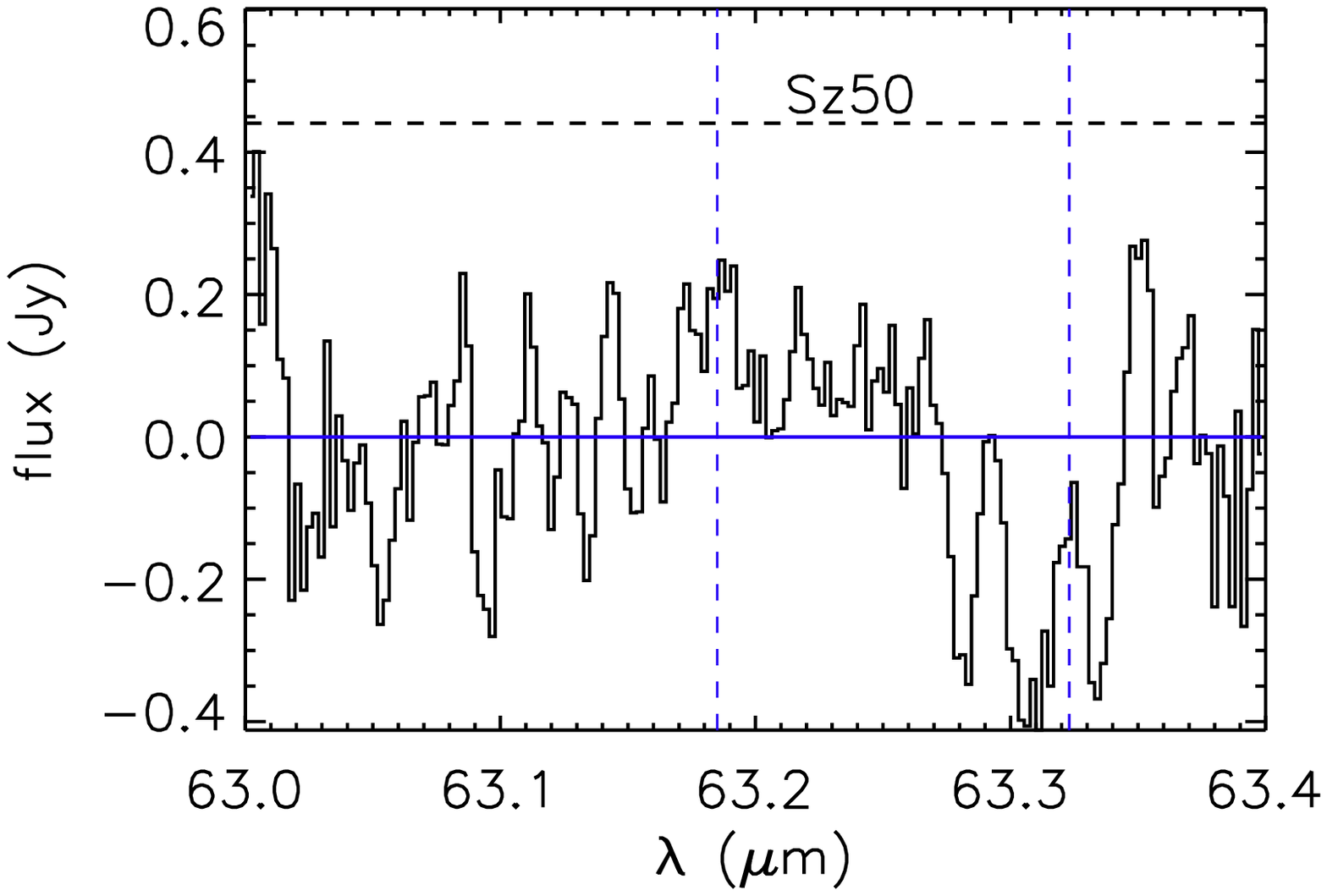}
\includegraphics[scale=0.3,trim=24mm 24mm 0mm 10mm,clip]{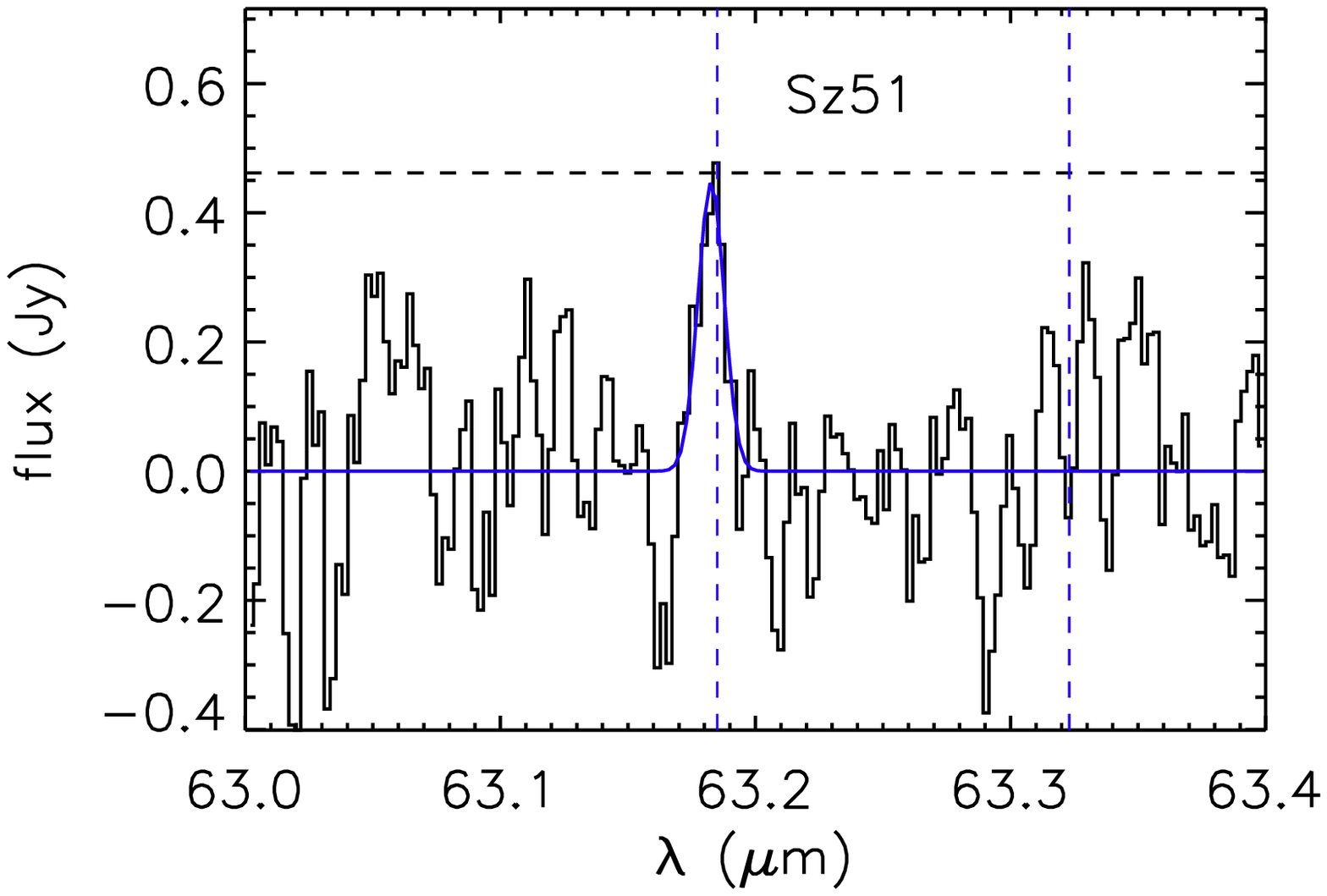}\\
\includegraphics[scale=0.3,trim=8mm 24mm 0mm 10mm,clip]{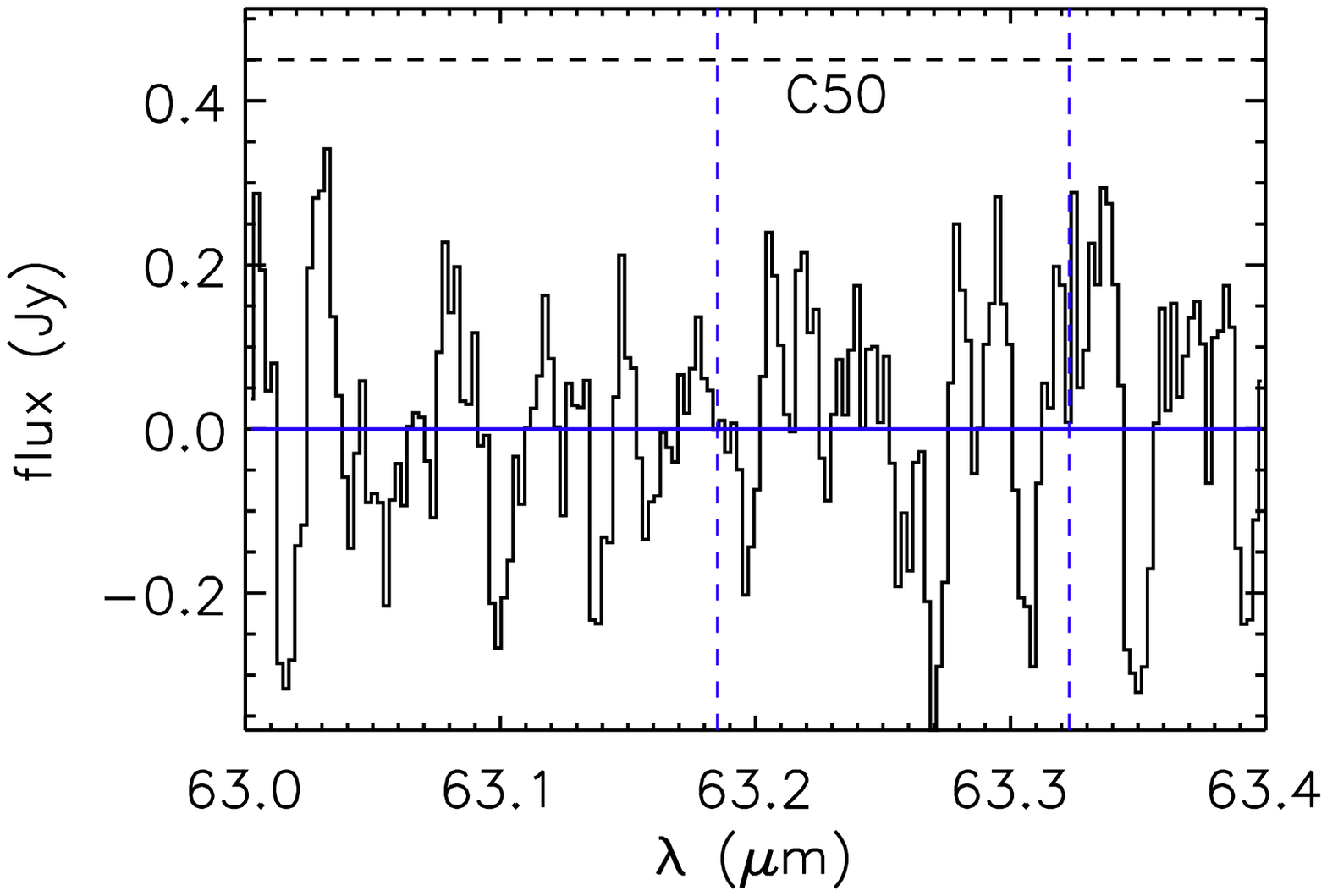}
\includegraphics[scale=0.3,trim=24mm 24mm 0mm 10mm,clip]{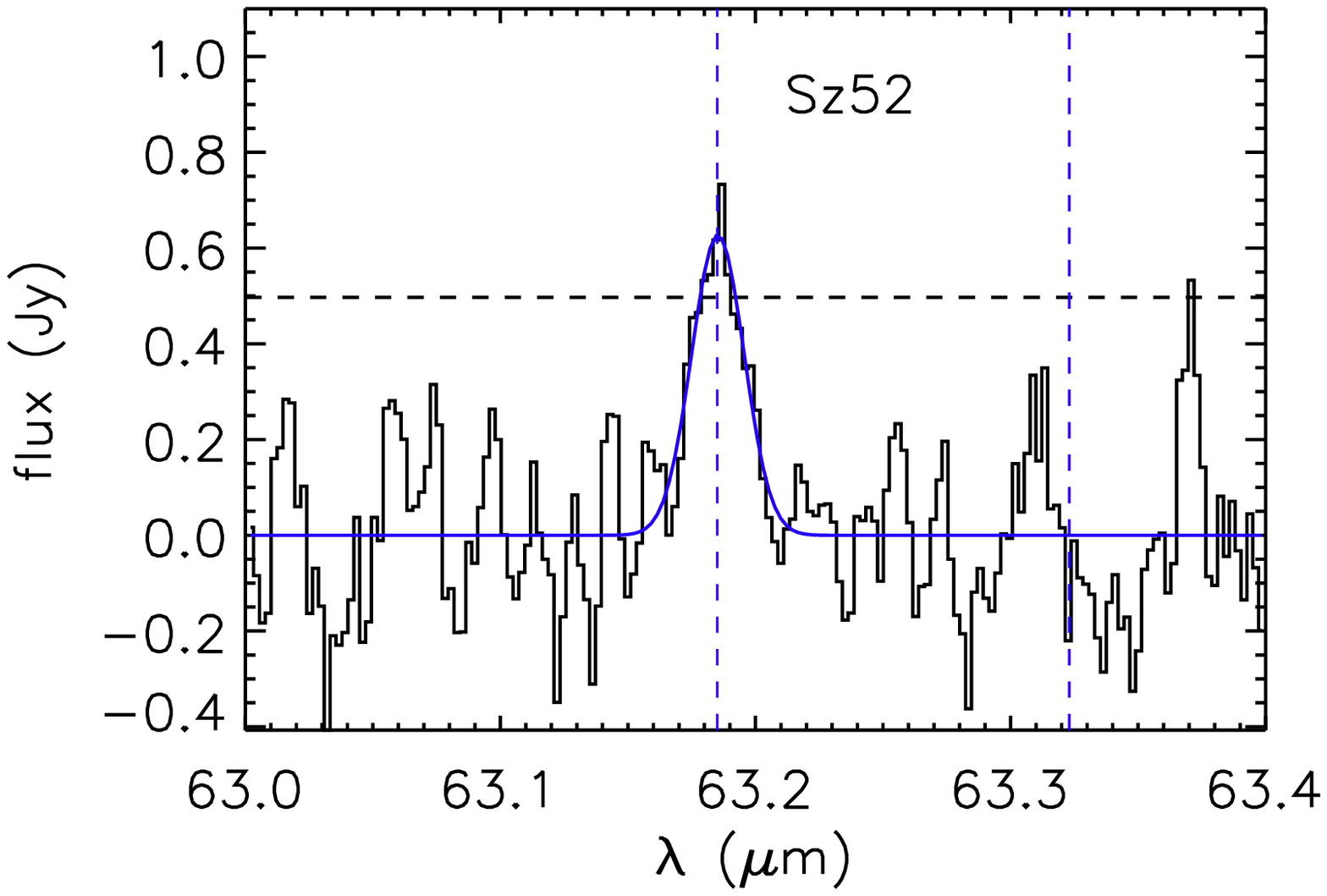}
\includegraphics[scale=0.3,trim=24mm 24mm 0mm 10mm,clip]{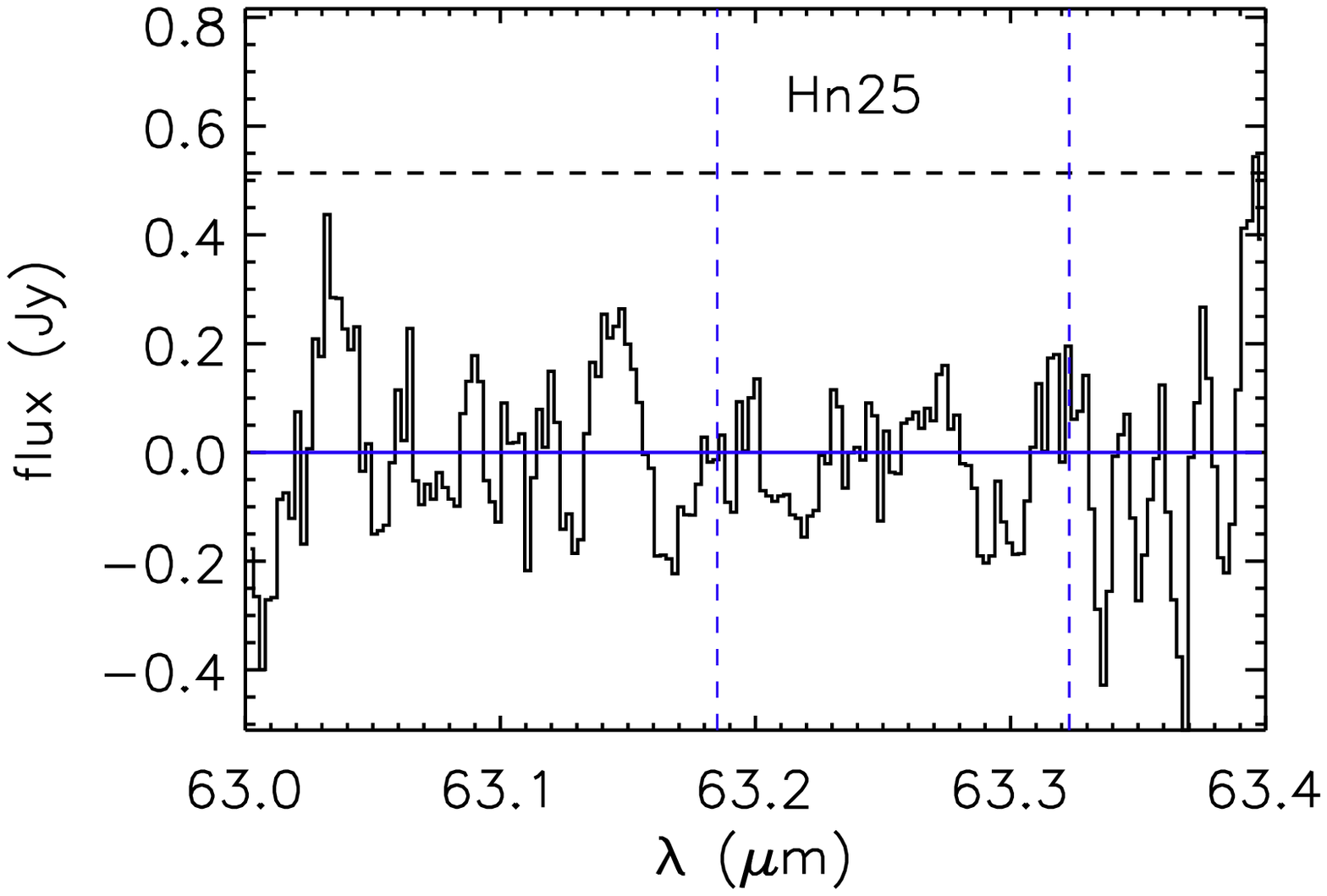} \\
\includegraphics[scale=0.3,trim=8mm 24mm 0mm 10mm,clip]{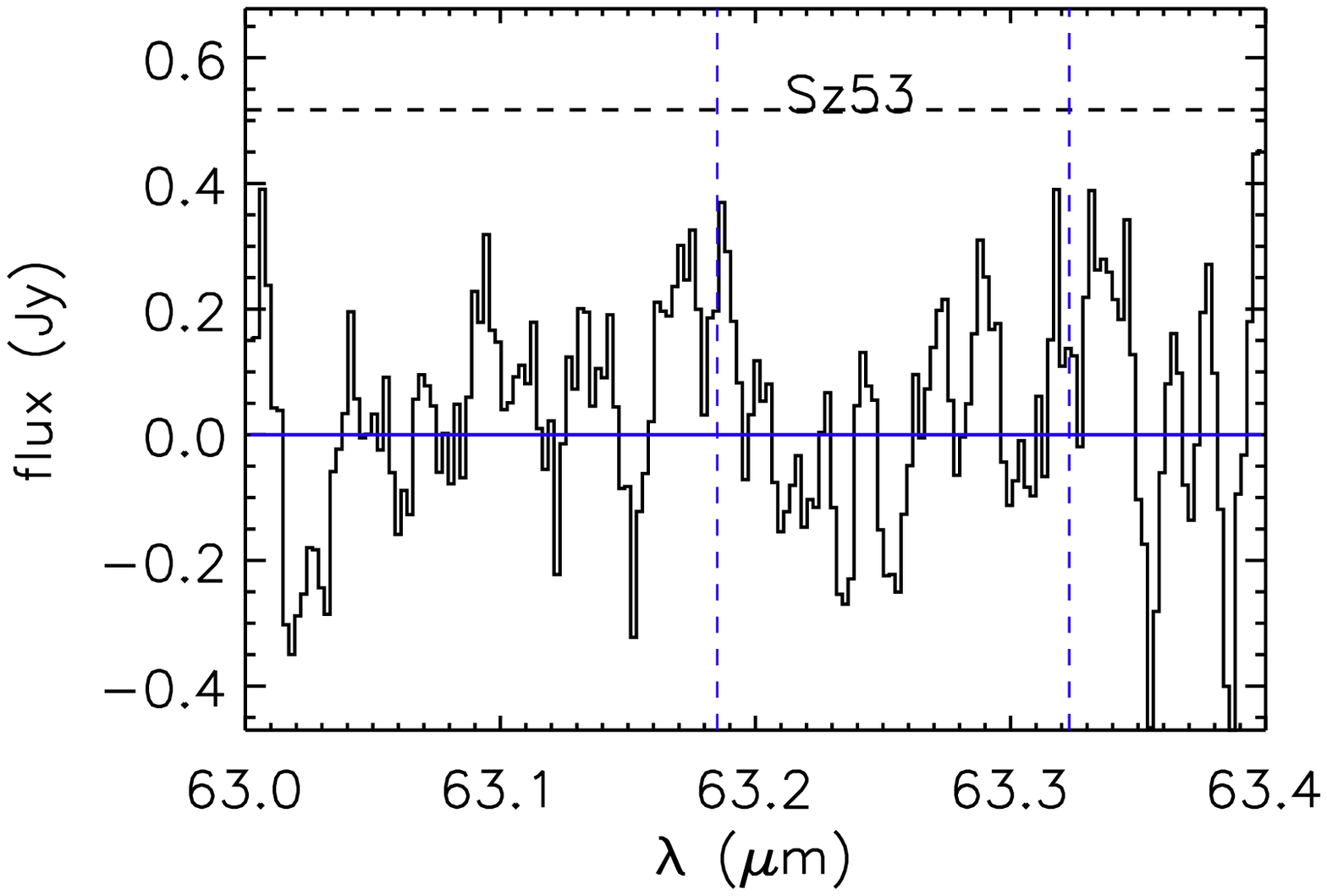}
\includegraphics[scale=0.3,trim=24mm 24mm 0mm 10mm,clip]{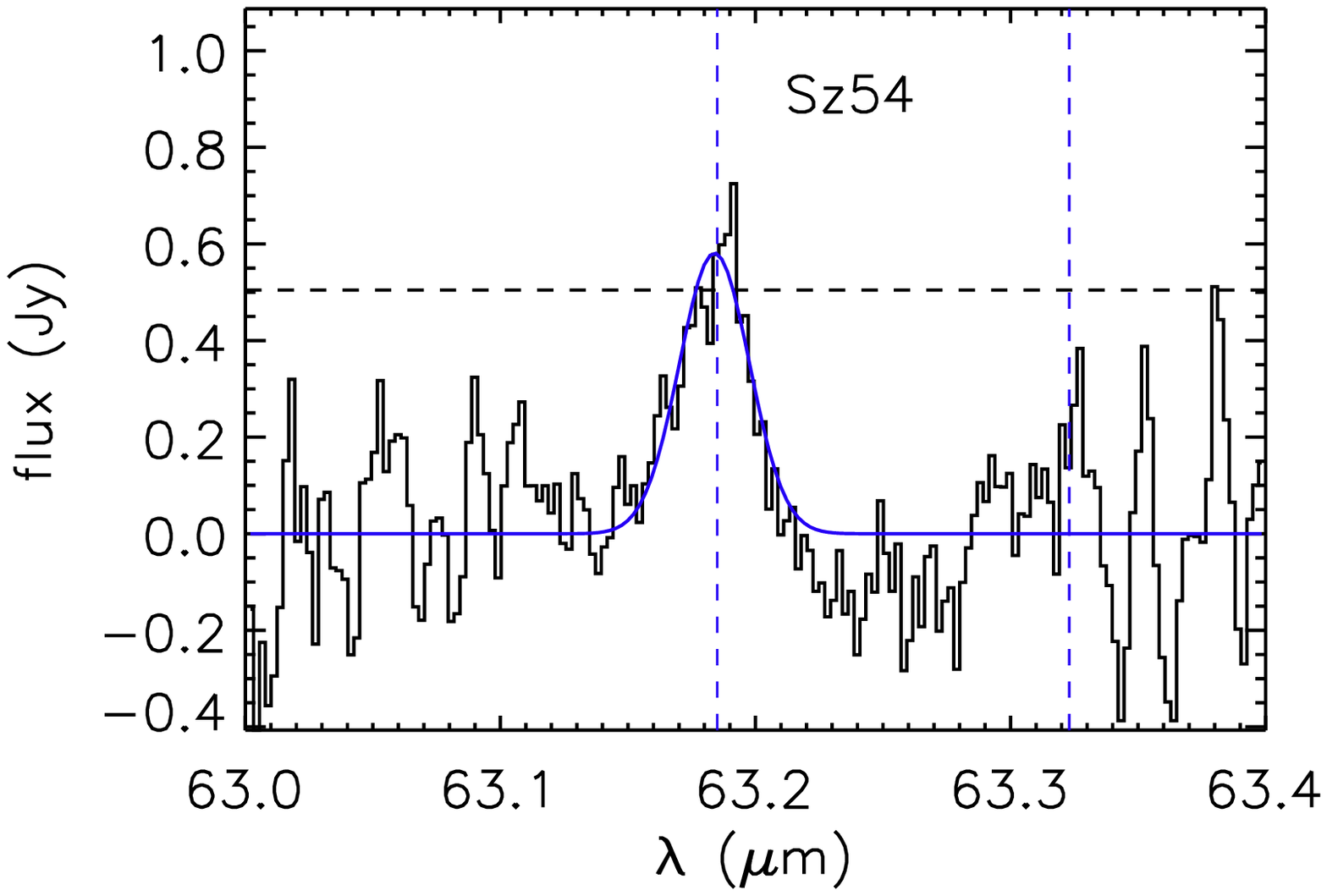}
\includegraphics[scale=0.3,trim=24mm 24mm 0mm 10mm,clip]{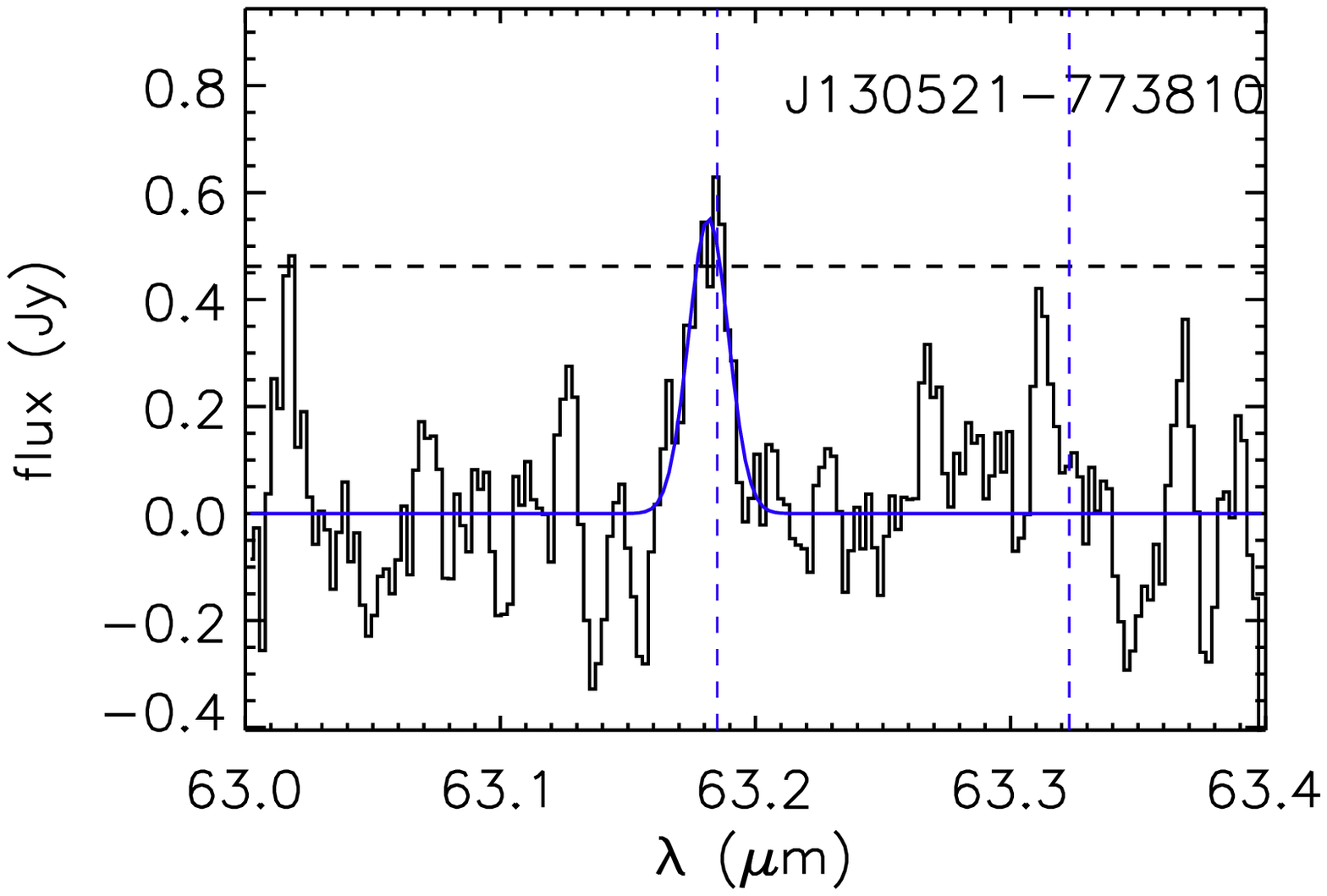} \\
\includegraphics[scale=0.3,trim=8mm 24mm 0mm 10mm,clip]{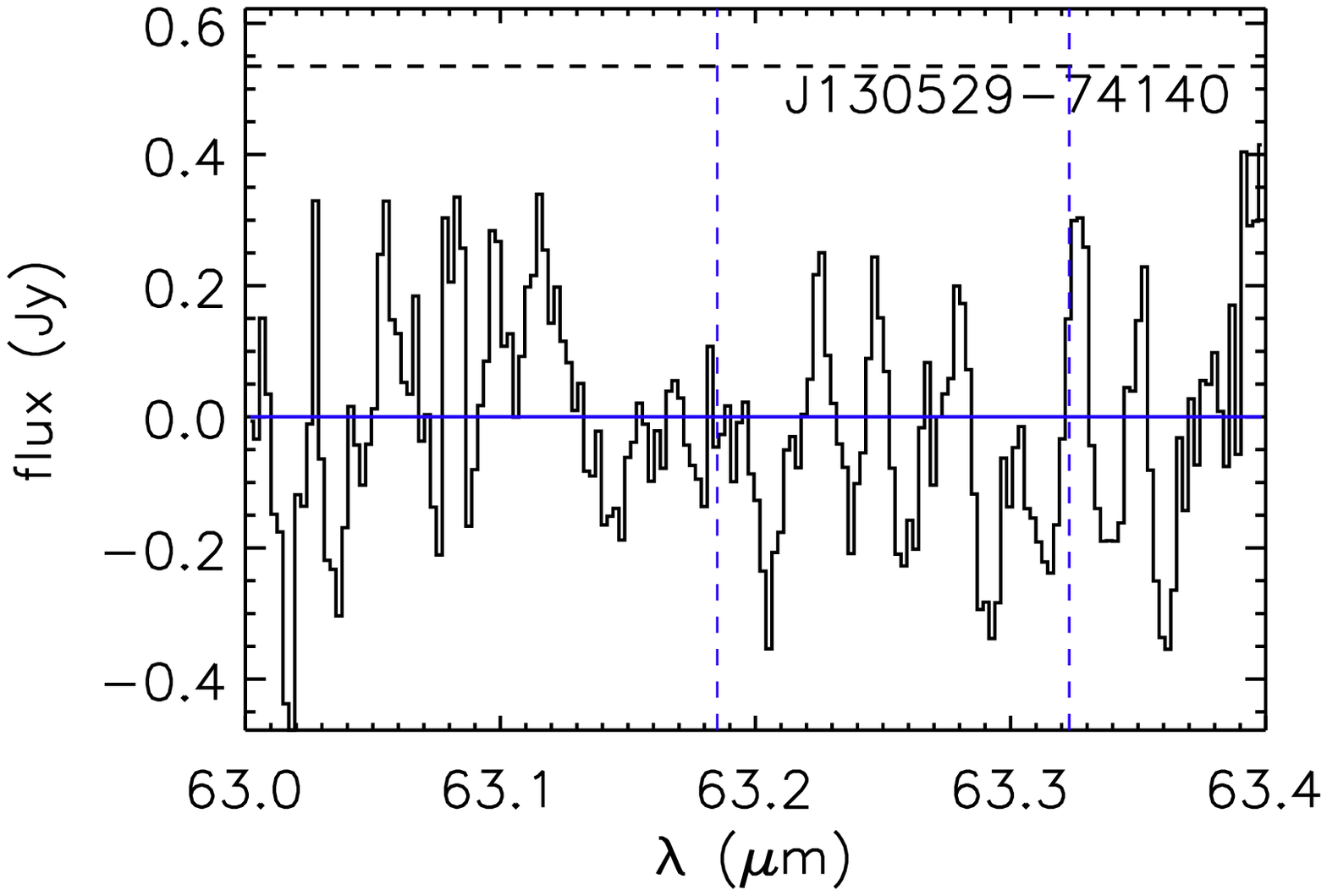}
\includegraphics[scale=0.3,trim=24mm 24mm 0mm 10mm,clip]{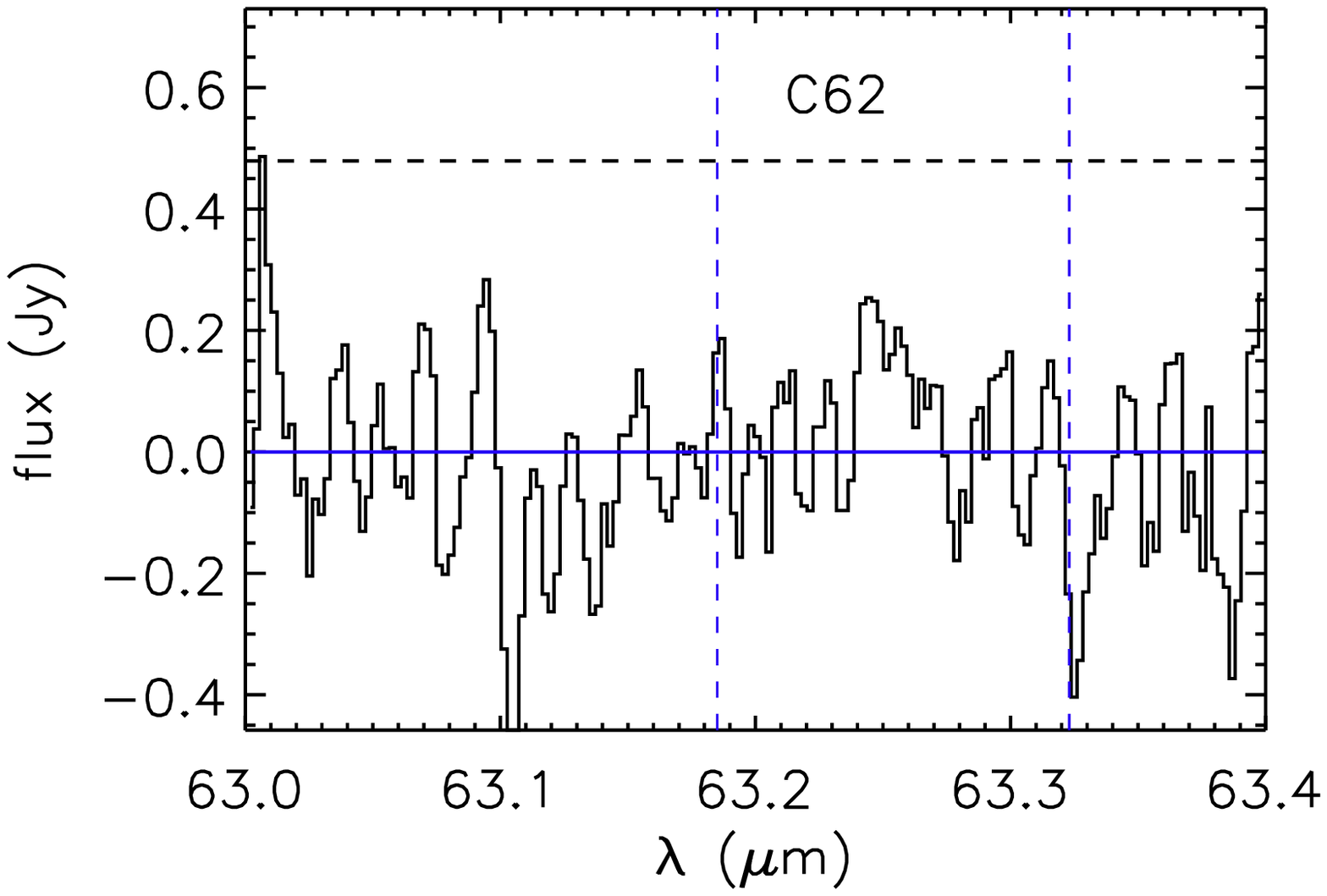}
\includegraphics[scale=0.3,trim=24mm 24mm 0mm 10mm,clip]{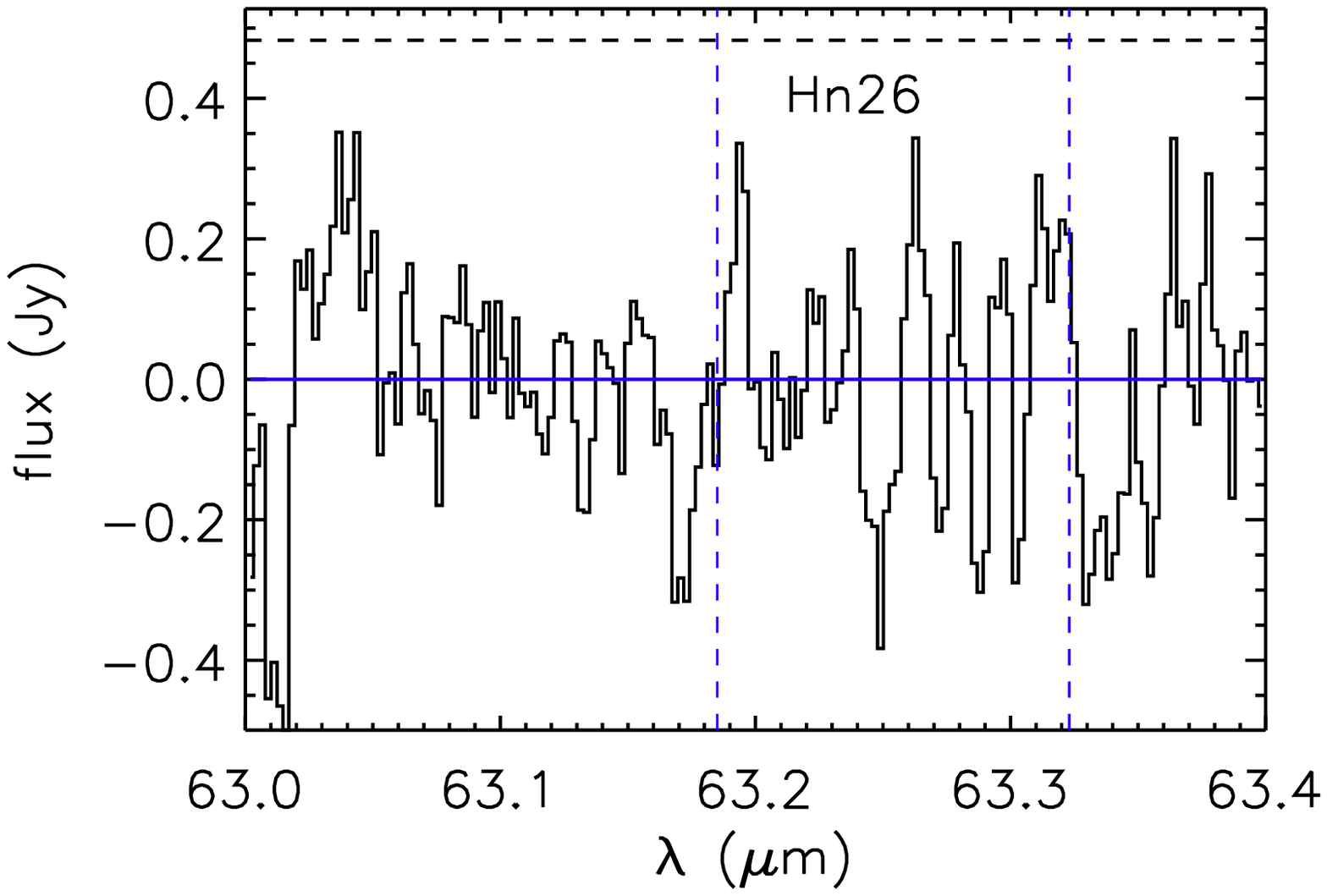} \\
\includegraphics[scale=0.3,trim=8mm 0mm 0mm 10mm,clip]{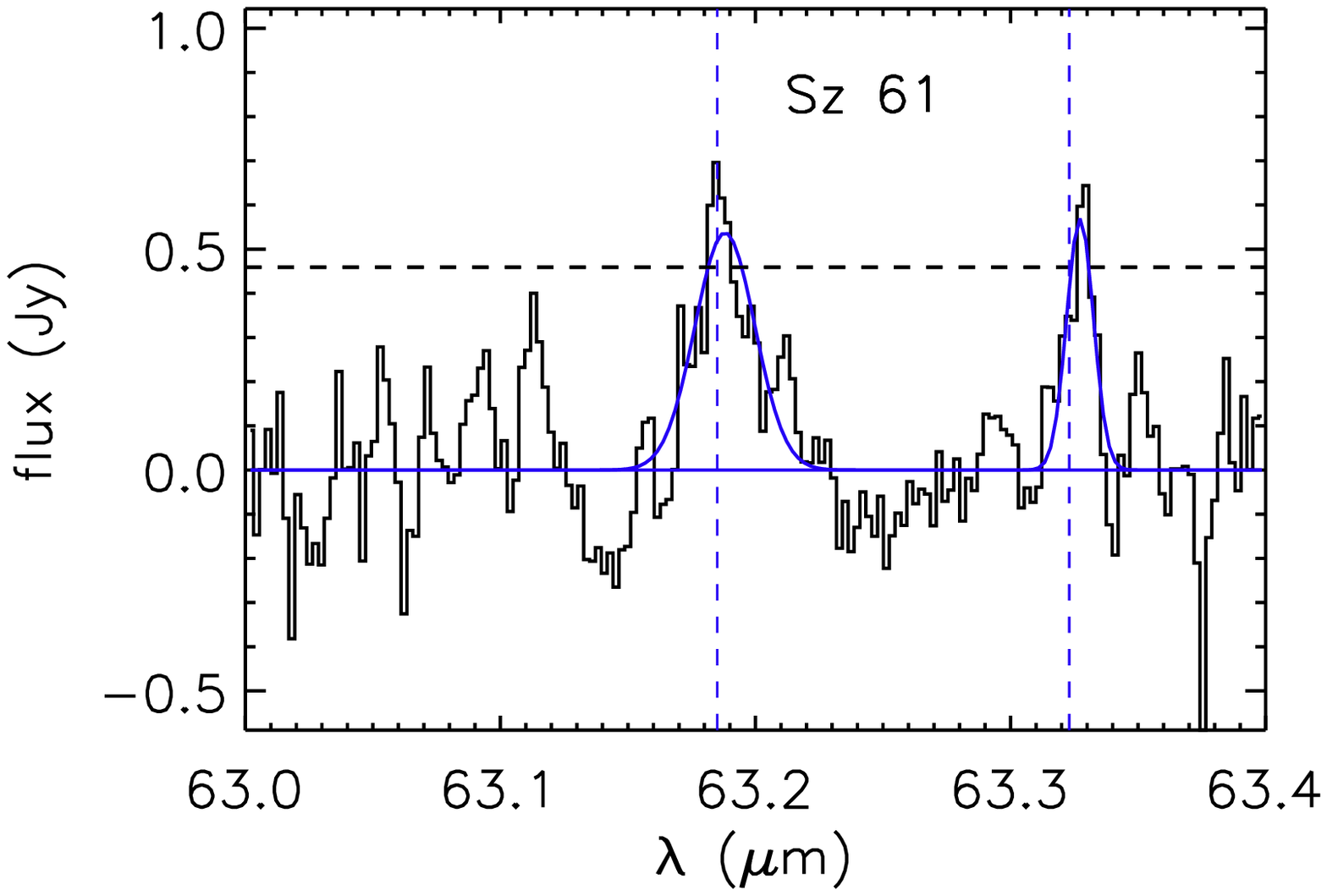}
\includegraphics[scale=0.3,trim=24mm 0mm 0mm 10mm,clip]{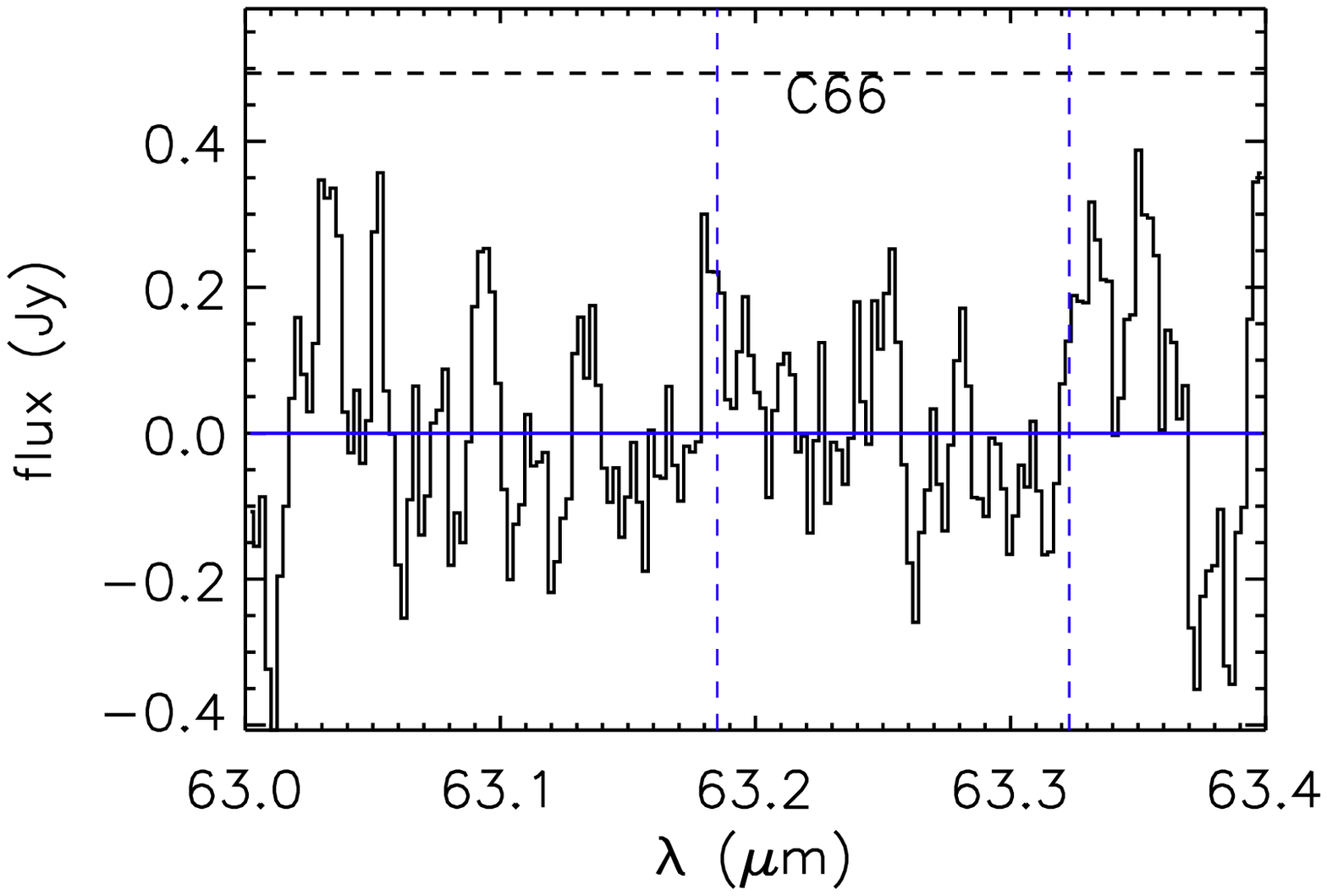}                                 
\includegraphics[scale=0.3,trim=24mm 0mm 0mm 10mm,clip]{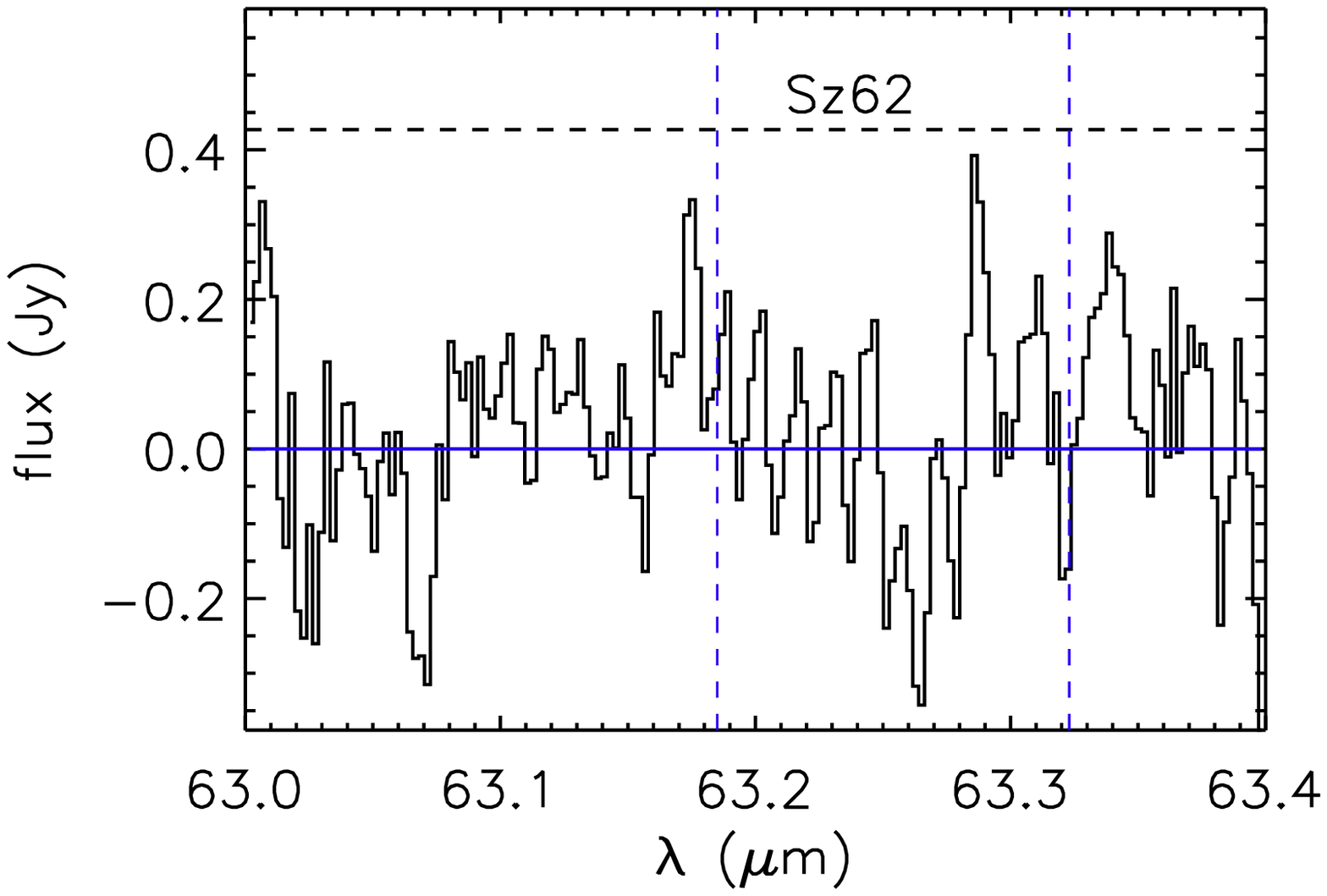}\\
   \caption{Continuum-subtracted PACS spectra at 63 $\rm \mu m$. The horizontal, black dashed line marks the position of the 3$\sigma$ detection limit for each observed spectrum. The blue solid line shows a Gaussian fit to the observed data. The vertical, blue dashed lines mark the position of  the line transitions of [OI] at 63.18 $\rm \mu m$ and o-$\rm H_{2}O$ at 63.32 $\rm \mu m$. We show the spectrum of DK Cha in Fig. \ref{DKCha_spec}. }
   \label{ChaII_OI_spec}
\end{center}
\end{figure*}

Cha II members were observed with \textit{Herschel}-PACS between 11 August 2011 and 28 September 2011. The scans were performed in chopping/nodding mode to account for the background and telescope contribution, with a small chopper throw ($\rm 1.5\arcmin$), using two Nod cycles, with a total integration time of 3316 s.

Data was reduced using HIPE 10 and the last version available of the calibration files (v5 for absolute calibration of the flux, leading to an accuracy of 10\%). Saturated and bad pixels were first masked and the spectra were corrected for the movement of the satellite. Bad data affected by chopper and grating movements were also flagged. A Q statistical test was performed to detect and mask glitches. The spectra were corrected for the instrumental response and dark current, and the background was subtracted by computing the difference between the Chop-On and Chop-Off positions. No emission in the off position in the central spaxel is detected in any of the observations, so the subtraction does not affect the line flux. For DK Cha, however, we observed emission in the Off position in spaxels 42 and 43 (with distances to the central spaxel of 50 arc sec and 76 arc sec, respectively; the details will be discussed in Section \ref{DKCha:ext}). The spectra were then divided by the relative spectral response function and flat-fielding was applied. An oversampling factor of 2 and  an upsampling of 3 were used to re-sample the spectra, almost recovering the native wavelength sampling. Finally, we computed the average between the two nod positions. 

The PACS spectroscopic FOV is $\rm 47\arcsec \times 47 \arcsec$, which, at the distance of Cha II translates to $\rm \sim 8000~AU \times 8000~AU$, with a typical distance between spaxels of $\rm 9.4 \arcsec$ ($\rm \sim$1600 AU). Shifts in the position of the star with respect to the central spaxel can lead to shifts in wavelength as well as flux loss (therefore, off-centre emission can appear red or blue-shifted with respect to the rest-frame wavelength).  To find the position of the source in the array of spaxels, we used the method described in Section 3.2.2 of \cite{Howard2013} to compare the distribution of flux in the spaxels with an array of models simulating different source displacements. The method was only applied to DK Cha and IRAS 12500-7658, since none of the other targets were detected in the continuum. 

Both sources are well centred within the central spaxels, with shifts smaller than $\rm 1 \arcsec$ in both cases, while the pointing accuracy for Herschel\footnote{http://herschel.esac.esa.int/twiki/bin/view/Public/SummaryPointing} was $\rm \sim 2\arcsec$. For the objects not detected in the continuum, the lines were only observed in the central spaxel. Therefore, we assume they are well centred within the pointing accuracy. We extracted the final spectrum from the central spaxel and applied the corresponding aperture correction to account for a $\rm \sim 30\%$ flux loss to the neighbouring spaxels. To avoid the noisy region at the edge of the spectra, only the wavelength range 63.0 to 63.4 $\rm \mu m$ was included in the subsequent analysis. We subtracted the continuum level by fitting straight lines to regions with no known spectral features. The excluded regions  were defined as $\rm \lambda_{line}-3\sigma < \lambda_{line} < \lambda_{line}+3\sigma$,
where $\rm \sigma $ is the instrumental standard deviation of the Gaussian line profiles ($\rm 7.6 \times 10^{-3} ~\mu m$) and $\rm \lambda_{line}$ is the rest-frame wavelength of each of the transitions present, which are [OI] $\rm ^{3} P_{1} \rightarrow$$\rm^{3}P_{2}$ at 63.185 $\rm \mu m$ and o-$\rm H_{2}O~8_{18}\rightarrow7_{07}$ at 63.323 $\rm \mu m$. For DK Cha, owing to the broad nature of the [OI] line, we excluded a larger region of 6$\sigma$ on each side of the profile. Since its central spaxel line profile is double peaked, making a single-Gaussian fit a poor description of the emission, we computed the line flux for this object as the integral of the continuum-subtracted flux. Furthermore, as reported by \cite{Green2013}, the [OI] emission for DK Cha is extended, so that co-adding all the 25 spaxels returns a more reliable measure of its total emission. 

We consider that a line is detected when the value of the peak is more than three times the value of the noise computed in the continuum. Line fluxes were computed as the integral of Gaussian fits to the observed profiles. Errors were computed as the integral of a Gaussian with peak equal to the noise level of the continuum ($\Delta f_{cont}$) and standard deviation that of the fitted profile of the line ($\sigma_{fit}$), i. e.:
\begin{equation}
	\Delta f_{line}=\pi ^{1/2} \Delta f_{cont} \sigma_{fit}
\end{equation}
 Re-sampling the spectra can result in an artificial increase in S/N due to correlated noise. To test the impact of re-sampling on line flux errors, we also reduced the spectra of detected objects with native instrument sampling and found no significant difference. The only exception was IRAS 12535-7623 where resampling the spectra resulted in a 16\% decrease in the errors, and therefore we report the errors as computed  in the native instrument-sampling spectra for this object. 

Upper limits were computed as the integral of a Gaussian with peak equal to the 3$\rm \sigma$ level of the continuum and standard deviation equal to the instrumental FWHM/2.3548 ($\rm FWHM= 0.018  ~\mu m ~at~ 63 ~\mu m$). The resulting line fluxes and upper limits, together with the lines FWHMs, are shown in Table \ref{lineFluxes}. The average 1$\sigma$ noise level for Cha II sources observed in the GASPS survey is $\rm 1.2\times 10^{-17}~erg/cm^{2}/s/\AA$, which translates to an average 1$\sigma$ line detection limit of  $\rm \sim 2.4 \times 10^{-18}~W/m^{2}$. 

\section{Results and discussion}\label{sec:resDisc}

\begin{figure*}[!t]
\begin{center}
 \hspace{0.51cm} \includegraphics[scale=0.52,trim=0mm 0mm 0mm 0mm,clip]{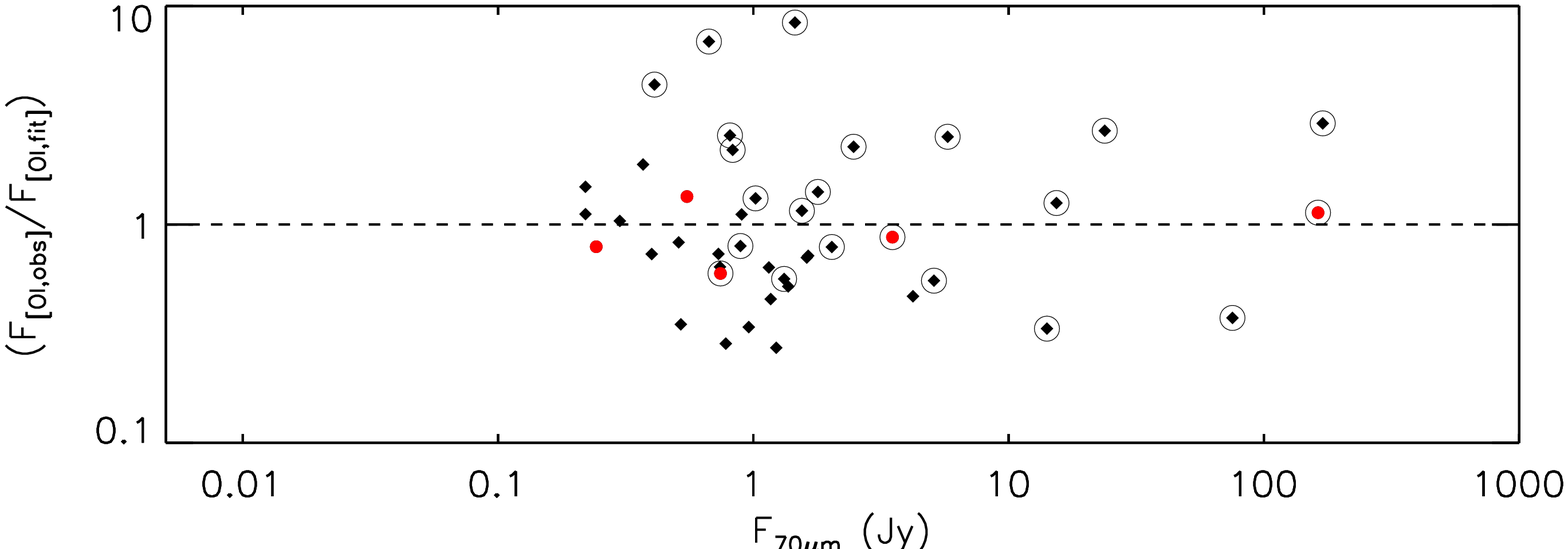}\\
  \includegraphics[scale=0.5,trim=0mm 0mm 0mm 10mm,clip]{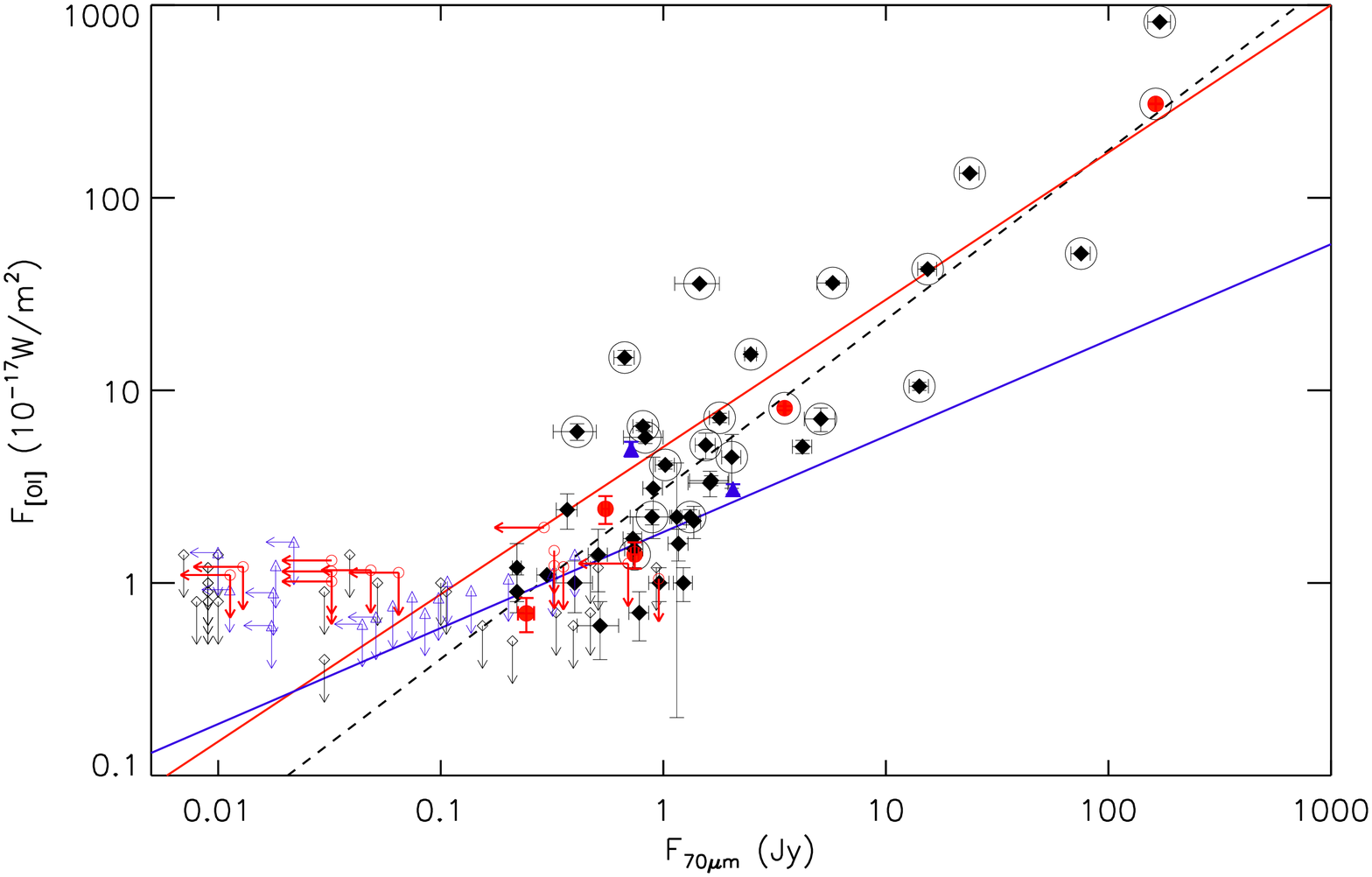}\\
        \vspace{0.3cm}
        \caption{Top: ratio of the observed flux to the flux derived from the fit to all the data. Bottom: [OI] fluxes at 63.18 $\rm \mu m$ versus the continuum fluxes at 70 $\rm \mu m$. Red filled circles are Cha II members with an [OI] detection at 63.18 $\rm \mu m$, while red empty circles are upper limits. Black filled diamond are Taurus members with an [OI] detection at 63.18 $\rm \mu m$, while black empty diamonds are upper limits. Blue upward filled triangles are Upper Scorpius objects from \cite{Mathews2013}, while empty ones are upper limits from the same work. Encircled symbols show the position of known outflow sources in Taurus (in black) and in Cha II (in red). The red line shows a fit to jet sources in Taurus and Cha II, while the blue  line shows a fit to non-jet sources in the same associations and the black dashed line shows a fit to all the data. The fluxes for objects in Cha II and Upper Scorpius have been scaled according to the distance to Taurus. }
   \label{ChaII_oi_cont}
\end{center}
\end{figure*}

\subsection{Gas detection and non-detections}\label{Sec:gasDet}
We detected [OI] emission in seven out of the nineteen objects observed, leading to an overall detection fraction of $\rm 0.37^{+0.12}_{-0.09}$ \citep[see][for a description of the method used to compute uncertainties in low number ratios]{Burgasser2003}, with line fluxes in the range $\rm 4\times 10^{-18}$ to $\rm 4\times 10^{-15}~W/m^{2}$. FWHMs for the [OI] line are in the range 0.012 for Sz 51 to 0.032 $\rm \mu m$ for J13052169-7738102. The FWHM for Sz 51 is below the instrumental value (0.018 $\rm \mu m$). However, the peak of the line is over the 3$\sigma$ detection limit, so we consider it a real detection.Upper limits are in the range (6--9)$\rm \times 10^{-18}~W/m^{2}$. As an example, the measured flux for Sz 53, which is not detected, is $\rm (3.2 \pm 2.3)\times 10^{-18}~W/m^{2}$. Among the detected objects, only DK Cha showed line emission in more than one spaxel. We discuss the details of its spatial distribution in Section \ref{DKCha:ext}. 
 
To infer line detection fractions for the different disc classes, we consider DK Cha as a Class I object, because it shows prominent envelope emission \citep{Kempen2006,Kempen2009}. We detected [OI] emission  in all Class I and flat-spectrum objects in the sample, with fluxes in the range $\rm (0.81-384)\times 10^{-17}~W/m^{2}$. Only four out of 16  Class II objects were detected ($\rm 0.25^{+0.13}_{-0.08}$ detection fraction), with fluxes  in the range $\rm (0.42-1.5)\times 10^{-17}~W/m^{2}$ and an average flux of $\rm (1.0 \pm 0.6)\times 10^{-17}~W/m^{2}$. For comparison, 32 out of 41 Class II objects in Taurus showed [OI] emission, leading to a detection fraction of $\rm 0.78^{+0.05}_{-0.08}$ while the  detection fraction for Class I sources in Taurus is the same as in Cha II \citep[six out of six objects observed, see][]{Howard2013}.
 
When the sample is divided by spectral type, we detected [OI] in only one out of 12 Class II M stars, leading to a detection fraction of $\rm 0.08^{+0.15}_{0.03}$. We observed three K stars belonging to Class II, and detected [OI] in all of them, leading to a detection fraction in the range 0.63--1.0 (when uncertainties are considered). Therefore, the detection fraction seems to correlate with the spectral type, although the sample is too small to bear strong conclusions.

All Cha II members in the sample show an IR excess, indicative of the presence of a circumstellar disc. Objects with similar IR excess can either show or not show an [OI] detection, indicative of the decoupled structure of the gas and dust. The result may point to a spread of gas-to-dust ratios for stars with the same age. Nevertheless, there are alternative explanations for the lack of [OI] emission, such as flat-disc geometries \citep[see e. g. ][]{Riviere2013}. 

Water emission at 63.32 $\rm \mu m$ was detected towards Sz 61, with a flux $\rm 5.6 \times 10^{-18}~W/m^{2}$, which leads to a water detection rate among gas rich discs (i.e., Class II discs with an [OI] detection) of $\rm 0.20^{+0.25}_{-0.08}$, compatible with the fraction of water-bearing gas rich discs in Taurus \citep[0.24, see][]{Riviere2012}. Other systems with stellar properties similar to Sz 61, such as Sz 51, Sz 54, and J130521.6-773810, show [OI] fluxes in the range (0.43--1.5)$\rm \times 10^{-17}~W/m^{2}$, similar to Sz 61. However, none of them show a water detection at 63 $\rm \mu m$, or show upper limits compatible with low $\rm H_{2}O$ line  fluxes, similar to Sz 61. The FWHM=0.012 $\rm \mu m$ of the line is slightly below the instrumental value of 0.018 $\rm \mu m$. If we assume the instrumental FWHM when computing the errors, then the detection becomes 2.6$\rm \sigma$ with a flux $\rm (5.6\pm2.2)\times 10^{-18}~W/m^{2}$. 

Water at 63 $\rm \mu m$ in Taurus was only detected in systems harbouring an outflow \citep{Riviere2012}, and water emission at 63 $\rm \mu m$ from \object{NGC~1333~IRAS~4B} was reported by \cite{Herczeg2012}, who conclude that the bulk of water emission in the far-IR is produced in the outflow. \cite{Antoniucci2011} report OI at 6300 $\rm \AA$ in emission with an equivalent width of -1.5 $\rm \AA$ for Sz 61, a value that is similar to those reported for Taurus sources with outflows by \cite{Hartigan1995}. The most likely conclusion is therefore that Sz 61 harbours a jet or a low-velocity outflow. Furthermore, \cite{Antoniucci2011} report an equivalent width of -5 $\rm \AA$ for the OI at 6300 $\rm \AA$ line towards Sz 53, and therefore it is also likely to drive an outflow. However, no [OI] nor $\rm H_{2}O$ at 63 $\rm \mu m$ was detected towards Sz 53.

\subsection{Correlation with continuum at 70 $\rm \mu m$}
\cite{Howard2013} report a correlation between the [OI] flux at 63.18 $\rm \mu m$ and the continuum level at 63 $\rm \mu m$. However, a number of ChaII sources were not detected in the continuum at 63 $\rm \mu m$, and so we decided to test a correlation against the continuum at 70 $\rm \mu m$ instead, using PACS fluxes from \cite{Spezzi2013}. The resulting plot is shown in the lower panel of Fig. \ref{ChaII_oi_cont}. We included for comparison Taurus and Upper Scorpius sources from \cite{Howard2013} and \cite{Mathews2013}, respectively. A correlation between the [OI] flux and the continuum flux at 70 $\rm \mu m$ is clear from the plot, with outflow sources typically showing larger [OI] fluxes: the average ratio of [OI] line flux to continuum flux at 70 $\rm \mu m$ in Taurus is 2.6 times greater for outflow sources ($\rm F_{OI}/F_{70} =1.1$) compared to non-outflow sources ($\rm F_{OI}/F_{70} =0.4$). The difference is more pronounced for bright continuum sources. 

To explore this correlation further, we first review the presence of outflows in Cha II sources. DK Cha has long been known to drive an outflow \citep{Hughes1989,Hughes1991,Knee1992}. \cite{Caratti2009} propose IRAS12500-7658 as the best candidate to drive the \object{HH~52}, \object{HH~53} and \object{HH~54} outflows. In the following, we consider that IRAS12500-7658 is in fact driving these outflows. According to our previous analysis in Section \ref{Sec:gasDet} Sz  53 and Sz 61 are probably driving an outflow, too. While DK Cha and IRAS12500-7658 lie in the high OI flux part of the diagram, Sz 61 lies in the crowded region where most Class II objects without outflows are located. Interestingly, Sz 53 is not detected in [OI] at 63 $\rm \mu m$.

To characterise the separation between outflow and non-outflow sources in the diagram, we performed a two-dimensional Kolmogorov-Smirnov test \citep[see][]{Press1992} using the continuum flux at 70 $\rm \mu m$ and [OI] flux at 63.18 $\rm \mu m$ as parameters for each population (outflow and non-outflow). The probability that both populations are drawn from the same distribution is only $\rm  2\times 10^{-5}$. To test the strength of this result further , we computed the ratio of the observed flux to the flux derived from a linear fit to the logarithmic distribution of OI fluxes versus 70 $\rm \mu m$ continuum fluxes. The result of this test is shown in the top panel of Fig. \ref{ChaII_oi_cont}. We also performed a one-dimensional Kolmogorov-Smirnov test to compare the distribution of ratios between outflow and non-outflow sources: the probability that both populations come from the same distribution is only $\rm 6.0 \times 10^{-3}$. Therefore we conclude that separation is real and both populations are indeed drawn from different distributions: on average, outflow sources have stronger continuum fluxes and higher line-to-continuum ratios than non-outflow sources. 

\subsection{Accretion in Cha II}

\begin{figure}[!t]
\begin{center}
%   \centering 
     \includegraphics[scale=0.5,trim=15mm 5mm 0mm 0mm,clip]{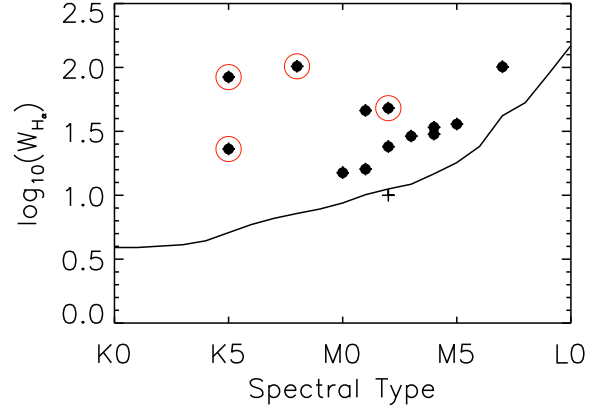}
   \caption{Left: Accretion in Cha II members. Plus symbols depict objects that might have ${\rm H}_{\alpha}$ emission coming only from purely chromospheric emission, while black dots are objects with ${\rm H}_{\alpha}$ in agreement with ongoing accretion. The solid line shows the saturation criterion by \cite{Barrado2003}. Red circles surrounding black dots denote objects with [OI] emission at 63.18 $\rm \mu m$ detected by PACS.}
   \label{ChaII_Acc}
\end{center}
\end{figure}

\begin{figure}[!t]
\centering
\hspace{0.25cm}
     \includegraphics[scale=0.75]{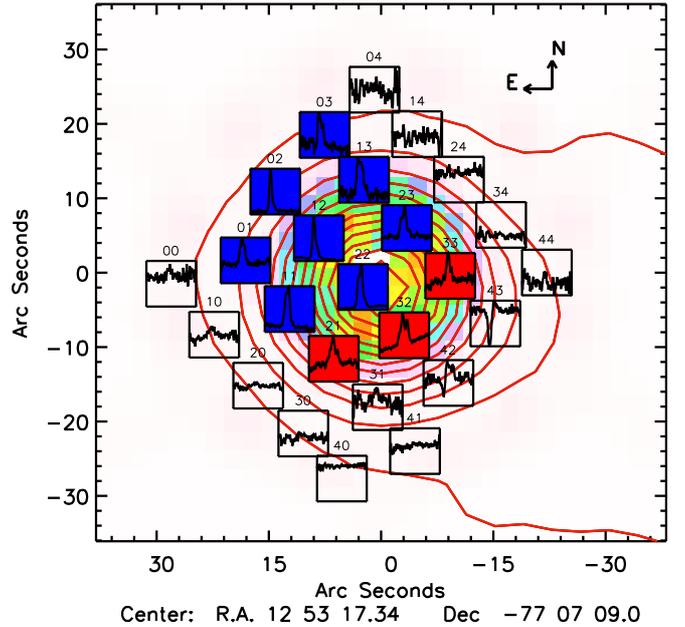}\\
     \vspace{1.0cm}
   \caption{PACS spectral FOV for DK Cha displayed over a MIPS image at 24 $\rm \mu m$ (colour-filled contours). LABOCA contours at 870 $\rm \mu m$ \citep{DeGregorio2014} (De Gregorio et al. submitted) are also shown in red. Spectra for sources with a blue-shifted high-velocity component are shown with blue background, while spectra  for sources with a red-shifted high-velocity component are shown with red background.}
   \label{DKCha_spec}
\end{figure}

\begin{figure*}[!t]
\centering     
     \includegraphics[scale=0.25,trim=10mm 25mm 10mm 0mm,clip]{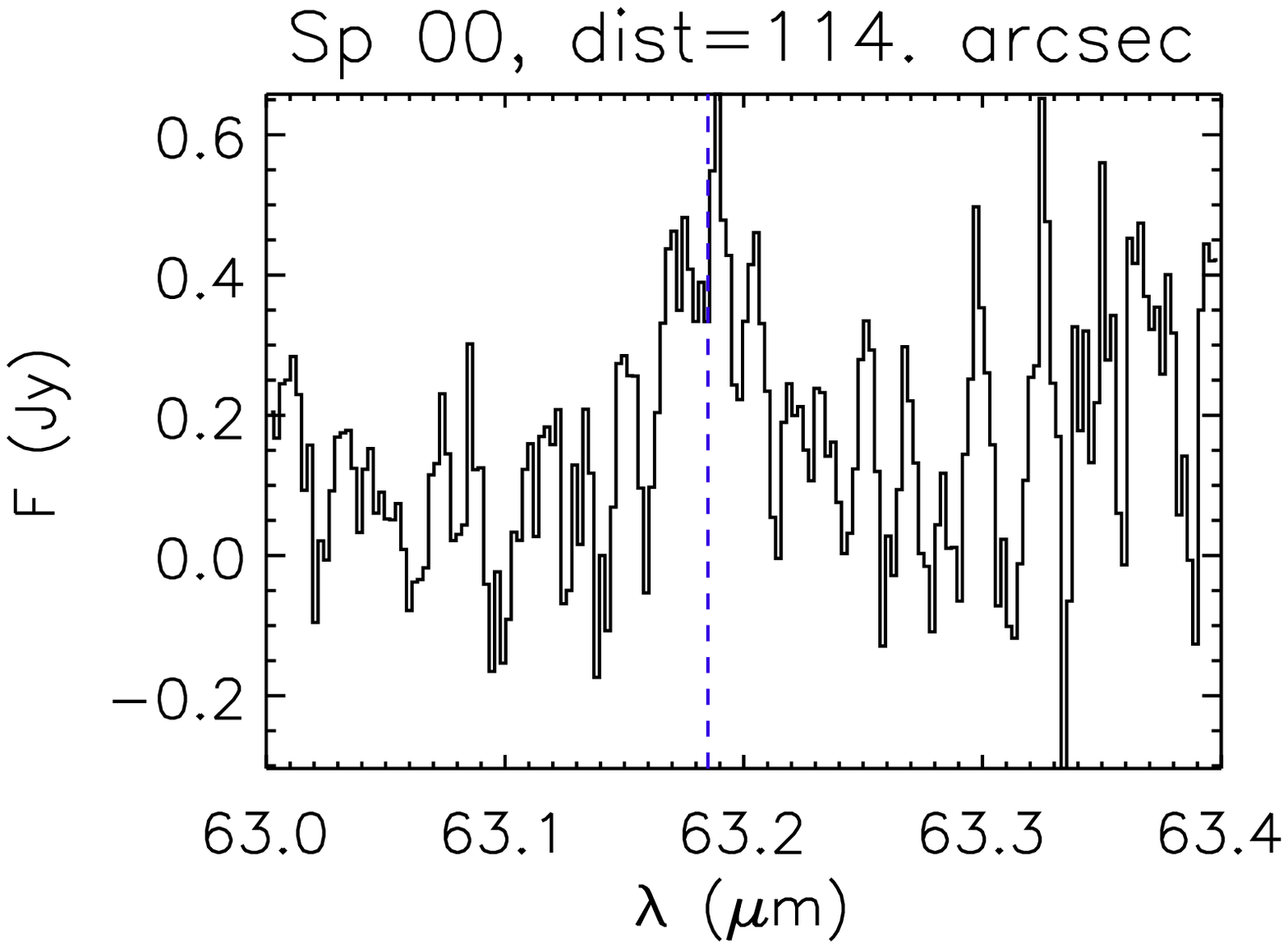}
     \includegraphics[scale=0.25,trim=30mm 25mm 10mm 0mm,clip]{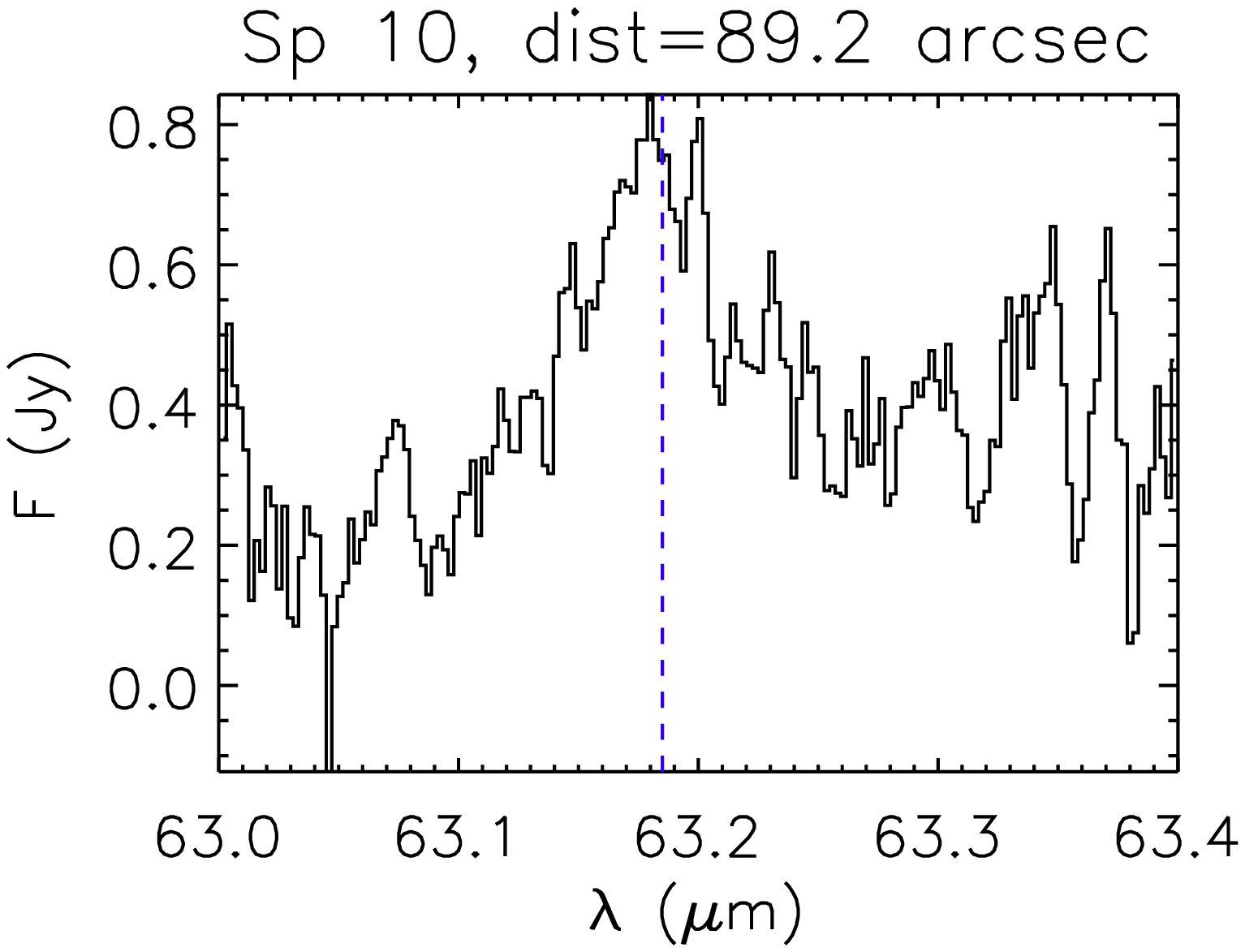}
     \includegraphics[scale=0.25,trim=30mm 25mm 10mm 0mm,clip]{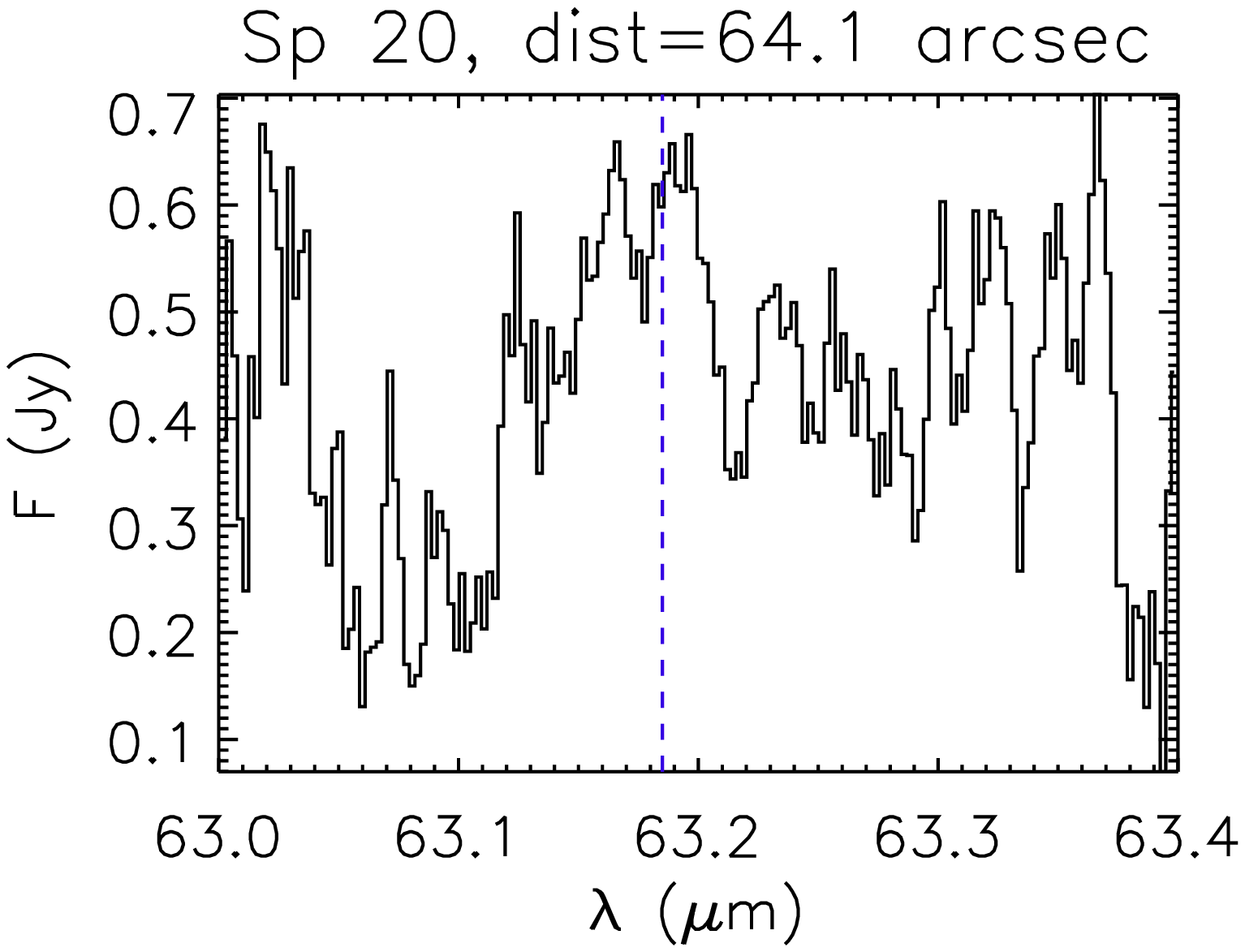}
     \includegraphics[scale=0.25,trim=30mm 25mm 10mm 0mm,clip]{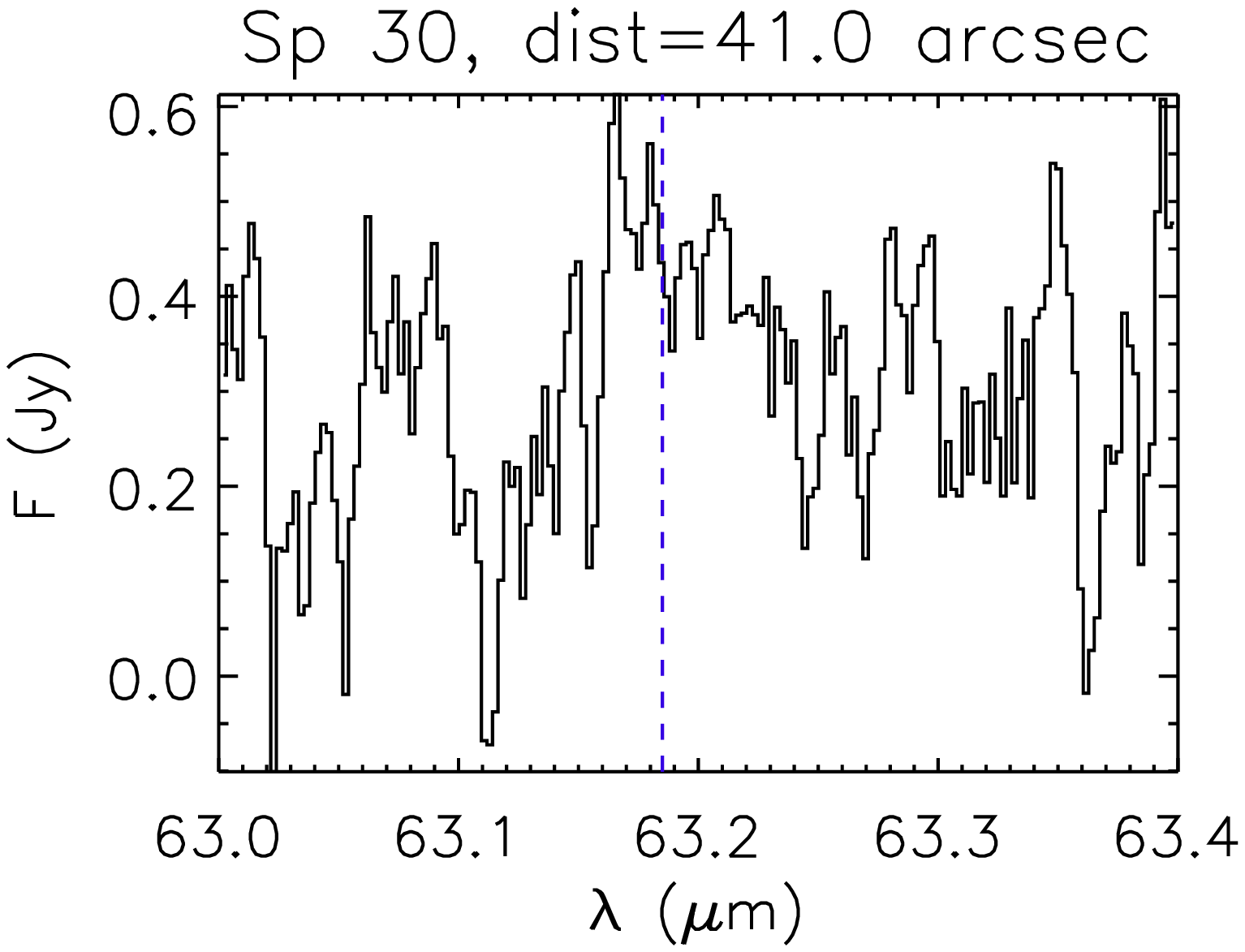}
     \includegraphics[scale=0.25,trim=30mm 25mm 10mm 0mm,clip]{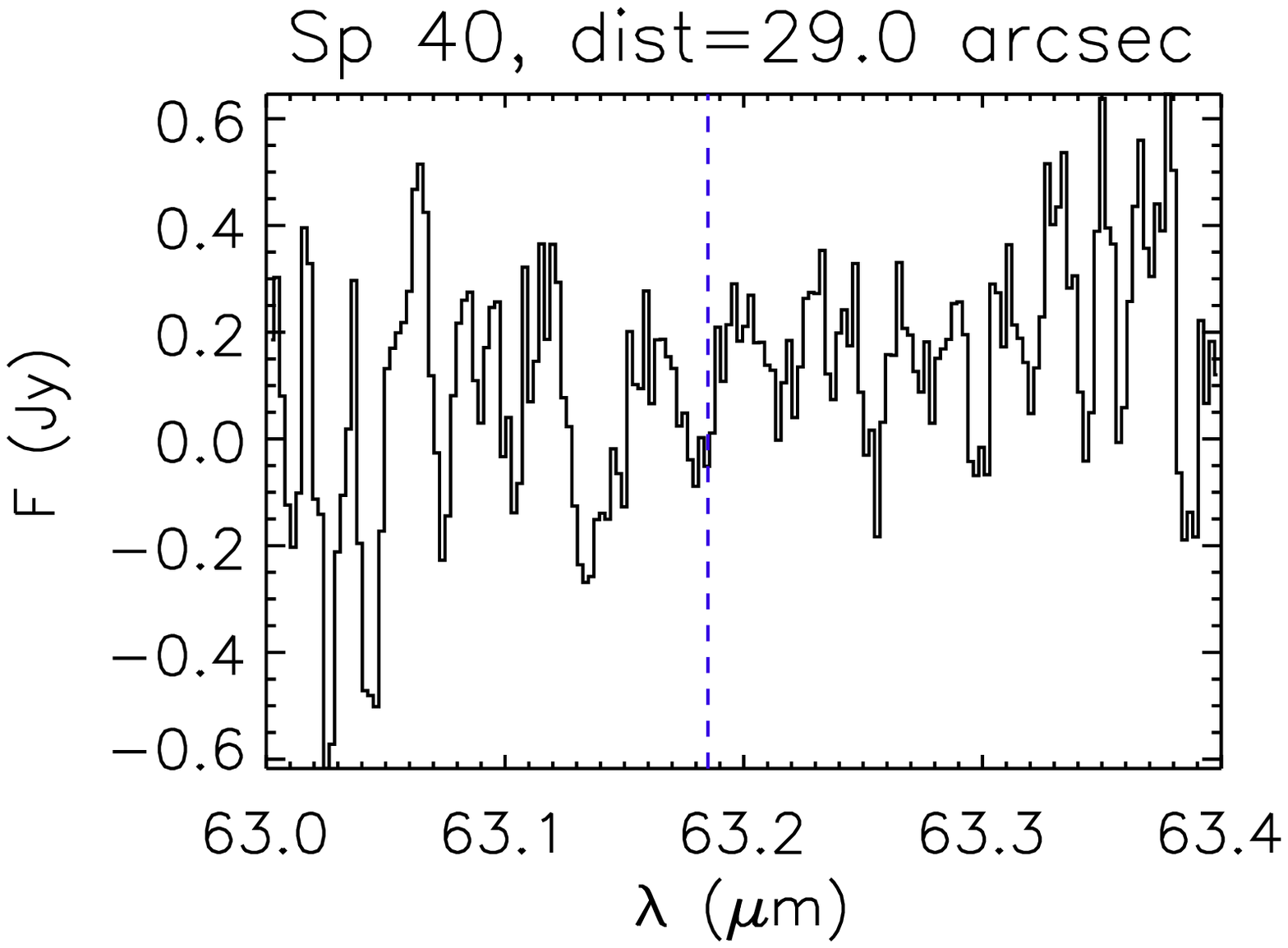}\\     
     \includegraphics[scale=0.25,trim=10mm 25mm 10mm 0mm,clip]{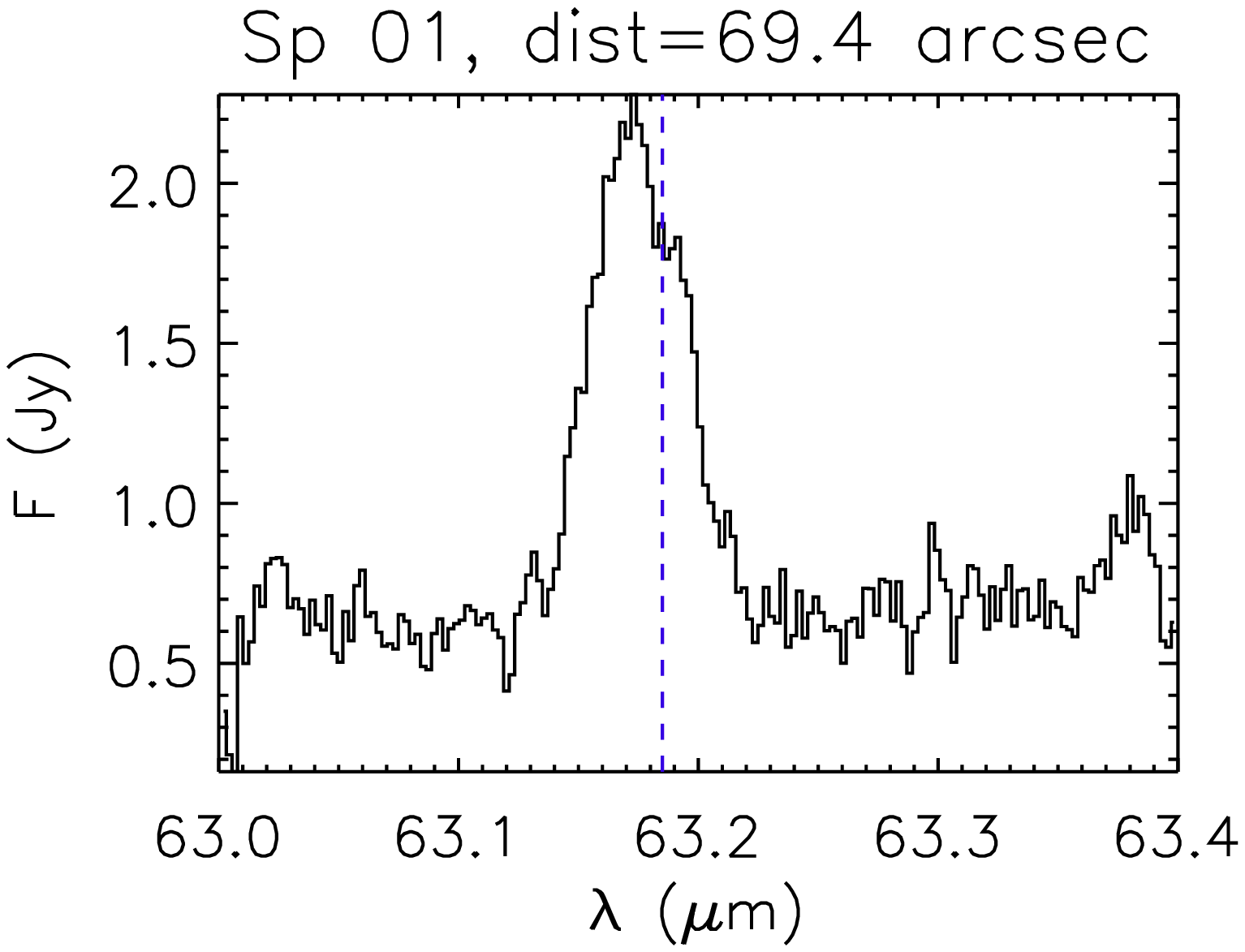}
     \includegraphics[scale=0.25,trim=30mm 25mm 10mm 0mm,clip]{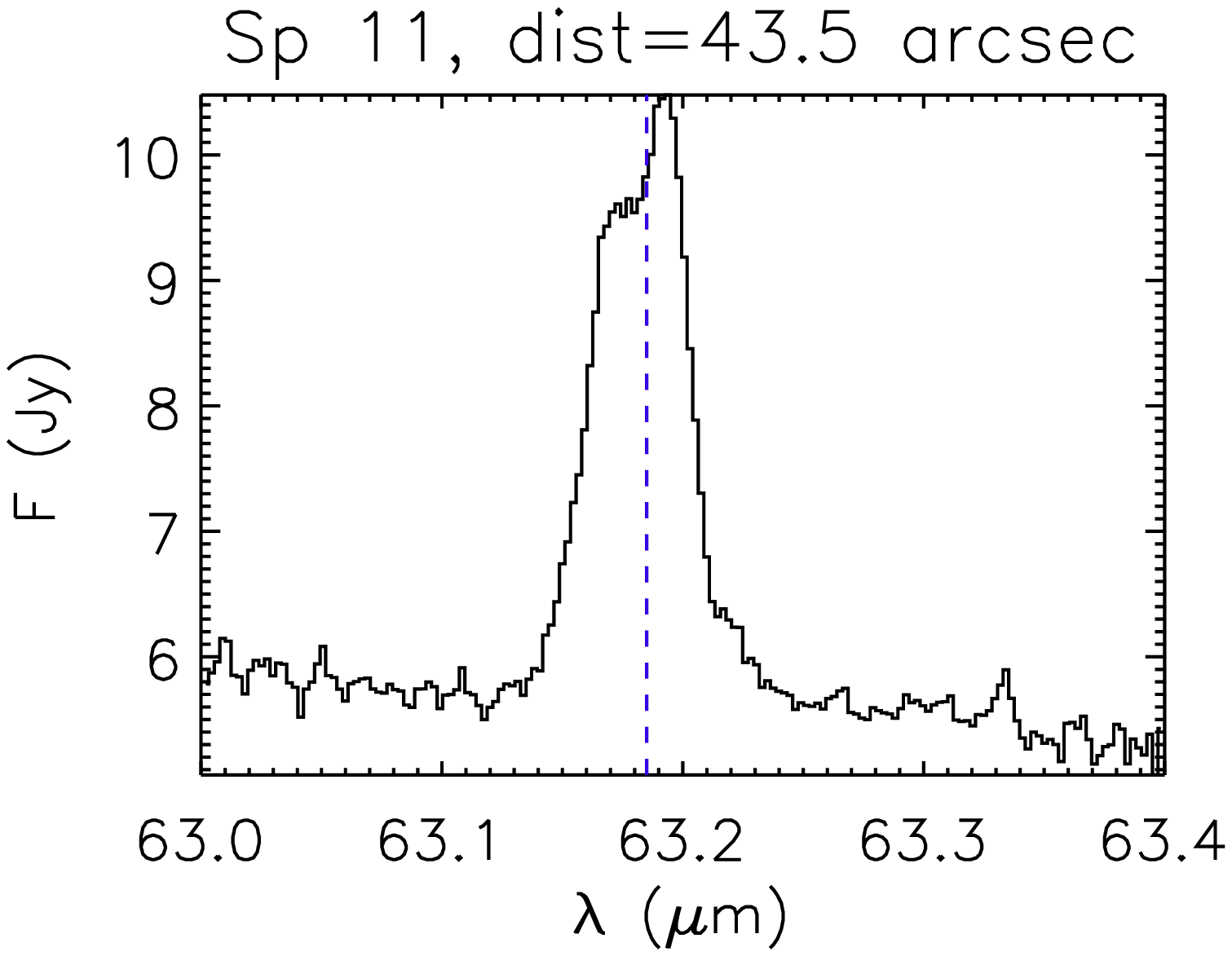}
     \includegraphics[scale=0.25,trim=30mm 25mm 10mm 0mm,clip]{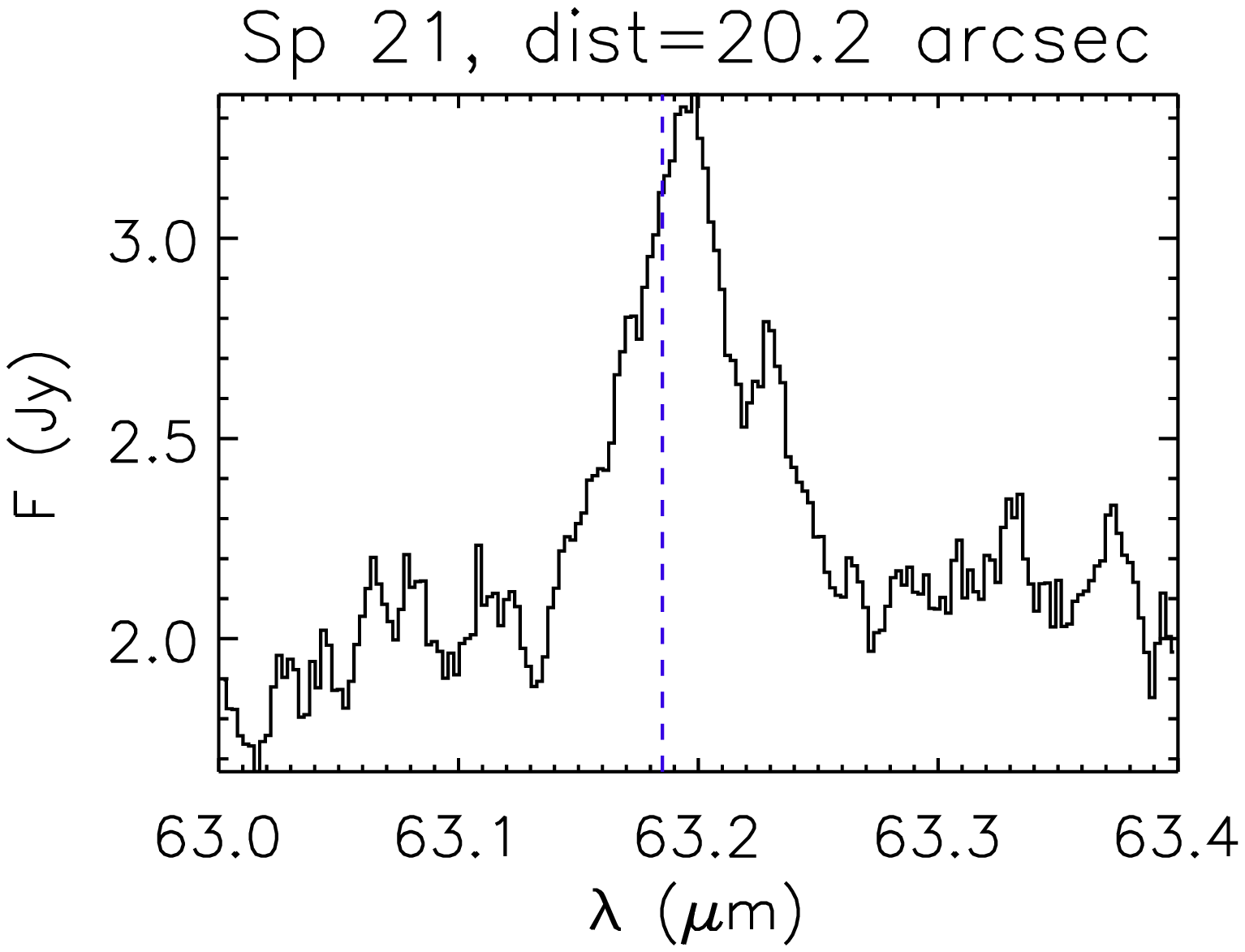}
     \includegraphics[scale=0.25,trim=30mm 25mm 10mm 0mm,clip]{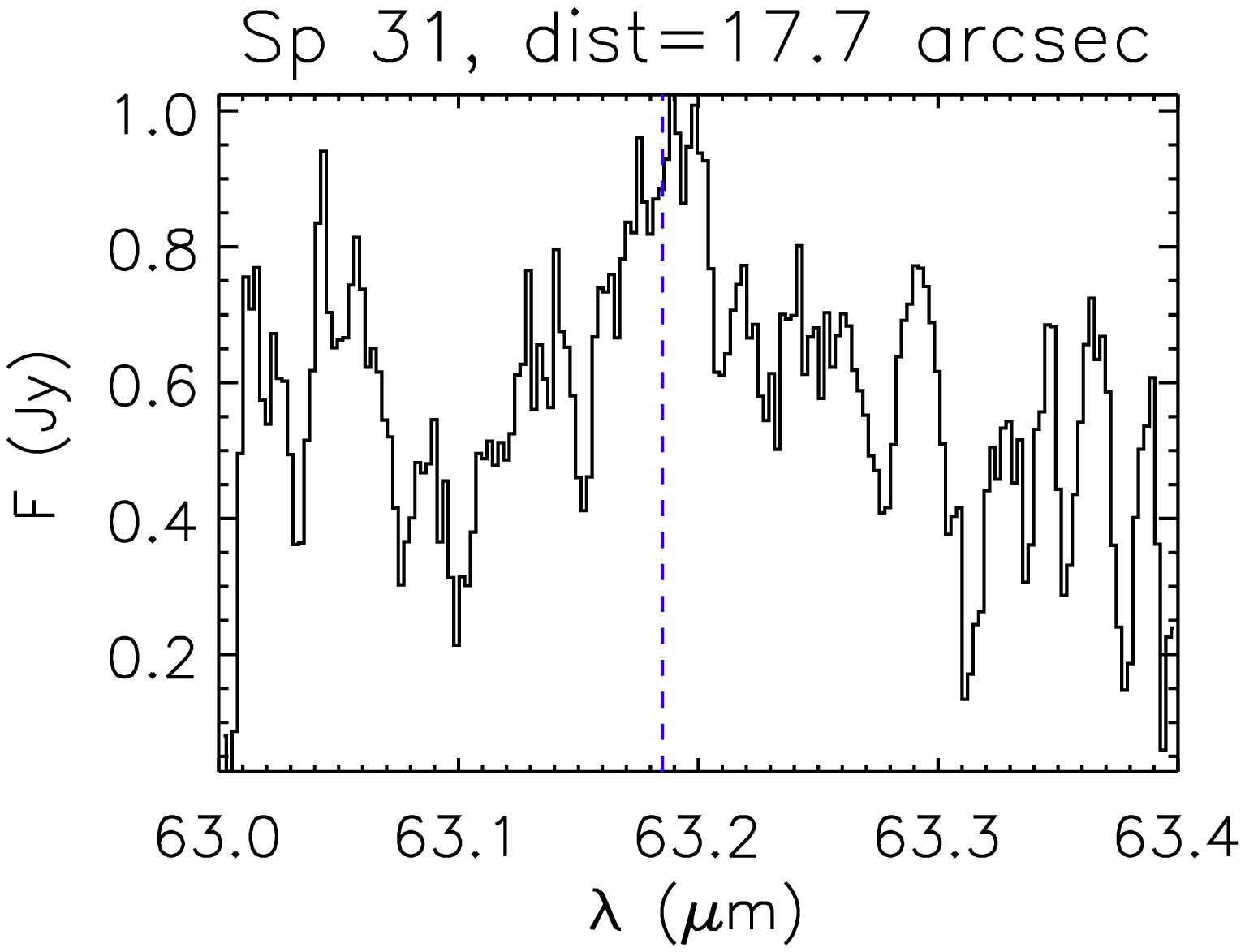}
     \includegraphics[scale=0.25,trim=30mm 25mm 10mm 0mm,clip]{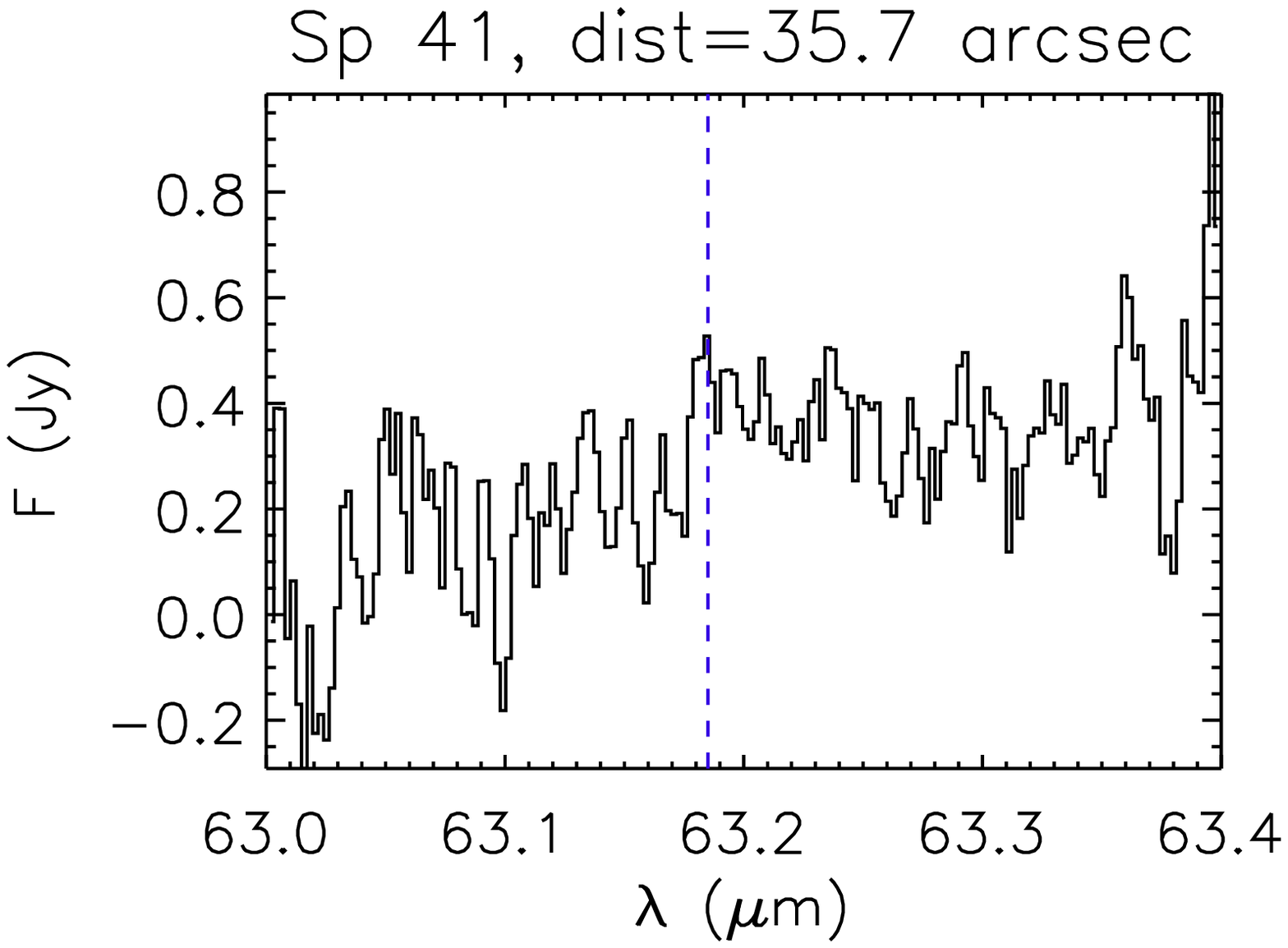}\\
     \includegraphics[scale=0.25,trim=10mm 25mm 10mm 0mm,clip]{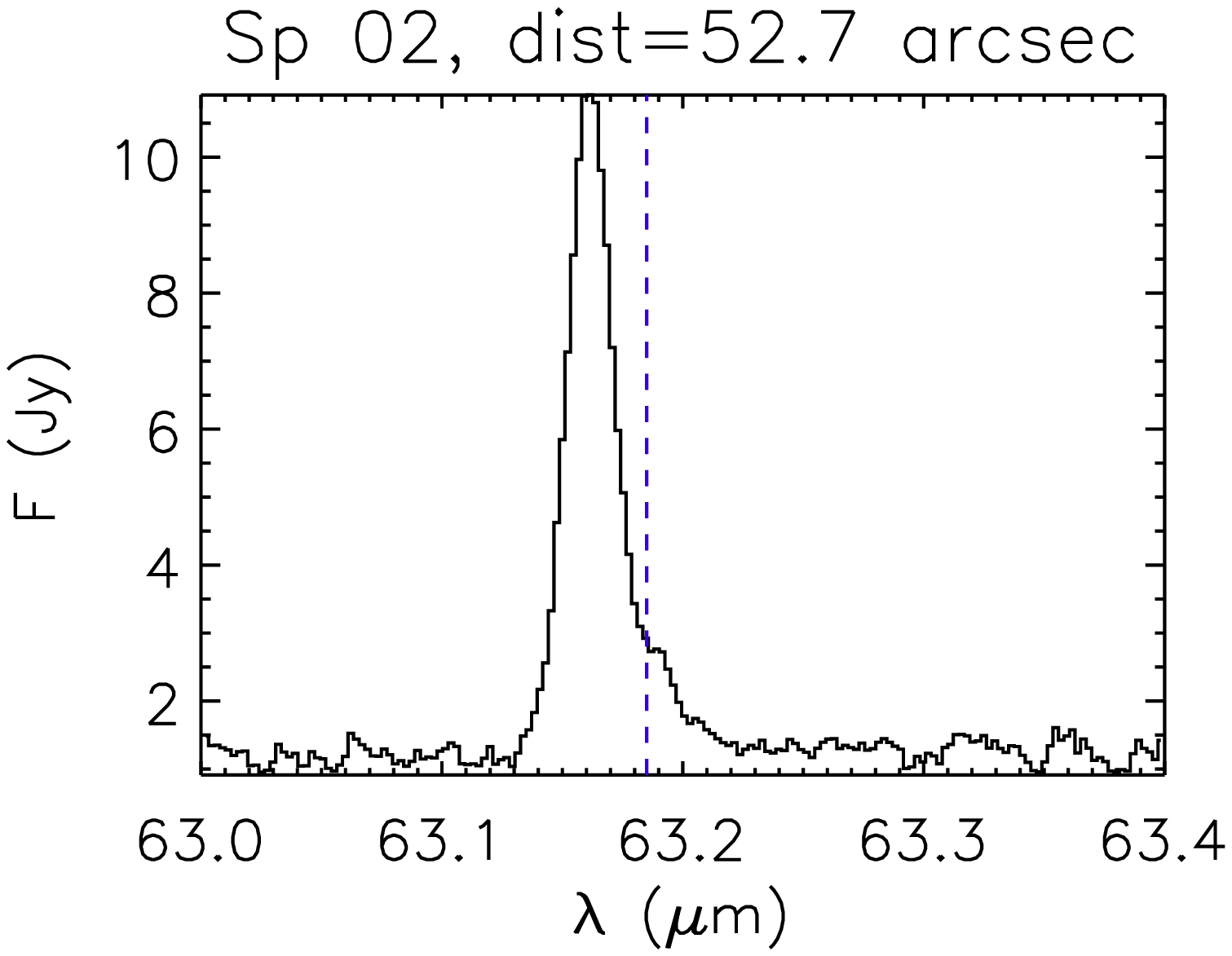}
     \includegraphics[scale=0.25,trim=30mm 25mm 10mm 0mm,clip]{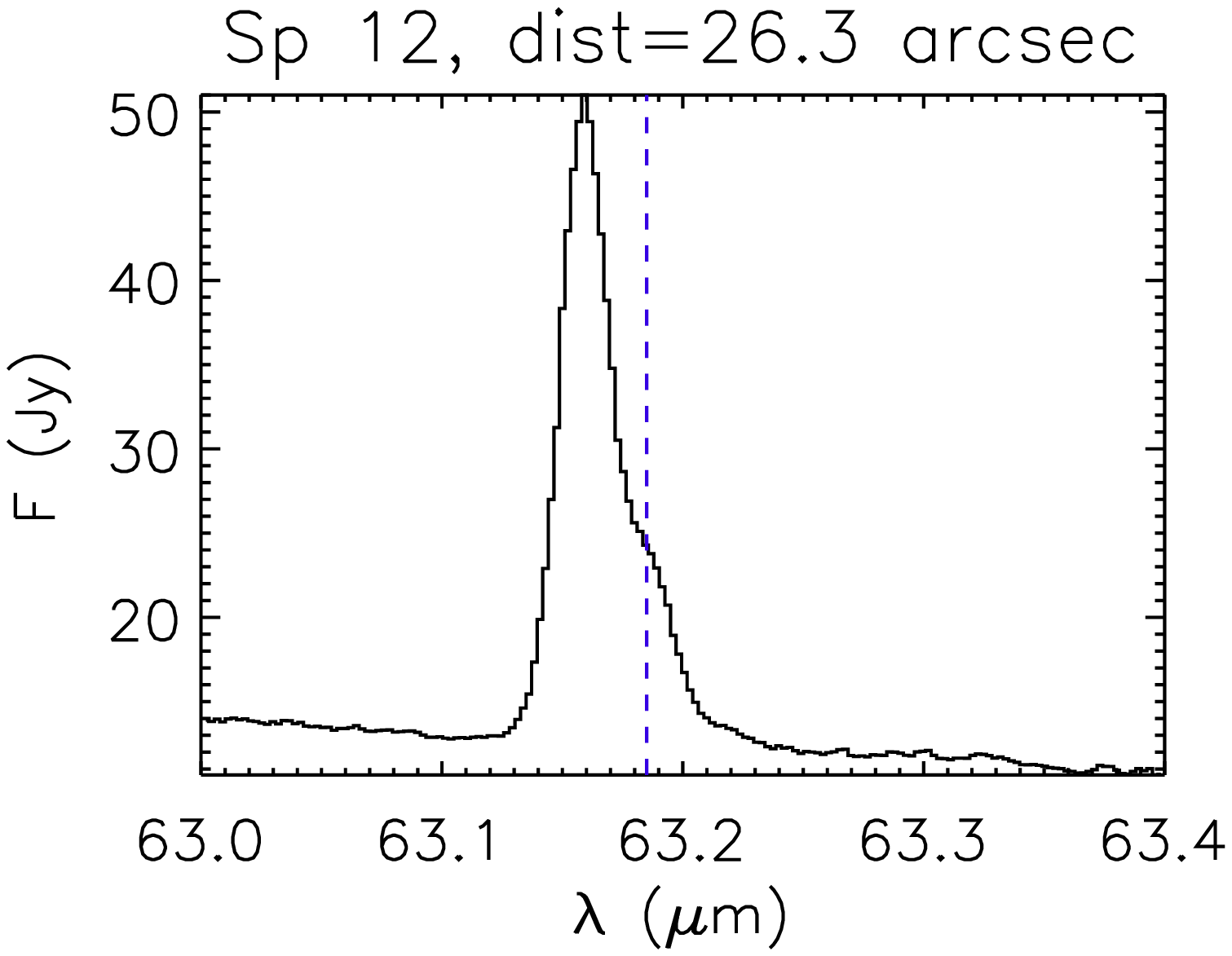}
     \includegraphics[scale=0.25,trim=30mm 25mm 10mm 0mm,clip]{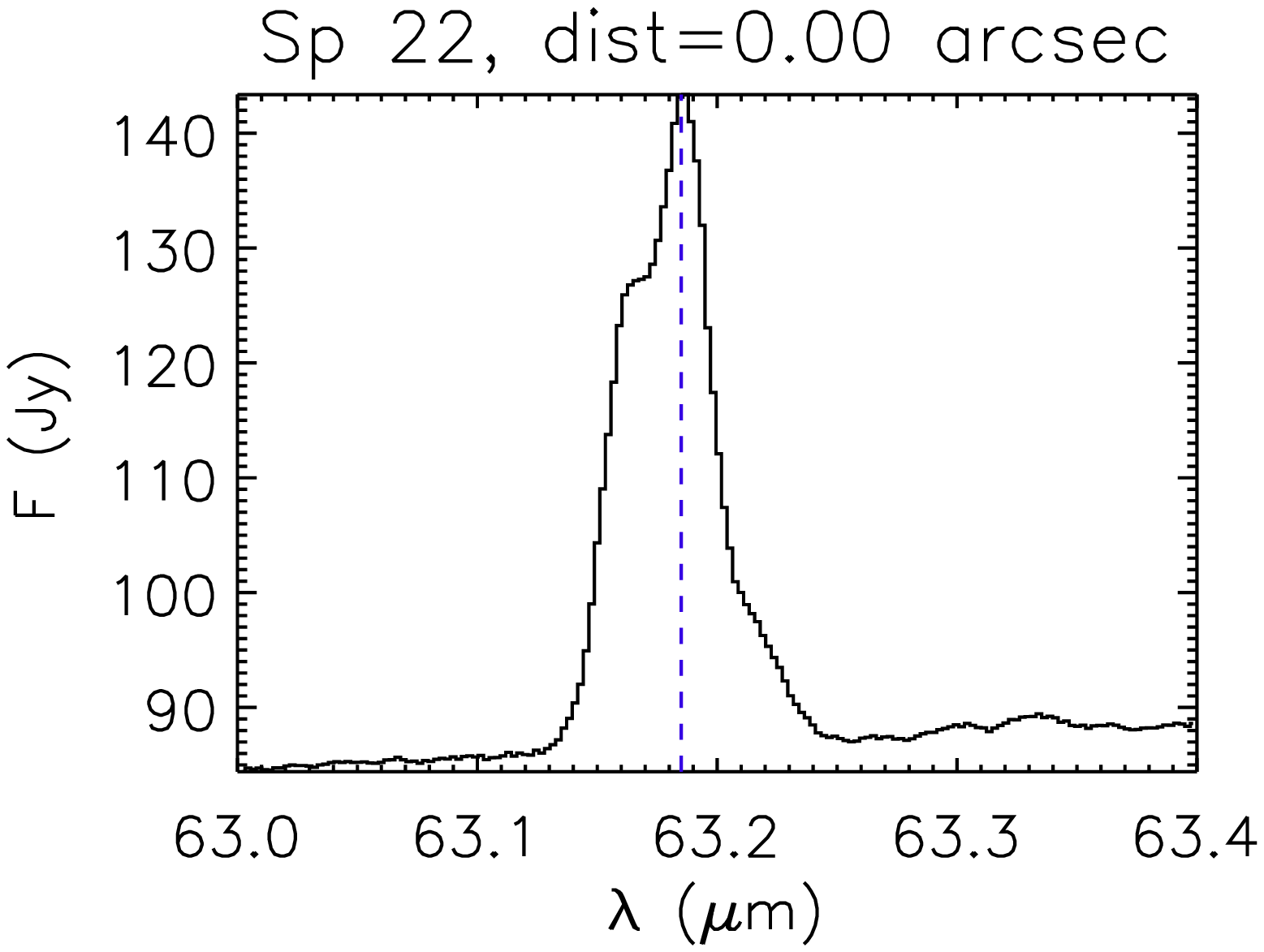}
     \includegraphics[scale=0.25,trim=30mm 25mm 10mm 0mm,clip]{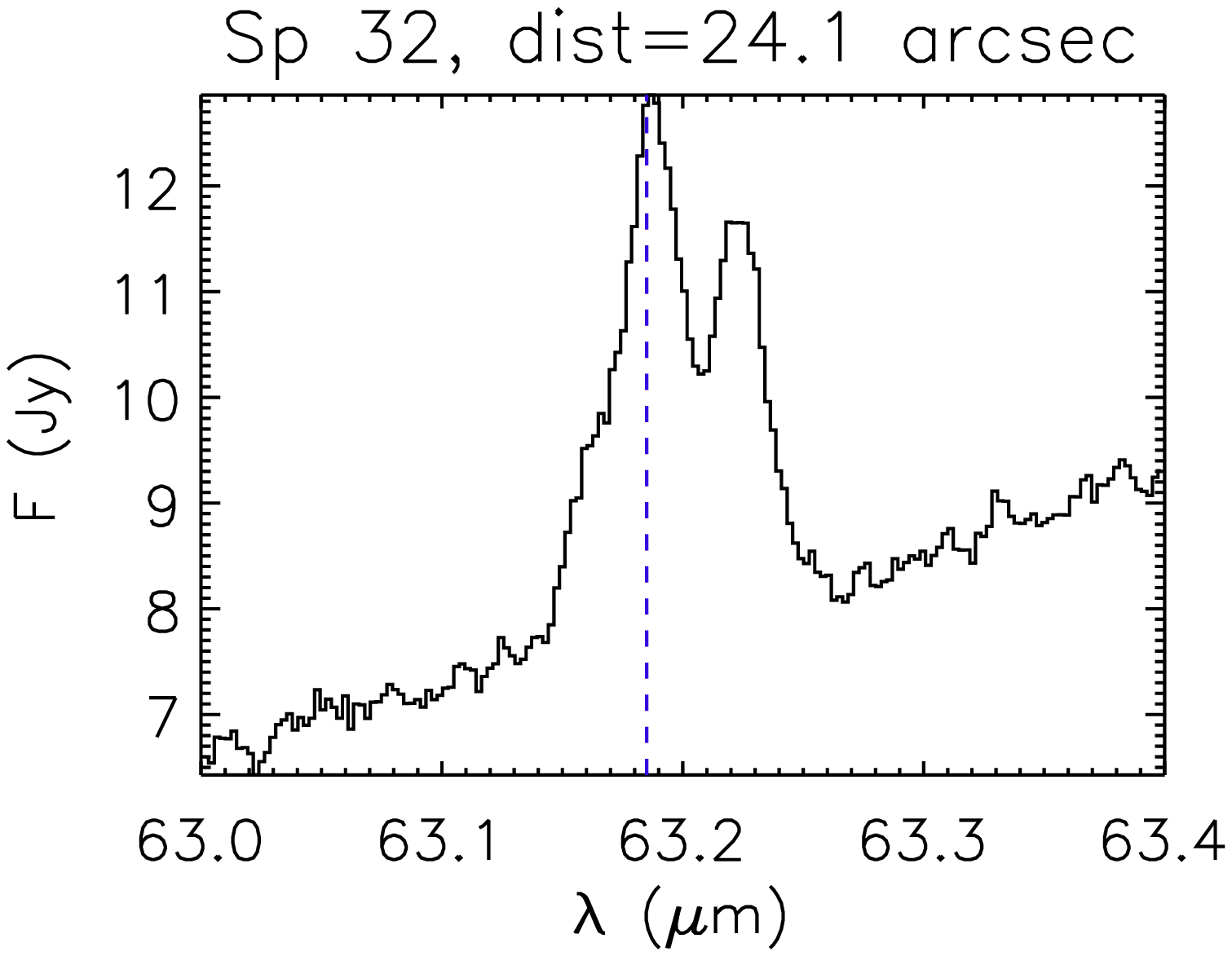}
     \includegraphics[scale=0.25,trim=30mm 25mm 10mm 0mm,clip]{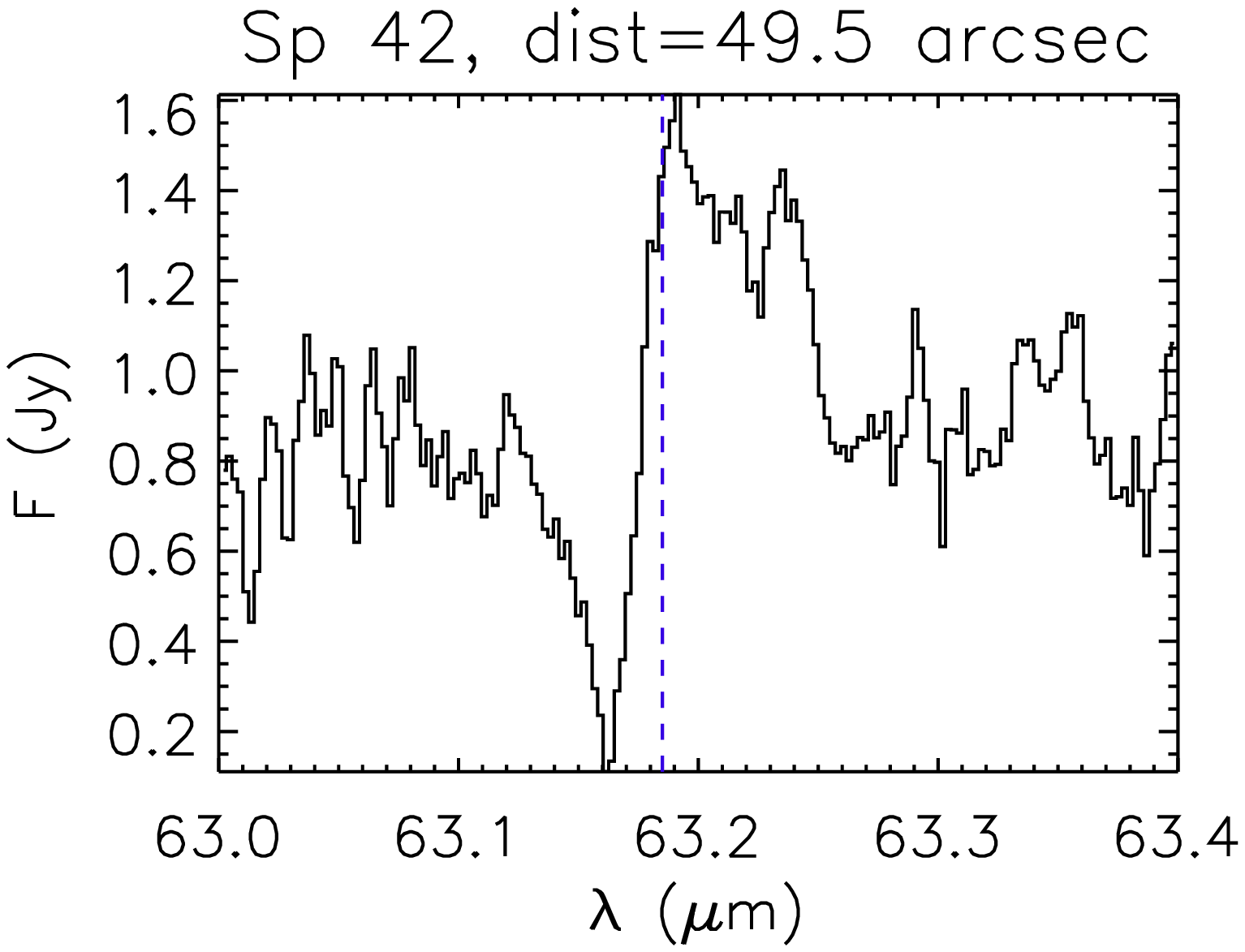}\\
     \includegraphics[scale=0.25,trim=10mm 25mm 10mm 0mm,clip]{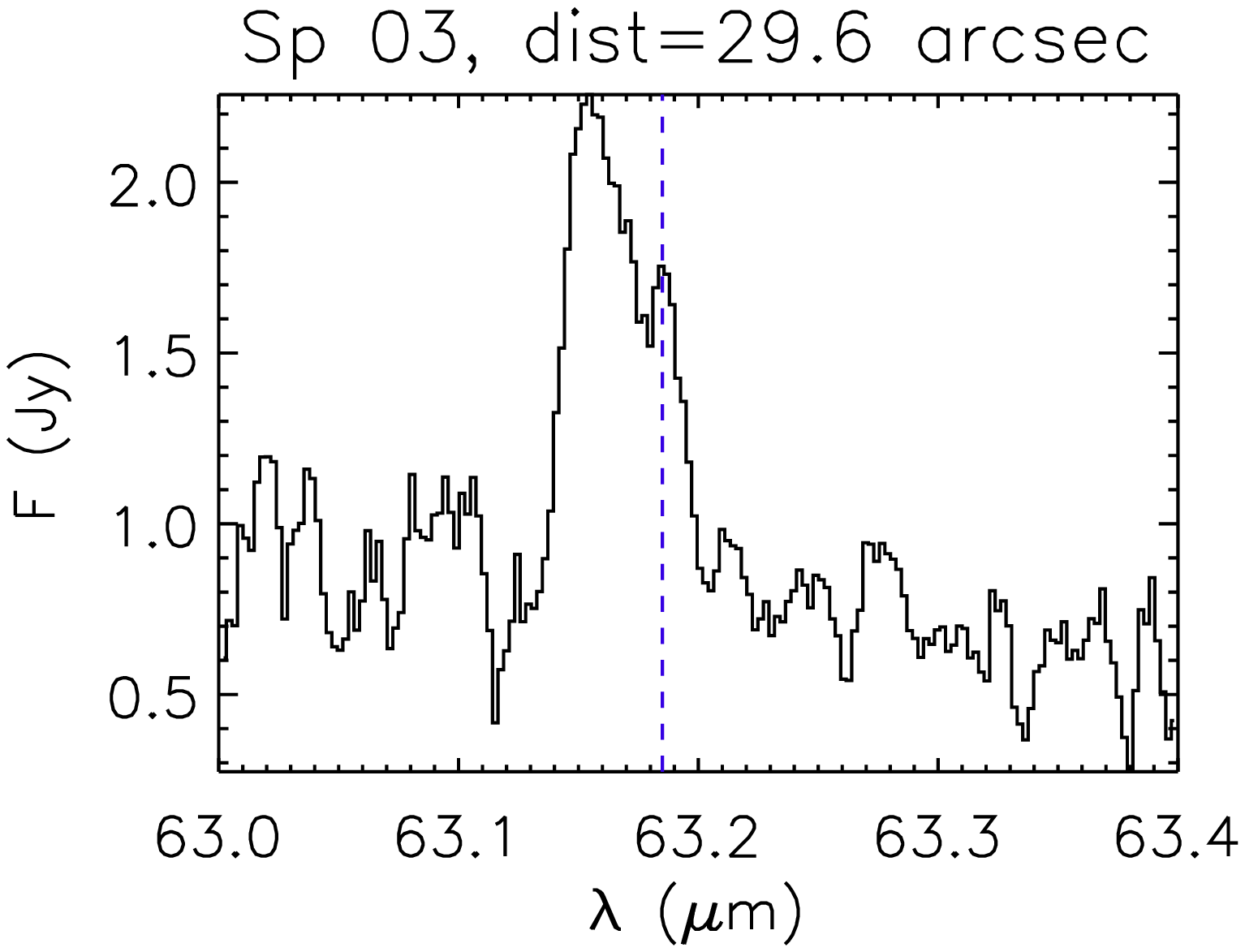}
     \includegraphics[scale=0.25,trim=30mm 25mm 10mm 0mm,clip]{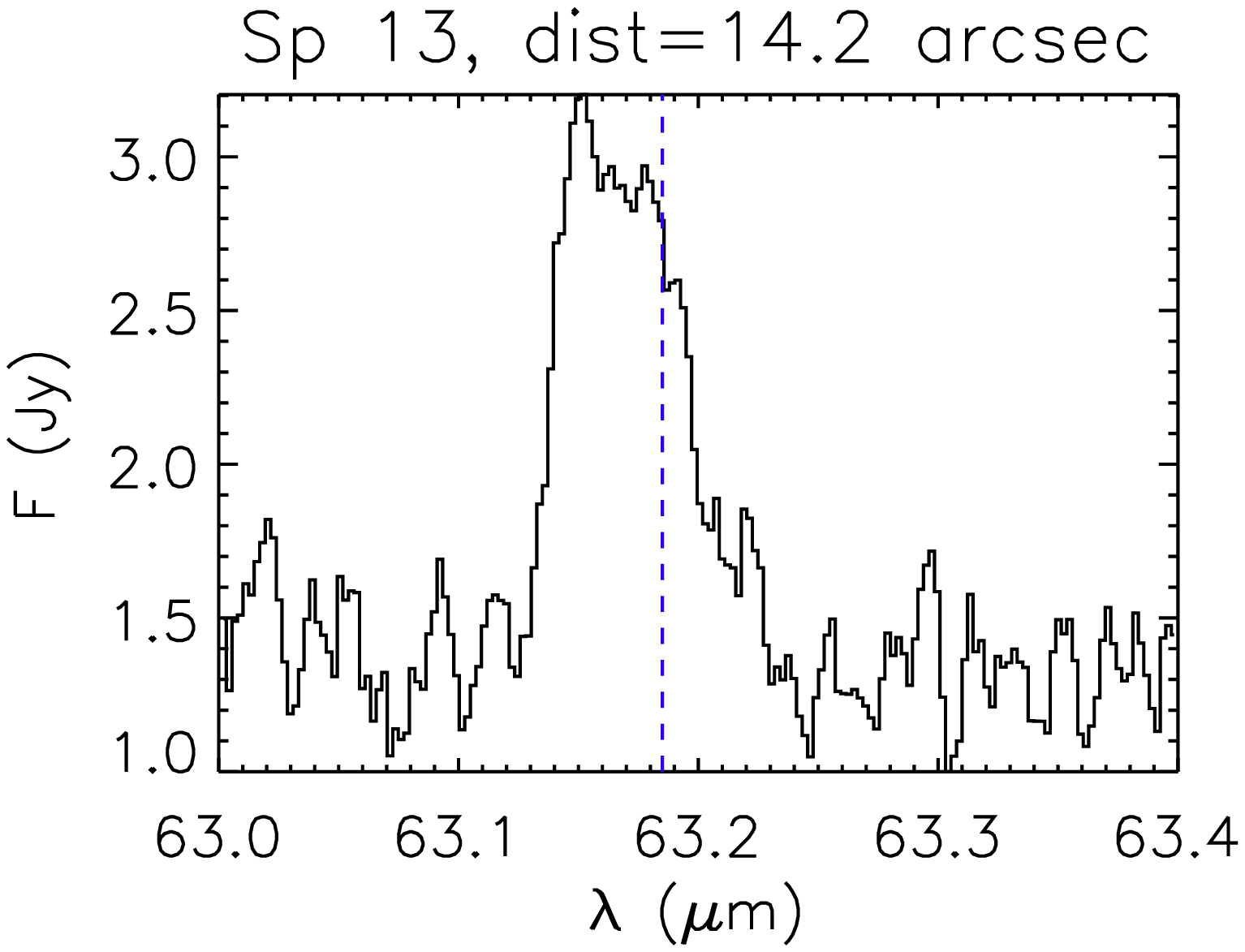}
     \includegraphics[scale=0.25,trim=30mm 25mm 10mm 0mm,clip]{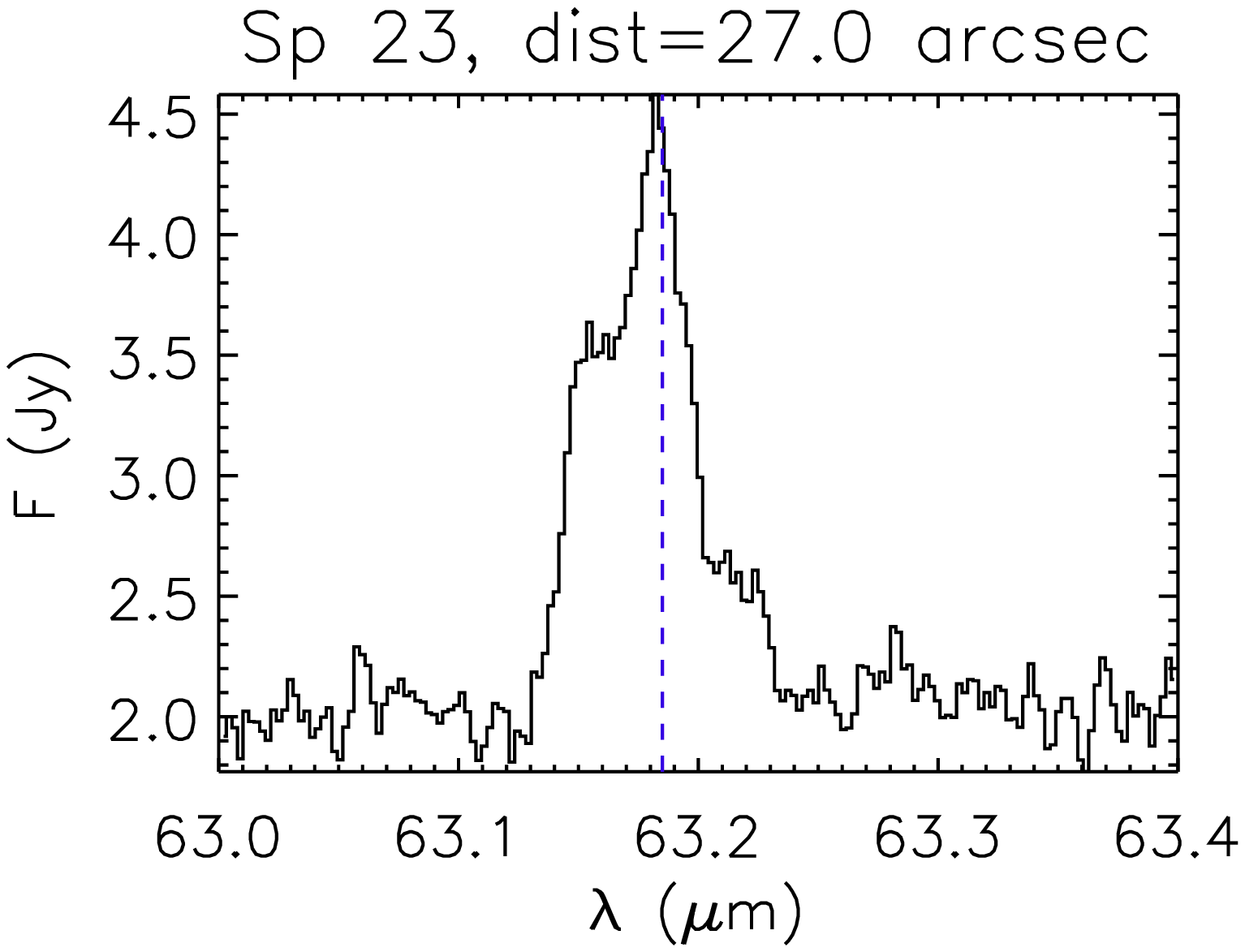}
     \includegraphics[scale=0.25,trim=30mm 25mm 10mm 0mm,clip]{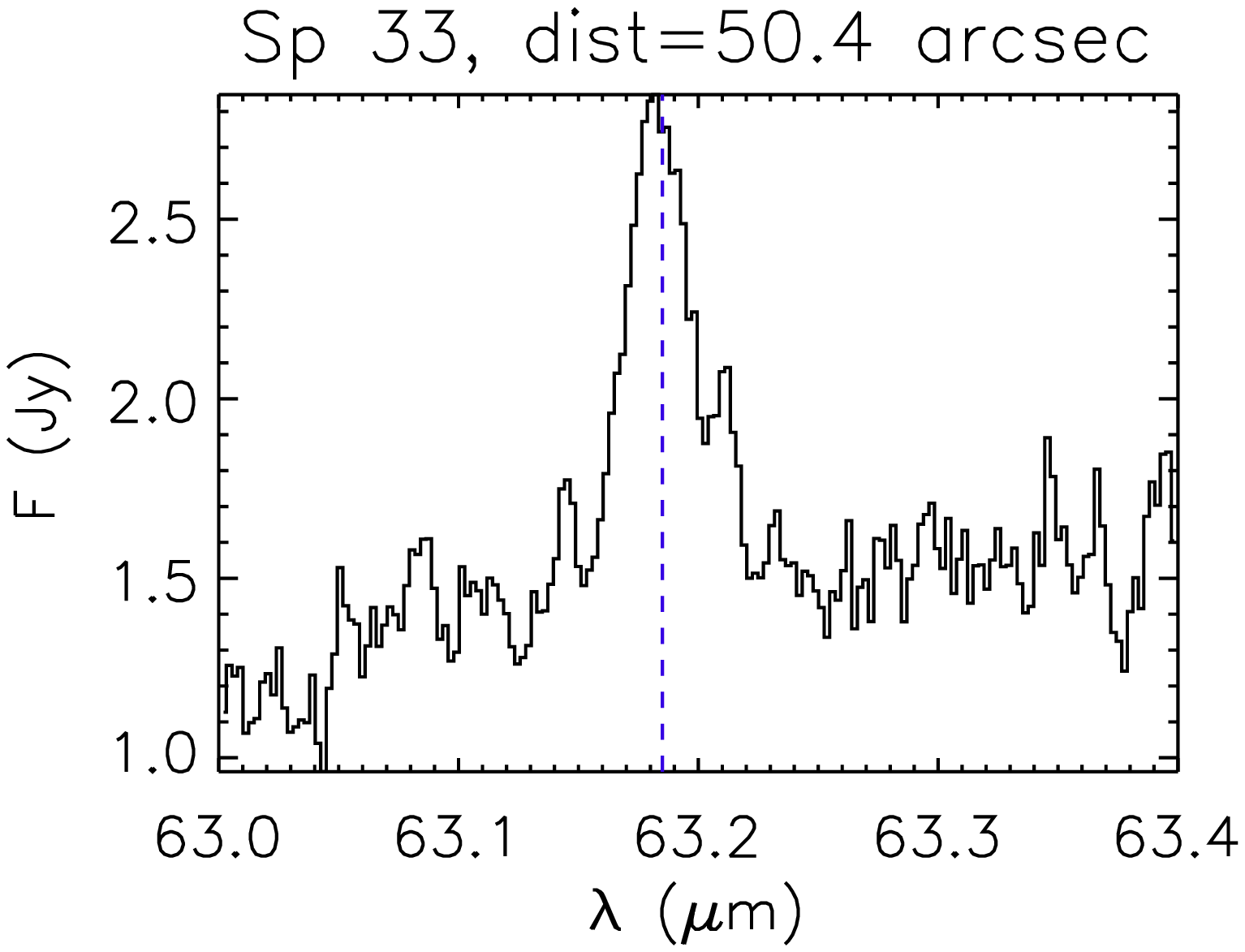}
     \includegraphics[scale=0.25,trim=30mm 25mm 10mm 0mm,clip]{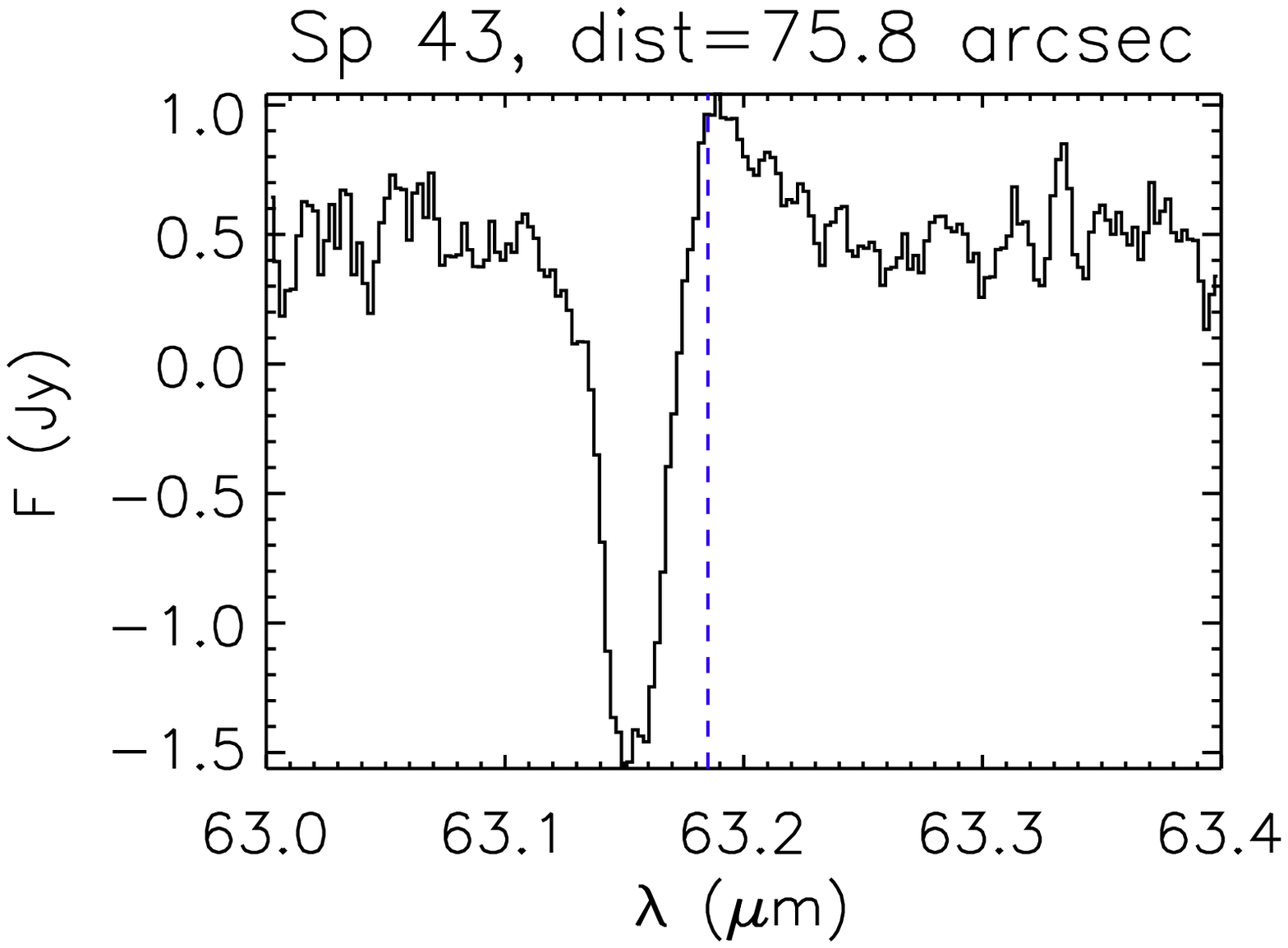}\\
     \includegraphics[scale=0.25,trim=10mm 5mm 10mm 0mm,clip]{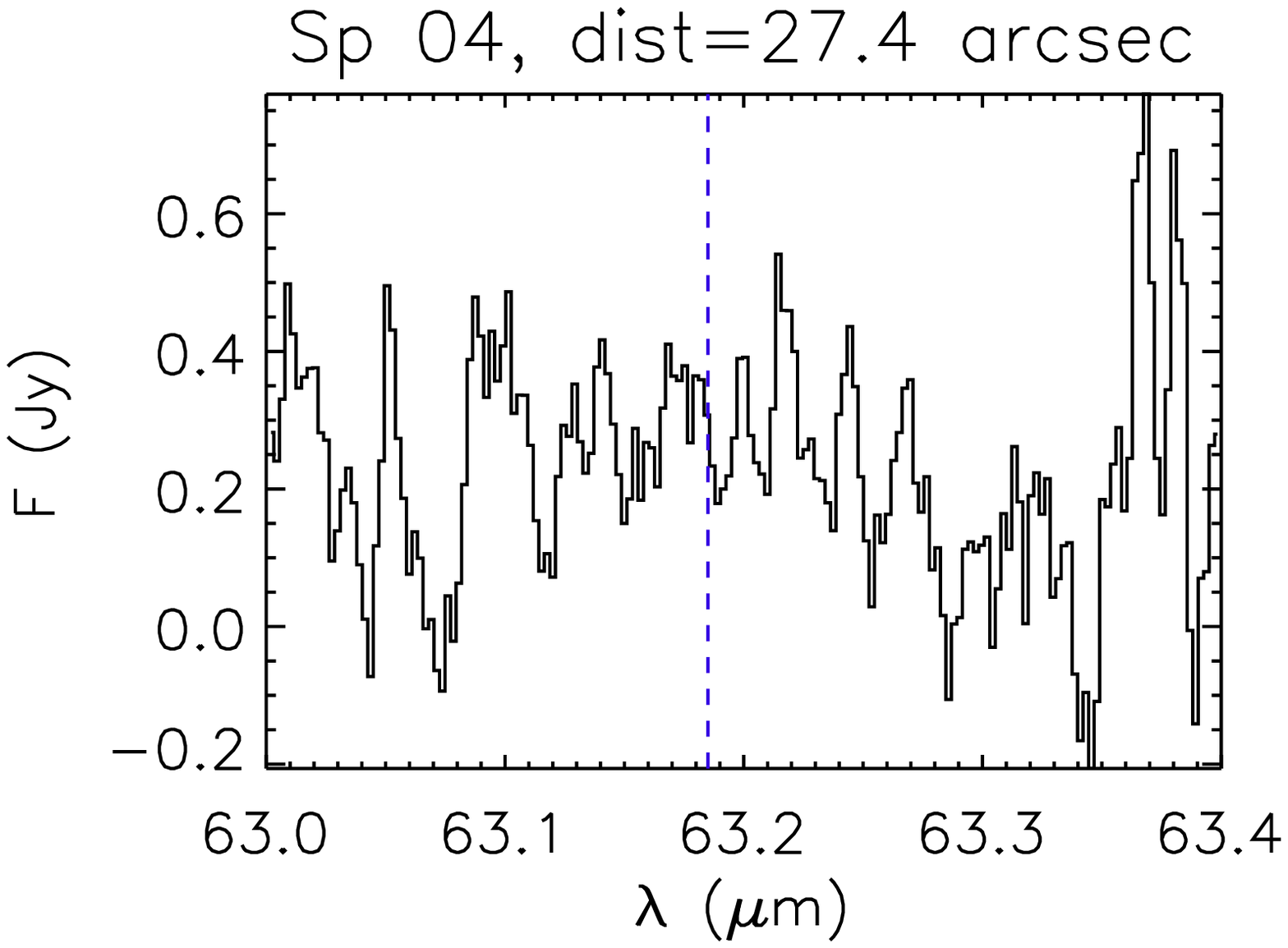}
     \includegraphics[scale=0.25,trim=30mm 5mm 10mm 0mm,clip]{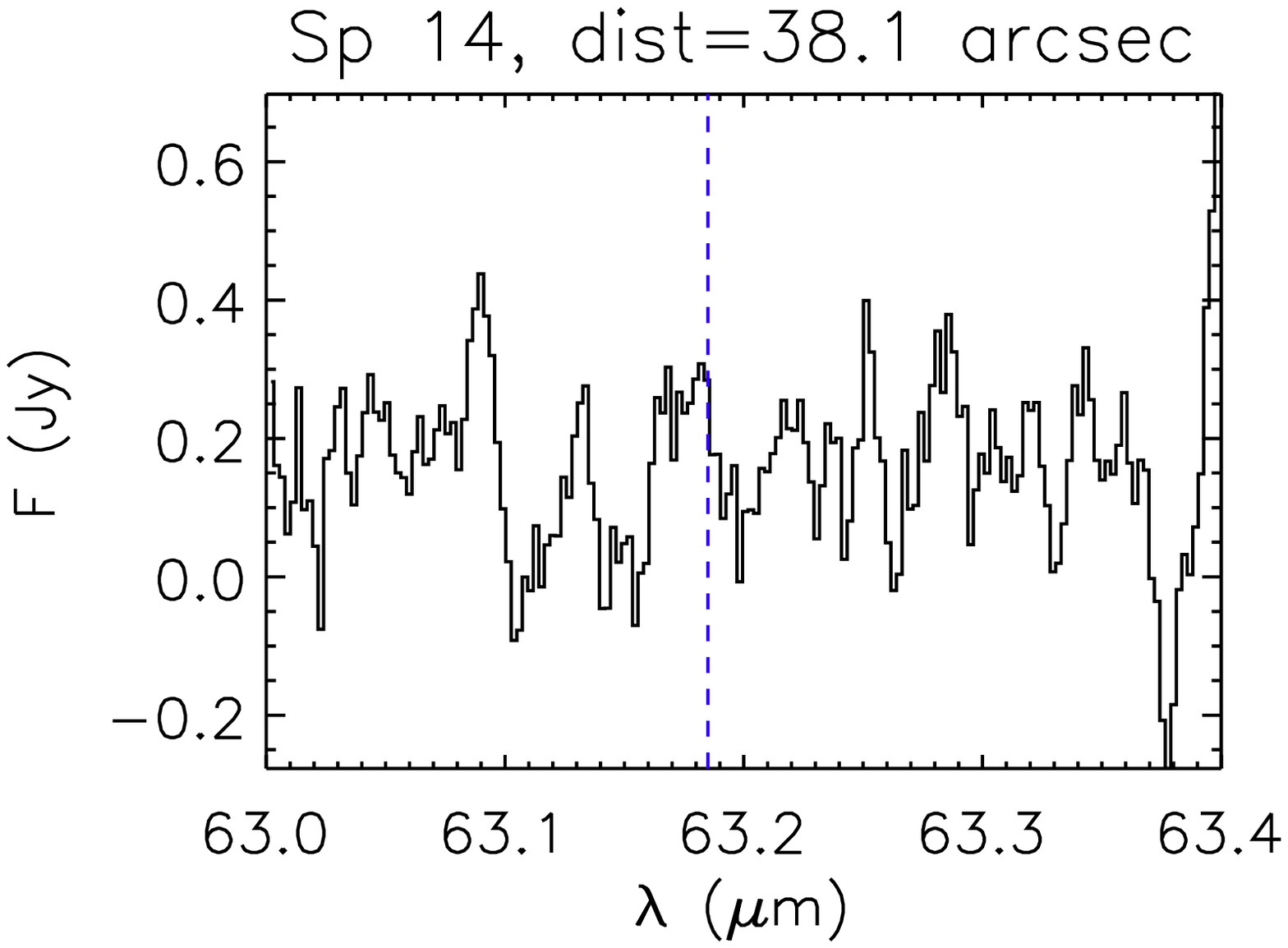}
     \includegraphics[scale=0.25,trim=30mm 5mm 10mm 0mm,clip]{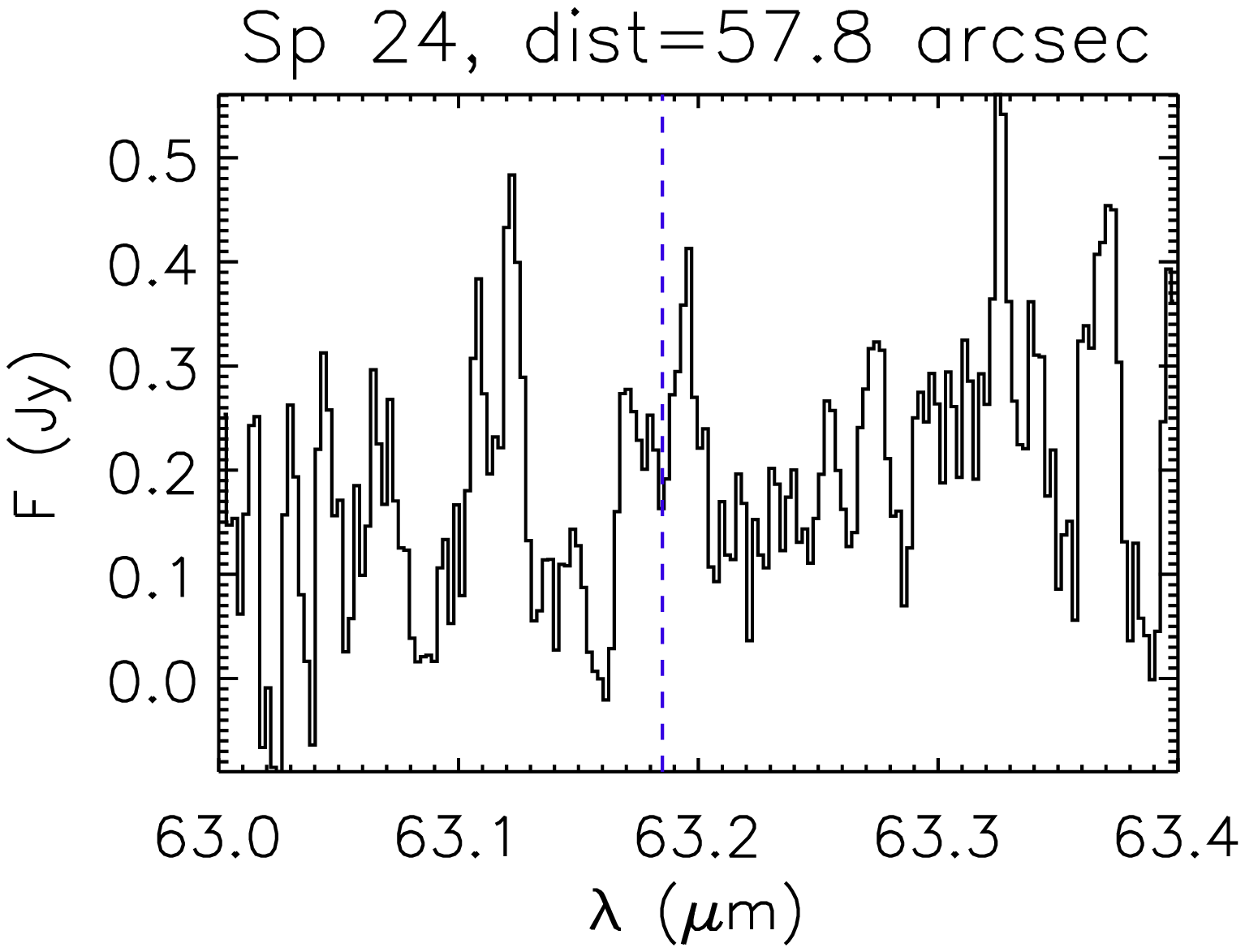}
     \includegraphics[scale=0.25,trim=30mm 5mm 10mm 0mm,clip]{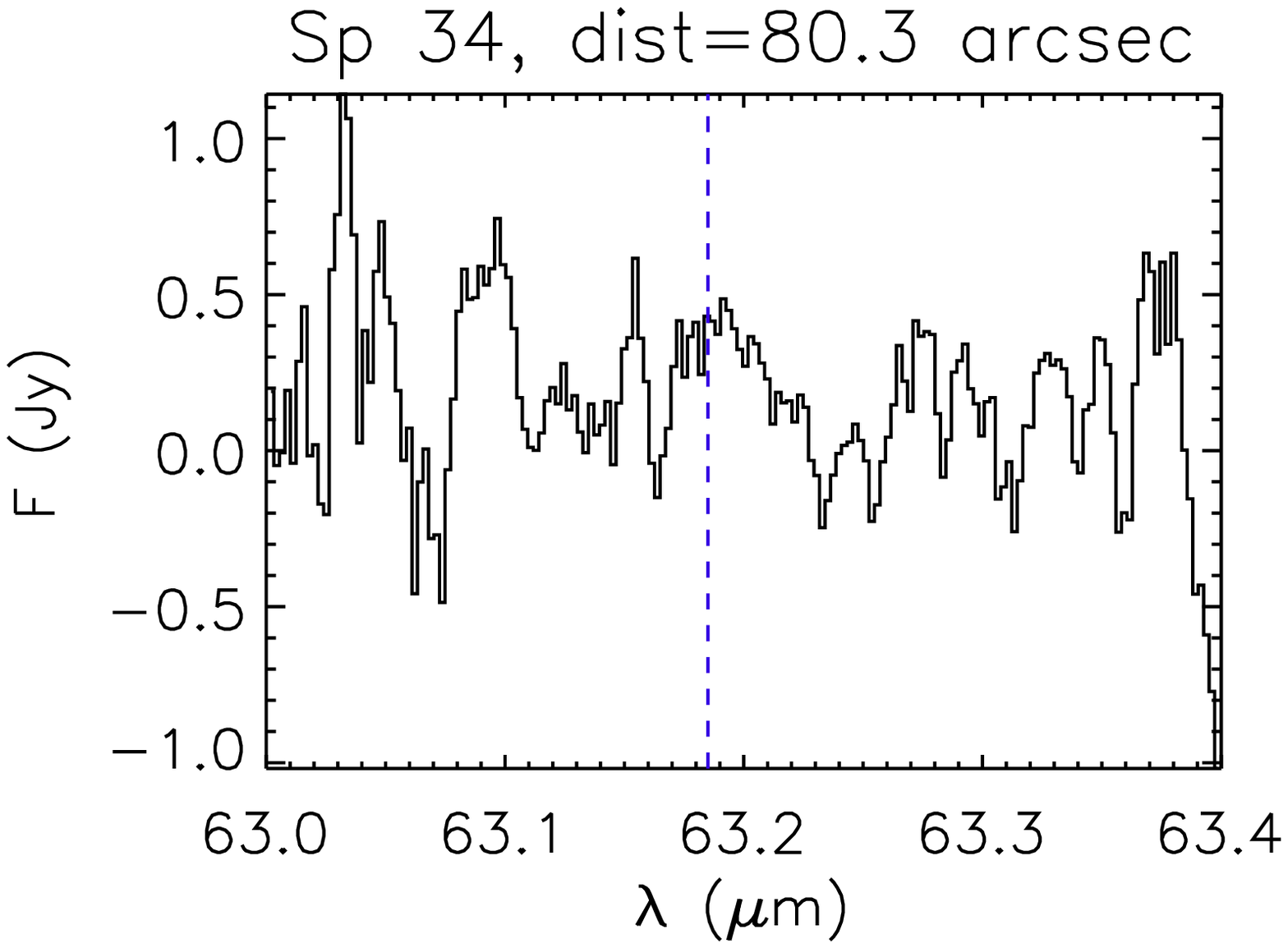}
     \includegraphics[scale=0.25,trim=30mm 5mm 10mm 0mm,clip]{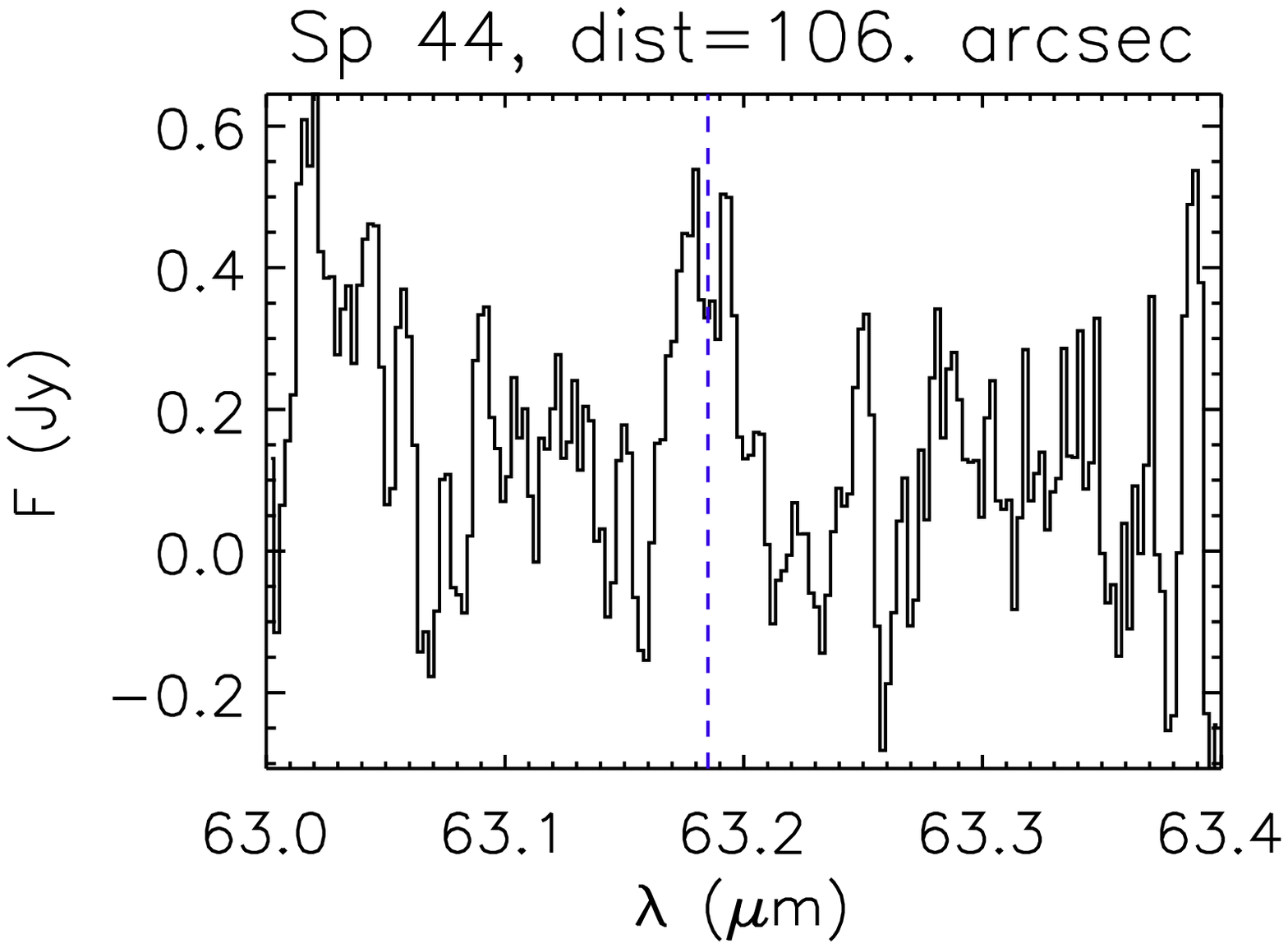}\\
     \vspace{.7cm}
   \caption{Individual spectra for the 25 spaxels for the DK Cha observation. The blue vertical line marks the position of the rest-frame wavelength of the [OI] $\rm ^{3}P_{1} \rightarrow ^{3}P_{2}$ transition.  Spaxel number and distance to the central spaxel are given at the top of each plot.}
   \label{DKCha_spec2}
\end{figure*}

\begin{table}[!t]
\caption{$\rm H_{\alpha}$ equivalent widths from the literature for Cha II members in the sample.}             
\label{Table:eqW}      
\centering          
\begin{tabular}{lcc}     % 6 columns 
\hline\hline       
Name & $\rm W(H_{\alpha})^{a}$ & Accretor$\rm ^{b}$ \\ 
\hline
-- & $\rm \AA$ & --\\ 
\hline                    
\object{DK~Cha}  & 88 & Y \\ 
\object{IRAS~12500-7658} & 20 & Y\\
\object{Sz~46N} & 16 & Y \\
\object{IRAS~12535-7623} & 15 & Y\\
\object{ISO~ChaII~13}   & 101 & Y  \\
\object{Sz~50} & 29 & Y \\
\object{Sz~51}$\dagger$ & 102 & Y  \\        
\object{[VCE2001]~C50} &  36 & Y \\     
\object{Sz~52}  & 48 & Y \\    
\object{Hn~25} & 24 & Y  \\        
\object{Sz~53}  & 46 & Y  \\        
\object{Sz~54} & 23 & Y  \\        
\object{J13052169-7738102}  & 29 & --  \\       
\object{J13052904-7741401}  & -- & ? \\             
\object{[VCE2001]~C62}     & 34  & Y \\                 
\object{Hn~26}$\rm ^{*}$ & 10  & N? \\                        
\object{Sz~61}  & 84 & Y \\                       
\object{[VCE2001]~C66} & 30 & Y \\                        
\object{Sz~62}  & 150 & Y  \\                       
\hline                  
\end{tabular}
\tablefoot{(a): $\rm H_{\alpha}$ equivalent widths taken from \cite{Spezzi2008}. (b): according to the criterion by \cite{Barrado2003}. All ClassI and Flat sources are listed as accretors by default. }
\end{table}

\cite{Barrado2003} show how the equivalent width of the $\rm H_{\alpha}$ line can be used to classify young stellar objects (YSO) as accreting or non-accreting depending on the spectral type, what is called the saturation criterion. In Fig. \ref{ChaII_Acc} we show the $\rm H_{\alpha}$ equivalent width versus spectral type for Class II sources in Cha II that were observed with PACS and have available $\rm H_{\alpha}$ measurements. The compilation of equivalent widths from the literature is shown in Table \ref{Table:eqW}.  Most Cha II members are classified as actively accreting, the only exception being Hn 26, which is in any case very close to the accretion threshold. The systems ISO-Cha II 13 and Sz 53 showed $\rm H_{\alpha}$ equivalent widths that are in clear agreement with active accretion, but no [OI] was detected. A similar case was discussed in \cite{Riviere2013} for \object{TWA~03A} (\object{Hen~3-600 A}), where a particularly flat geometry was invoked to explain [OI] emission below the detection limit. Another group of objects (namely IRAS 12535-7623, Sz 46N, Hn 25, Sz 50, [VCE2001] C62, [VCE2001] C66, [VCE2001] C50) show $\rm H_{\alpha}$ equivalent widths that are just above the limit for accretors, but do not show [OI] emission. While some of them could also be explained in terms of flat discs, an alternative explanation would be that the $\rm H_{\alpha}$ measurements were taken during a period of extreme stellar flaring activity \citep[see][for examples of disc-less stars with extreme $\rm H_{\alpha}$ activity]{Bayo2012}. However, it is very unlikely that all of them correspond to periods of intense flaring. Besides, if accretion is highly episodic, no correlation should be found between gas emission and accretion indicators.

\subsection{DK Cha line emission}\label{DKCha:ext}

\begin{figure*}[!t]
\centering
\includegraphics[scale=0.85,trim=-10mm 0mm 0mm 0mm,clip]{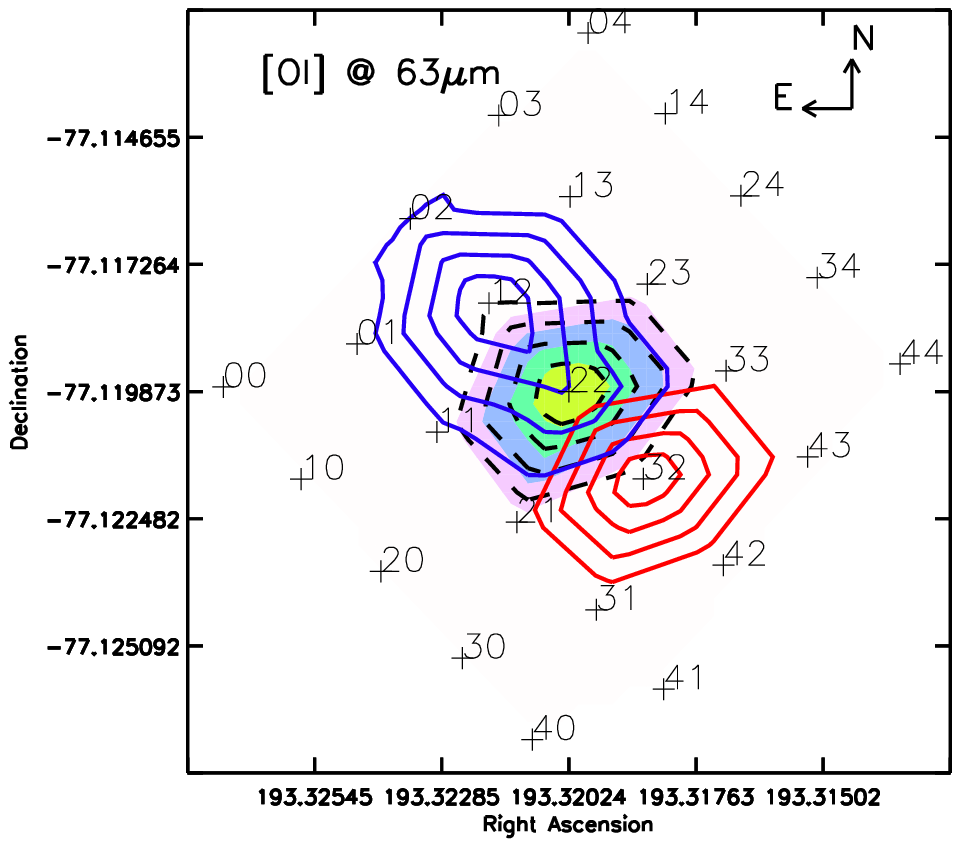}\includegraphics[scale=0.85,trim=12mm 0mm 0mm 0mm,clip]{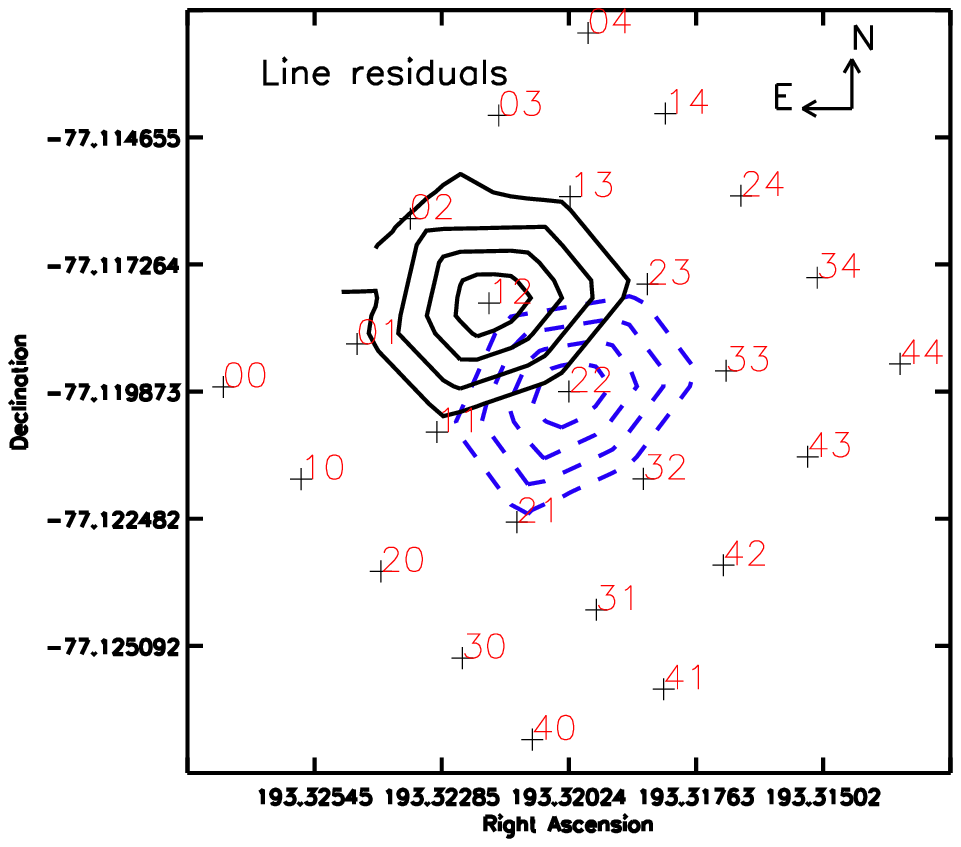}\\
\caption{Left: contour plot for line and continuum emission. Black dashed lines are contours for the low-velocity component. The high velocity component is shown in blue for  blue-shifted emission and red for red-shifted emission. The continuum distribution is shown with filled, coloured contours. Right: contour plot for line residual emission \citep[see Appendix B in][]{Podio2012}. The blue dashed lines are contours for the continuum emission, while the solid black ones are contours for the line residuals. In both panels, the position of individual spaxels is shown as a net of black crosses labelled according to the spaxel ID. The contour levels shown are at the 80\%, 60\%, 40\% and 20\% of the peak level in each case.}
   \label{Fig:DKCha_line_to_cont_ratio}
\end{figure*}

It has been reported that DK Cha has extended [OI] by \cite{Green2013}. In our observations, line emission is detected toward 12 out of 25 spaxels, covering an area of $\rm \sim 38\arcsec \times 28\arcsec$, or $\rm \sim 6500 \times 4800~AU^{2}$. Extended emission from jets with similar velocities ($\rm \sim 100~km/s$) was previously detected with PACS \citep[see e. g.][]{vankempen2010,Karska2013}.  After co-adding the 25 spaxels, we computed a flux of $\rm (3.84\pm0.01)\times 10^{-15}~W/m^{2}$ in good agreement with the 3.65$\rm \times 10^{-15}~W/m^{2}$ flux computed by \cite{Kempen2010}.

In Fig. \ref{DKCha_spec}, we show the spatial distribution of fluxes and in Fig. \ref{DKCha_spec2}, we represent the individual spectra. In 11 of the 25 spectra, we observe two components at different radial velocity: a high velocity component that can be either red- or blue-shifted plus a low velocity component (with a velocity that is smaller than the PACS spectral resolution of 88 km/s at 63 $\rm \mu m$). Although the spectral resolution is low, we can use the separation of both components to tackle the origin of the emission. The range of velocities is 131 to 234 km/s for the red-shifted high-velocity emission, -70 to -177 km/s for the blue-shifted high-velocity component and -51 to 52 for the low-velocity component. The average separation is $\rm \langle \Delta v \rangle = (136 \pm 31)~km/s$. In the central spaxel we see one component at -126 km/s, and a second one at $\rm \sim$-2 km/s. 

We went on to tentatively identify a red-shifted component in the wing of the line, at a velocity of $\rm \sim$160 km/s. In the spaxels surrounding the central one, we see emission from two components, with a separation in velocity greater than the spectral resolution. The three spaxels southwest of the star show a high velocity component that is redshifted, with velocities in the range 126-222 km/s. The eight spaxels northeast of the star show a high velocity component that is blue-shifted, with velocities in the range 72-176 km/s. Spaxels on the southwest edge (IDs 42 and 43) show [OI] in absorption. However, visual inspection of the source-on and source-off spectra reveals no absorption in the on-position, but instead an emission in the Nod B off-position is detected in both spaxels. The emission in the Nod B off-position is considerably blue-shifted. Furthermore, the Nod B off-position is aligned with the position of the blue-shifted jet as we discuss later on in this section (see Fig. \ref{Fig:DKCha_line_to_cont_ratio}, left panel). A possibility is then that the source of pollution in the Nod B off-position for spaxels 42 and 43 is the jet itself, and the observed absorption is a reduction artefact. No emission is observed in the Nod B off-position for the other spaxels.

\cite{Podio2012} propose a simple method of detecting extended line emission that makes use of the line-to-continuum ratio in the different spaxels compared to that in the central one. When the line originates in an extended region, or in a region that is offset with respect to compact continuum emission, we expect the line-to-continuum ratios to be higher in the outer spaxels than in the central one. See Appendix B in \cite{Podio2012} for a detailed description of the method. We applied the test to the residual of the total line emission. We show the resulting line residual contour in the right-hand panel of Fig. \ref{Fig:DKCha_line_to_cont_ratio}. The distribution of residuals  is clearly shifted to the north-east with respect to the region containing the continuum emission. To test whether the continuum emission is extended, we compared its spatial distribution with that of a model PSF at 63 $\rm \mu m$. Both the observed continuum and the model PSF show very similar spatial distributions, and therefore we conclude that continuum emission is not extended at 63 $\rm \mu m$.

In the left panel of Fig. \ref{Fig:DKCha_line_to_cont_ratio} , we show  the spatial distribution of the different velocity components, with the blue- and red-shifted high-velocity components coloured accordingly, together with the distribution of the low velocity component and the continuum. Observations of [OI] emission at 6300 $\rm \AA$ in T Tauri stars usually show two velocity components, where the low-velocity component is brighter and less extended than the high-velocity one \citep{Kwan1988,Hartigan1995}. According to \cite{Kempen2009}, the DK Cha disc is seen almost face-on ($\rm i<18^{\circ}$), i. e., we see the disc through the outflow cone. We attribute the high-velocity component to jet emission in a system with low-inclination. Interestingly, the high-velocity component in the central spaxel ($-126\pm88$ km/s) and the average velocity of the high velocity component in blue-shifted spaxels ($\rm -124 \pm 95 ~km/s$) are roughly consistent with the velocity for [OI] emission at 6300 $\rm \AA$ (-102 km/s) derived by \cite{Hughes1991}, which was attributed to outflow emission. The optical and near-IR emission spectrum of DK Cha was studied in detail by \cite{GarciaLopez2011}, who detected several permitted, forbidden and molecular emission lines and propose a simple model of an expanding spherical envelope of gas to reproduce the observations. The same wind might be the origin of the high-velocity component detected with PACS, although the velocity of the wind derived by \cite{GarciaLopez2011} is too high ($\rm V_{max}=300~km/s$) when compared with velocities that we observe. 

The origin of the low-velocity component is harder to face. The spatial distribution for this component coincides with the continuum distribution (see left panel of Fig. \ref{Fig:DKCha_line_to_cont_ratio}). A possible origin for the emission is the envelope. However, the temperature in the envelope is too low to passively excite the line ($\rm E_{up}=228~K$). \cite{Visser2012} studied CO and $\rm H_{2}O$ emission in a sample of embedded objects, including DK Cha, and conclude that a combination of passive heating of the envelope, plus UV heating of the cavity walls and C-type shocks along the cavity walls could explain the observations. Another possibility is that the low-velocity [OI] component is due to a low-velocity wind like those proposed by \cite{Hartigan1995} to explain [OI] emission at 6300 $\rm \AA$ in T Tauri stars. 
In \cite{Rigliaco2013}, the low-velocity component at 6300 $\rm \AA$ is explained as the result of a photo-evaporative wind from a disc layer where FUV photons can dissociate OH.
There could be also a contribution from the disc as seen through the cone of the jet, and therefore the low-velocity component flux from the central spaxel is an upper limit to emission from the disc.

\subsection{Stacking of spectra for non-detected sources}
\begin{figure}[!t]
\begin{center}
%   \centering
     \includegraphics[scale=0.6,trim = 5mm 5mm 0mm 0mm, clip]{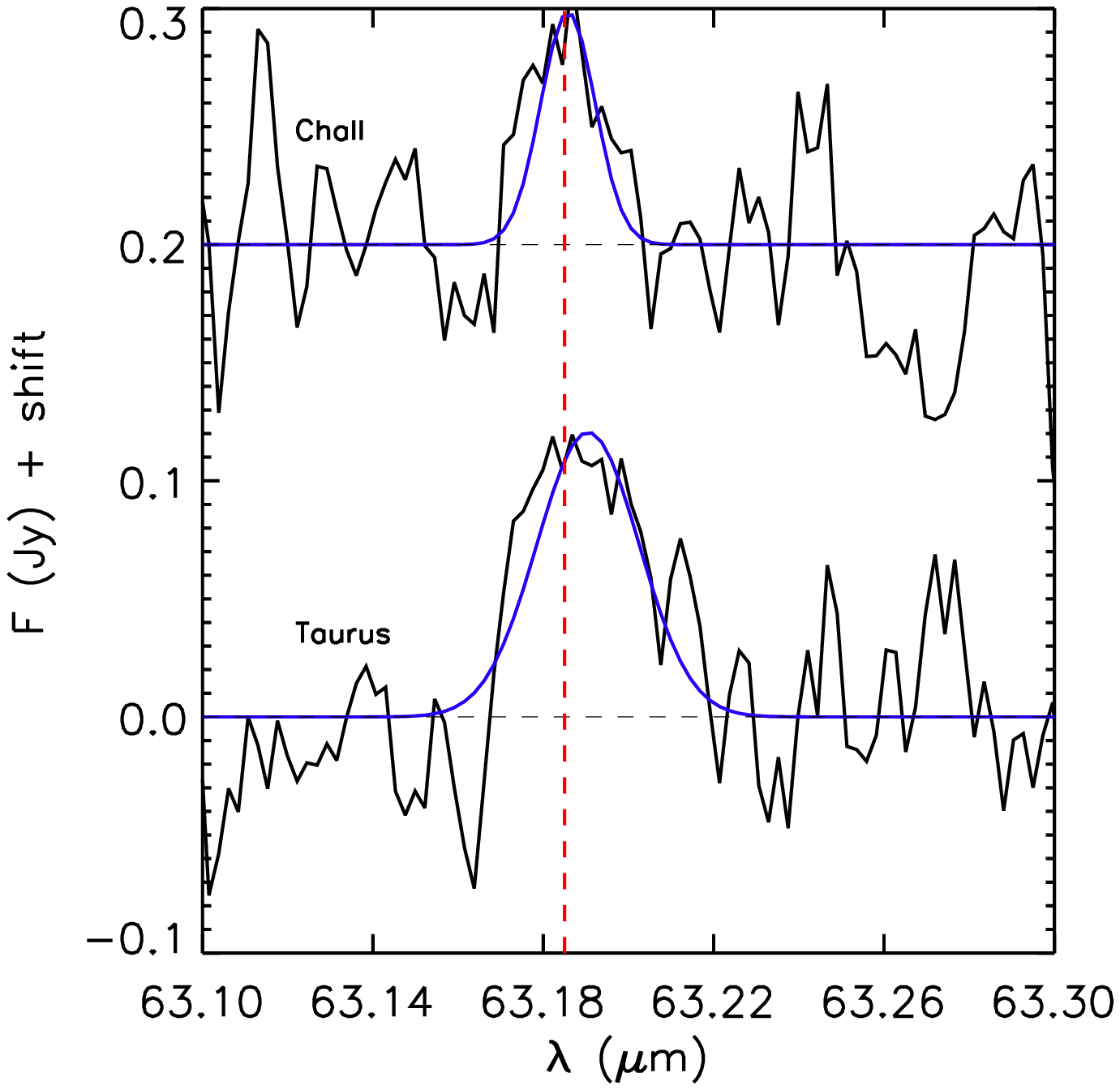} \includegraphics[scale=0.6,trim = 5mm 0mm 0mm 7mm, clip]{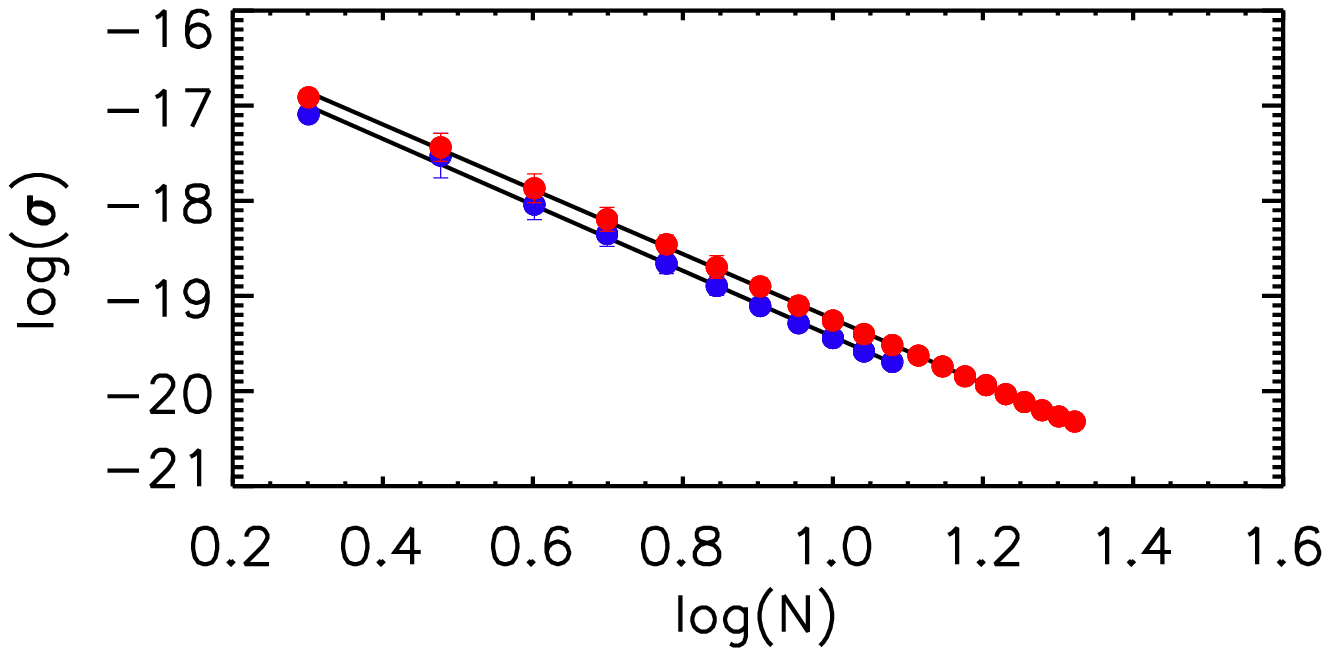}\\ 
     \caption{Top: stacking of PACS spectra at 63 $\rm \mu m$ for objects with no detection in Taurus and Cha II. The blue line shows a Gaussian fit to the emission features. The red dashed vertical line marks the position of the rest frame wavelength for the  [OI]  $\rm ^{3}P_{1} \rightarrow ^{3}P_{2}$ transition at 63.185 $\rm \mu m$. Bottom: continuum noise in the resulting stacked spectra for different numbers of co-added spectra in Taurus (in red) and Cha II (in blue). The average noise and noise error are computed as the average and standard deviation of the N realisations of the bootstrap test (see text).}
   \label{ChaII_oi_stack}
\end{center}
\end{figure}

We proceeded to stack the spectra of individually undetected sources to enhance the detectability of marginal amounts of [OI] emission. Background subtracted spectra were co-added, with a weight equal to one over its noise in the continuum ($\rm \propto \sigma_{cont}^{-2}$). The resulting  spectrum is shown in Fig. \ref{ChaII_oi_stack}. For comparison, we also performed the stacking for Taurus sources \citep{Howard2013}. All the stars used to produce the stacked spectra have infrared excesses, which is indicative of the presence of dust. To prevent any bias that the spectral type can introduce in the results, we included only M stars in our analysis.

By stacking spectra, the noise in the resulting spectrum is supposed to be Gaussian (i. e., inversely proportional to the square root of the number of spectra stacked). Following \cite{Delhaize2013}, we studied the evolution of the noise in co-added spectra versus the number of spectra co-added to assess the influence of non-Gaussian noise. To do the study, we first performed a bootstrap analysis. For each number of possible combinations of spectra, n, we define N as:
\begin{equation}
N=n \times ln(n)^{2}
\end{equation}
\citep[see][]{Feigelson2012} and create N random samples of n elements (n spectra from the sample), allowing for repetition. The n spectra were then stacked for each of the N realisations of the bootstrap test, and the noise in regions with no possible contamination from the lines was computed. The average noise of all the N realisations was used as the real estimation of the noise of the spectrum resulting in stacking n spectra. The bootstrap is helpful for making our analysis strong against selection effects. The resulting evolution is shown in the bottom panel of Fig. \ref{ChaII_oi_stack} and demonstrates that our analysis is not influenced by non-Gaussian noise.

A Gaussian fit was performed to the stacked spectra. The results from these fits are shown in Table \ref{lineFluxes_stack}, with the number of stacked spectra in each association. There is a marginal 2.6$\sigma$ detection with $\rm F_{[OI]}=1.6 \times 10^{-18} ~W/m^{2}$ for Cha II. The Taurus stack produces a $3\sigma$ detection, $\rm F_{[OI]}=2.8\times 10^{-18} ~W/m^{2}$. These values are very similar, after scaling to a common distance, and suggests that similar small amounts of gas are present in the individually non-detected sources in both Taurus and Cha II. Sensitive observations to search for gas using, for example the Atacama Large Millimetre Array, should prove fruitful.

\begin{table}[!t]
\caption{Line fluxes for stacked spectra of different associations}             
\label{lineFluxes_stack}      
\centering          
\begin{tabular}{lllll}     % 6 columns 
\hline\hline       
Association & Age & $\rm f_{[OI]}$ & FWHM & Number \\ 
\hline
-- & (Myr) & ($\rm 10^{-18}W/m^{2}$) & ($\rm \mu m $) & --\\ 
\hline                    
Taurus$\rm ^{1}$ & 1--3 & 2.8$\rm \pm$0.9 & 0.026 & 21 \\ 
Cha II$\rm ^{2}$ & $\rm 4 \pm 2$ & 1.6$\rm \pm$0.6 & 0.019 & 12 \\       
\hline                  
\end{tabular}
\tablefoot{Ages from: (1)  \cite{Kucuk2010}; (2) \cite{Spezzi2008} }
\end{table}

\section{Summary and conclusions}\label{ref:SumConc}
Using \textit{Herschel}-PACS we have spectroscopically observed 19 Class I and II objects that belong to the Cha II star forming region. The observations intended to detect [OI] gas emission from circumstellar discs to study gas properties in the association. The main results from our survey follow:
\begin{itemize}
        \item[1)] We detected [OI] emission at 63.18 $\rm \mu m$ in seven out of 19 objects observed, leading to a detection fraction of $\rm 0.36^{+0.13}_{-0.08}$, which is slightly smaller than the fraction of $\rm 0.56^{+0.05}_{-0.06}$ found for Taurus.
        \item[2)] We detected water emission at 63.32 $\rm \mu m$ towards Sz 61, with the highest ratio compared to [OI] flux to date. We concluded that Sz 61 is likely to drive an outflow, based on previous observations of [OI] at 6300 $\rm \AA$.
        \item[3)] Cha II sources, like Taurus members, follow a correlation of [OI] and continuum emission at 70 $\rm \mu m$. In this correlation, outflow sources typically show larger [OI] flux for the same continuum flux.
        \item[4)] DK Cha shows extended [OI] emission with two different components: a low- and a high-velocity one with blue- and red-shifted lobes. We attributed the high-velocity component to a jet. The low-velocity component in the central spaxel  is attributed to a combination of disc emission seen through the cone of the jet in an almost face-on disc, plus a contribution from the envelope and stellar/disc winds.       
        \item[5)] The stacking of Cha II spectra for non-detected sources leads to a marginal 2.6$\sigma$ detection that results in a mean flux of $\rm F_{[OI], ChaII} = 1.6 \times 10^{-18}~W/m^{2}$, similar to non-detected sources in Taurus when distances are considered.
        
\end{itemize}

\acknowledgements
We thank the referee, Dr. Gregory J. Herczeg, for a very detailed report that helped to significantly improve the quality of the paper. We thank Dr. Alessio Caretti o Garetti for a fruitful discussion about the outflows associated to DK Cha and IRAS 12500-7658. I. K. and P. R. M. acknowledge funding from an NWO MEERVOUD grant. P. R. M. and D. B. also acknowledge funding from AYA2012-38897-C02-01. C. E. and P. R. M. acknowledge funding from AYA2011-26202. LP has received funding from the European Union Seventh Framework Programme (FP7/2007-2013) under grant agreement n. 267251. J.P.W. is supported by funding from the NSF through grant AST-1208911.

\bibliographystyle{aa} % style aa.bst 2
\bibliography{biblio.bib}
\end{document}